\documentclass[useAMS,usenatbib]{mn2e}
\usepackage{setspace}
\usepackage{amsmath,amssymb,amsfonts}
\usepackage{graphicx}
\usepackage{epsfig}
\usepackage{epstopdf}
\usepackage{rotating}
\usepackage{natbib}
\usepackage{multirow}
\usepackage[T1]{fontenc}

\title[Wide-field studies of three early-type galaxies] 
  {The SLUGGS Survey\footnotemark[0]\thanks{http://sluggs.swin.edu.au/}:  The globular cluster systems of three early-type galaxies using wide-field imaging}
\author[S. Kartha et al.]{Sreeja S. Kartha$^{1}$ \thanks{E-mail: skartha@astro.swin.edu.au} 
, Duncan A. Forbes$^{1}$ \thanks{dforbes@astro.swin.edu.au}, Lee R. Spitler$^{1,2,3}$, 
Aaron J. Romanowsky$^{4,5}$, 
 \newauthor Jacob A. Arnold$^{5}$, Jean P. Brodie$^{5}$ \\ 
 $^{1}$ Centre for Astrophysics \& Supercomputing, Swinburne University, Hawthorn VIC 3122, Australia\\
 $^{2}$ Macquarie University, Macquarie Park, Sydney, NSW 2113, Australia \\
 $^{3}$ Australian Astronomical Observatory, PO Box 915, North Ryde, NSW 1670, Australia\\
 $^{4}$ Department of Physics and Astronomy, San Jos\'{e} State University, One Washington Square, San Jose, CA 95192, USA\\
 $^{5}$ University of California Observatories, 1156 High St., Santa Cruz, CA 95064, USA}

\begin{document}
 
\date{Released 2013 October 08}

\pagerange{\pageref{firstpage}--\pageref{lastpage}} \pubyear{2013}

\def\LaTeX{L\kern-.36em\raise.3ex\hbox{a}\kern-.15em
    T\kern-.1667em\lower.7ex\hbox{E}\kern-.125emX}

\newtheorem{theorem}{Theorem}[section]

\maketitle

\label{firstpage}

\begin{abstract}
We present the results from a wide-field imaging study of  globular cluster (GC) systems in three early-type galaxies. Combinations of Subaru/Suprime-Cam,  CFHT/MegaCam and HST/WFPC2/ACS data were used to determine the GC system properties of  three highly flattened galaxies NGC 720, NGC 1023 and NGC 2768. This work is the first investigation of the GC system in  NGC 720 and NGC 2768 to very large galactocentric radius ($\sim$ 100 kpc). The three galaxies have clear blue and red GC subpopulations. The radial surface densities of the GC systems are fitted with S\'{e}rsic profiles, and detected out to 15, 8 and 10 galaxy effective radii respectively. The total number of GCs and specific frequency are determined for each GC system. The ellipticity of the red subpopulation is in better agreement with the host galaxy properties than is the blue subpopulation, supporting the traditional view that metal-rich GCs are closely associated with the bulk of their host galaxies' field stars, while metal-poor GCs reflect a distinct stellar halo.  With the addition of another 37 literature studied galaxies, we present a new correlation of GC system extent with host galaxy effective radius. We find a dependence of the relative fraction of blue to red GCs with host galaxy environmental density for lenticular galaxies (but not for elliptical or spiral galaxies). We propose that tidal interactions between galaxies in cluster environments might be the reason behind the observed trend for lenticular galaxies.
\end{abstract}


\section{Introduction}

Globular clusters (GCs) are present in almost all large galaxies and are good tracers of host galaxy properties \citep{Brodie2006}. They are very compact objects and thus able to withstand the powerful events of galaxy evolution. They are expected to  form during the initial proto-galactic collapse and in gas-rich merging events; as a consequence they trace the field stars that form along with them \citep{Brodie1991, Forbes1996, Cote1998}.  The luminosity and compact size of GCs make them the brightest and most easily identifiable  individual objects out to large ($\sim$ 200 kpc)  galactocentric radii around galaxies \citep{Spitler2012}. This makes them a convenient probe to study galaxy formation at large radii where the surface brightness of the host galaxy stars rapidly drops with increasing radius. 
  
GC systems can be studied using accurate photometry, from which a bimodal nature of the colour distribution is identified  \citep{Ashman1992,Forbes1997,Kundu2001,Peng2006,Harris2009a,Sinnott2010,Liu2011}. Bimodality indicates two subpopulations in a galaxy \citep{Brodie2012}. In some cases, the colour distribution is even found to be trimodal, e.g. in the case of  NGC 4365 \citep{Blom2012} and NGC 4382  \citep{Peng2006}. The components of the bimodal colour distributions are identified in terms of metallicity; metal-rich and metal-poor corresponding to red and blue subpopulations respectively \citep{Usher2012}. The presence of these subpopulations indicates that there were multiple episodes of star formation and metal enrichment in the past. 

To explain the bimodality in the colour distribution in the context of host galaxy formation, three broad scenarios have been put forward. \citet{Ashman1992} proposed that the colour bimodality is the result of a gas-rich merger of disk galaxies. They suggested that the blue GCs are intrinsic to the spiral galaxies, while red GCs are formed during the merger. \citet{Forbes1997} suggested an in-situ formation scenario, in which the blue GCs are formed first in the initial collapse with limited field star formation. A quiescent period follows, then red GCs are formed in a metal-rich environment along with the bulk of the stars in the galaxy.  Accretion of blue GCs may also contribute. A third scenario was proposed by \citet{Cote1998,Cote2000,Cote2002} in which the red GCs are inherent to the host galaxy (similar to \citealt{Forbes1997}), while the blue GCs are accreted via mergers or tidal stripping.  Signatures of these different stages of galaxy evolution are best preserved in galaxy outer halos rather than in the complex inner regions and hence, an investigation of GCs in outer halos gives a unique opportunity to trace the formation and evolution of host galaxies.  

In this paper, we present the results from a wide-field imaging study of GC systems in three early-type galaxies: NGC 720 (E5), NGC 1023 (S0) and NGC 2768 (E/S0). A more detailed discussion about individual galaxy characteristics is given in Section 1.1. The data presented in this paper are a part of an ongoing larger survey, the SAGES Legacy Unifying Globulars and GalaxieS (SLUGGS)\footnote{http://sluggs.swin.edu.au/}, which aims to understand the assembly history of early-type galaxies with the aid of imaging, spectroscopy and simulations of galaxy formation. The survey,  still underway, undertakes a large scale study of 25 early-type galaxies within a distance of 30 Mpc. 

With the aid of wide-field imaging data, we can study the global properties of individual GC systems and hence the association with their host galaxies. These global properties include radial surface density, colour and azimuthal distributions, total number of GCs and specific frequency.  The full radial extent of large GC systems can only be completely investigated with wide-field imaging data.   From the radial surface density distributions of blue and red GCs, the characteristics of the subpopulations such as  their extent and concentration (centrally or extended) can be investigated. A similar slope between host galaxy starlight and red GC surface density suggests a coeval formation \citep{Bassino2006b, Faifer2011, Strader2011, Forbes2012a}.  The  dark matter halo component of a galaxy is associated with the blue GC subpopulation \citep{Forte2012, Forbes2012a}, which shows their connection with the hidden dark matter (proposed by \citet{Cote1998}). \citet{Forbes2012a} found a good agreement between galaxy diffuse X-ray emission and the surface density of the blue GCs for  nine ellipticals.  

The two dimensional spatial distribution of GC systems can be constructed with imaging data.  Estimation of position angle, ellipticities and two dimensional sub-structures can be carried out. Most previous studies carried out using smaller  telescopes (e.g. \citealt{Rhode2010,Young2012}), are unable to probe very far down the GC luminosity function and thus yield too few GCs to properly separate the system in red and blue subpopulations. Literature studies of galaxies like NGC 4636 \citep{Dirsch2005} and NGC 1316 \citep{Gomez2001} show that the azimuthal distribution of red GCs closely matches that of the spheroid/bulge of the host galaxy. Such observations support the idea that the bulk of galaxy stars have a coeval origin with the red GC subpopulation \citep{Wang2013}.  The total number of GCs can only be determined  accurately from a complete radial surface density distribution. An advantage of wide-field imaging taken in good seeing conditions is a more accurate determination of specific frequency with reduced errors (e.g. for NGC 4365 S$_N$ varies from 3.86 $\pm$ 0.71 \citep{Peng2008} from small field of view of HST imaging to 7.75 $\pm$ 0.13  \citep{Blom2012} with wide-field Subaru data. 

With the global properties of a sample of GC systems, we are also equipped to study their global relations with the host galaxies.  A relevant question to study is the (in)dependence of GC formation efficiency with different environments. Recently, \citet{Tonini2013} constructed a theoretical model to investigate GC bimodality. She predicted that the GC bimodality is a direct outcome of hierarchical galaxy assembly. Also she predicted that a larger fraction of blue GCs  can be found in early-type galaxies residing in higher density environments. However, using ACSVCS data \citet{Cho2012} studied the variation in the fraction of red GCs in field and cluster environments. They found that the fraction of red GCs was enhanced from field to high density environment. \citet{Spitler2008} also studied the dependence of mass normalized blue GC number with environment for a sample of early-type galaxies. They concluded that  the GC formation efficiency depends primarily on galaxy mass and is nearly independent with respect to galaxy environment. In this paper, we also try to analyse these different results regarding the dependence of  GC formation efficiency on environment. 
 
In short, this paper presents the results from a wide-field imaging study of the GC systems in three early-type galaxies, their global properties and their connection with the host galaxy properties. Also we have explored the correlations of  global properties of GC systems (including GC systems of other well studied early-type galaxies) with host galaxy mass, galaxy effective radius and local environment density. 

This paper is organised as follows. Our three sample galaxies are briefly presented in the following subsection. Section 2 steps through observations, data reductions, photometry and the selection of GCs. Section 3 explores the different GC system properties such as radial density, colour distributions, azimuthal distribution, total number of GCs and specific frequency. Analysis of GC subpopulations and their connection with host galaxy properties are also described in Section 3. Section 4 discusses the relationship of GC system extent with galaxy  stellar mass, effective radii and environment for a sample of $\sim$ 40 galaxies. Section 5 concludes with the main results and their implications for GC formation scenarios.

\begin{table*}
\caption{Basic data for the target galaxies. Right Ascension and Declination (J2000) are from NASA/IPAC Extragalactic Database (NED). The galaxy types are discussed in Section 1.1. The distance for  NGC 720 is obtained from NED,  NGC 1023 and NGC 2768 are from \citet{Cappellari2011}. Total V band magnitudes are obtained from \citet{de1991}. The extinction correction for V band is calculated from \citet*{Schlegel1998}. The absolute total magnitude is derived from the V-band magnitude, distance and the extinction  correction. Position angle and ellipticity of the galaxy major axis are given in the last columns and are obtained from HyperLeda \citep{Paturel2003}.}
\begin{tabular}{lcclcccrrc} 
\hline
Name & RA & Dec & Type & D &V$_T$&A$_v$& M$_v^{\sc T}$ & PA & $\epsilon$\\ 
     & (h:m:s) & ($^o$:$'$:$\arcsec$)& & (Mpc)  &(mag)& (mag) &(mag) & ($^o$)& \\ 
\hline\hline
NGC 720   & 01:53:00.5 & $-$13:44:19   &E5  &23.4&10.18  & 0.05& $-$21.68& 142 & 0.47\\ 
NGC 1023 & 02:40:24.0  & +39:03:48   & S0 & 11.1 & 9.35  &0.20 & $-$21.08 & 87 & 0.58\\ 
NGC 2768 & 09:11:37.5 &+60:02:14 & E/S0 & 21.8  &9.87  &0.14& $-$21.91  & 93 & 0.60\\ 
\hline
\end{tabular} 
\label{data}
\end{table*}

\subsection{Sample galaxies}
 Our three galaxies of intermediate luminosity are taken from the ongoing SLUGGS survey (Brodie et al. 2013 in prep.) of 25 galaxies within 30 Mpc and are among the most elongated galaxies in the survey. The three galaxies reported here are among the most flattened in the SLUGGS survey and hence useful to search for trends between the flattening (ellipticity) of the GC system and the host galaxy. Table \ref{data} records the basic data for the sample galaxies and an individual description for each galaxy follows.
\subsubsection{NGC 720}
NGC 720 is an X-ray bright, relatively isolated elliptical galaxy.  The morphological classification is an E5 \citep{de1991}. NGC 720 has been well studied in X-rays by \citet{Buote1994,Buote1996,Buote1997} and \citet{Buote2002}. The X-ray studies showed an isophotal twist which is absent at optical wavelengths.  NGC 720 is found to be a strong X-ray source with filaments extending from the nucleus of the galaxy and curving towards the south \citep{Buote1996}.  \citet{Kissler1996} studied the GC system of NGC 720 out to a galactocentric distance of  4.37 arcmin (30 kpc). They did not study the properties of GC subpopulations, only the total system. They found the GC system to resemble the host galaxy light distribution in terms of ellipticity, position angle and surface density. In contrast, the properties of the GC system did not match those of the X-rays. \citet{Forbes2012a} found a similar slope for the X-ray surface brightness profile and the surface density of the blue GC subpopulation of NGC 720. 

\subsubsection{NGC 1023}
NGC 1023 is a nearby S0 galaxy  at a distance of 11.1 Mpc \citep{Cappellari2011} An interesting aspect of this lenticular galaxy is its bluer eastern companion, NGC 1023A.  HI maps of NGC 1023 show a high concentration of neutral hydrogen gas around NGC 1023A \citep{Sancisi1984}. \citet{Capaccioli1986} did not detect any traces of emission lines in the spectrum of NGC 1023, indicating no current star formation. \citet{Larsen2000} studied the central GCs of NGC 1023 using HST WFPC2 imaging. They found 221  GCs and  a bimodal colour distribution. They also found the presence of  red extended (effective radii > 7 pc) GCs, naming them `faint fuzzies'. \citet{Cortesi2011}  have used the PNe to analyze the kinematics of NGC 1023. They found that the kinematics of the galaxy resembles a spiral galaxy, supporting the theory of transformation of S0 galaxies from spiral galaxies.  \citet{Young2012} studied the GC system of NGC 1023 using imaging data from the 3.5-meter WIYN telescope and estimated the total number of GCs to be 490 $\pm$ 30, with S$_N$ = 1.7 $\pm$ 0.3. They also found a statistically significant bimodal colour distribution for the GC system.
\subsubsection{NGC 2768}
NGC 2768 is catalogued as a lenticular galaxy in the Carnegie Atlas of Galaxies \citep{Sandage1994} and an elliptical E6 in the Third Reference Catalogue of Bright Galaxies (RC3, \citealt{de1991}). \citet{Crocker2008} traced the interstellar medium of NGC 2768 from CO emission, finding a  molecular polar disc, which suggests a merger history for NGC 2768.  \citet{Kundu2001} studied the GC system of NGC 2768 using single HST/WFPC2 pointing and found a statistically significant bimodal colour distribution. \citet{Pota2013} did a kinematic study of the GC systems of 12 early-type galaxies including NGC 2768. They found GC  bimodality in ({\it R$_c-$z}) colour. They also found that the rotation velocity of red GCs matches the galaxy stars, supporting coeval formation. \citet{Usher2012} have carried out a study of CaT metallicity distribution of NGC 2768 GCs, but did not find bimodality in the CaT metallicity for the GCs. The available photometry for the galaxy was poor and they obtained spectra only for a few GCs, which they propose as the reason for not detecting bimodality in metallicity. \citet{Forbes2012b} analyzed the kinematics, combining PNe, GCs and galaxy starlight. They found similarity in the radial density distribution between red GCs, galaxy bulge PNe and galaxy starlight, strengthening the idea of coeval evolution.  Kinematic studies of these three components up to 4 R$_e$ showed a good agreement between them. 

\begin{figure*}
\includegraphics[scale=.5] {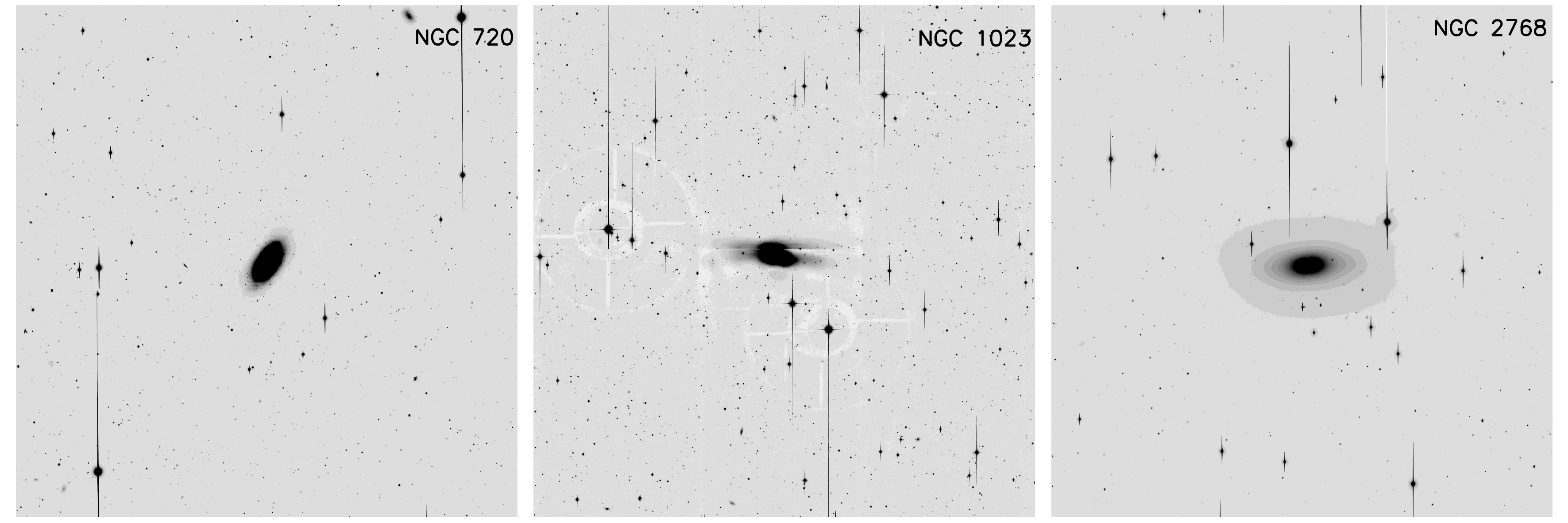}  
  \caption{Wide-field images of three galaxies in {\it i} band filter taken from the ground based telescopes. Each image covers on sky an area of 10 square arcmin. NGC 720 and NGC 2768 were observed using Subaru telescope while NGC 1023 was taken from the CFHT archive. North is up and East on the left.  }
\label{sky1}
\end{figure*}

\section{Data}
\subsection{Observations and reduction techniques}
The imaging data for NGC 720 and NGC 2768 were taken using the Suprime-Cam \citep{Miyazaki2002} imager mounted on the 8-meter Subaru telescope.  The instrument includes ten 2048 x 4096 CCD detectors  with  a pixel scale of 0.202 arcsec and a field of view of 34 x 27 square arcmin. Multiple exposures were taken in a dithered pattern to avoid the blank regions due to gaps between CCDs. The observation log is tabulated in Table \ref{obs}.  

The Suprime-Cam Deep Field Reduction package 2, SDFRED2 \citep{Ouchi2004} is utilised to carry out the pre-processing of the Suprime-Cam data. The pipeline includes scripts for flat fielding, distortion and atmospheric dispersion corrections. The pre-processed images were aligned and combined to form the mosaic image using a combination of softwares SExtractor \citep{Bertin1996}, Scamp \citep{Bertin2006} and Swarp \citep{Bertin2002}\footnote{http://www.astromatic.net/software/}. The SExtractor run on the individual CCD images selects point sources with a three sigma threshold above the background level. The relative positions between the selected objects were matched with an astrometric reference catalogue (USNO or SDSS) using the Scamp software to generate the astrometric solution. Using the Swarp software and the astrometric solution, the multiple CCD images were aligned and stacked to produce the mosaic image. 

\begin{table}
\caption{Log of observations. }
\begin{tabular}{p{0.7cm}p{0.6cm}p{1.8cm} p{0.7cm}p{1.2cm} p{1.4cm}} 
\hline
Galaxy & Filter &Obs. date &  Seeing& Telescope & Exp. time\\
    NGC  &           & HST$^\ast$                               &($"$)      &                     & (s)\\
\hline\hline
720   &{\it g}& 2008 Nov. 28 & 0.88 &   Subaru  &  1770\\ 
          &{\it i}     &   2008 Nov. 28  &   0.98&                  &1370\\
\hline
\multirow{2}{*}{1023} &{\it g} & 2004 Sep. 10 &  0.71&  CFHT   & 1232\\
	&  {\it i} &  2004 Sep. 11& 0.73 &               &1100  \\  
\hline	  
\multirow{3}{*}{2768} &{\it g}& 2011 Jan. 03& 0.95 & Subaru  & 4320\\
          &{\it r} &        2011 Jan. 04&        0.77 & &1860\\
          &{\it i} &       2011 Jan. 04&     0.75& & 1296\\ 
\hline
\end{tabular}
\newline
$\ast$ Hawaii-Aleutian Standard Time
\label{obs}
\end{table}

We have obtained a second photometric dataset for NGC 2768 from the Hubble Legacy Archive. The data (HST ID: 9353) consist of one pointing taken in F435W({\it B}), F555W({\it V}) and F814W({\it I}) filters using the Advanced Camera for Surveys (ACS) instrument installed on Hubble Space Telescope.  The Wide Field Channel mounted on ACS consists of two 2048 x 4096 CCDs with 0.049 arcsec pixel scale and  3.37 x 3.37 square arcmin field of view. Jordi et al. (2006) have given the transformation equations to convert the {\it B, V, I} magnitudes to the SDSS photometric system.  The {\it B, V, I} magnitudes for all of the NGC 2768 objects are converted to {\it g, r, i} magnitudes.

We also acquired the central GC radial surface density distributions for NGC 720 from Escudero et al. (2013, in prep.). This data set was observed in {\it g, r, i} filters using the Gemini Multi-Object Spectrographs (GMOS, \citealt{Hook2004}).  NGC 720 was observed along with five other galaxies published in \citet{Faifer2011}.  A detailed description about the observations and data reduction is given in the same publication. 

The wide-field imaging data for NGC 1023 were acquired from the Canada France-Hawaii Telescope (CFHT) archive.  Observations were taken with the MegaCam \citep{Boulade2003} imager. The detector consists of a 9 x 4 mosaic of 2048 x 4612 CCDs with a scale of 0.187 arcsec giving a field of view of 0.96 x 0.94 square degree.  A series of  images taken in  {\it g} and {\it i} filters  was processed through the MegaCam image stacking pipeline named MegaPipe \citep{Gwyn2008}. MegaPipe includes the pre-processing (bias and dark subtraction, flat fielding) of the images. The pipeline carries out an astrometric and photometric calibration for the MegaCam images. The individual CCD images were then mosaiced with Swarp software.   Figure \ref{sky1} shows the wide-field images of NGC 720, NGC 1023 and NGC 2768 observed in the {\it i} band filter using the ground based telescopes.

\subsection{Photometry}

We modelled the galaxy light for the three galaxies and subtracted it from the corresponding mosaic image with the {\tt IRAF} task {\tt ELLIPSE}  keeping the centre, PA and ellipticity as free parameters.  Here we remind the reader that the galaxy light subtracted images are only used to to improve source detection and not for any photometric analysis. The {\tt ELLIPSE} parameters (PA and ellipticity) derived from the task match well with the values mentioned in Hyperleda (given in Table \ref{data}). Sources on images were identified and aperture photometry was carried out using the  source finding software, SExtractor.  SExtractor identifies a probable source only if it has a minimum of 5 adjacent pixels with  a threshold level of three sigma above the local background. SExtractor estimates the total instrumental magnitude for the detected sources using the Kron radius \citep{Kron1980} in automatic aperture magnitude mode.  For this, magnitudes within aperture sizes of 1 to 7 pixels, equivalent to 0.2 to 1.4 arcsec, are estimated for all the detected sources in the respective mosaic images. Depending on the seeing values for the respective filters, the extraction radius is determined and hence we obtain instrumental magnitudes.  These instrumental magnitudes are corrected for the light outside the extraction radius and finally SExtractor provides a list of point sources with positions and aperture corrected magnitudes. We selected $\sim$ 20 bright stars within the colour range of 0.7 < ({\it g$-$i}) < 1.3 in the individual galaxy images  and obtained their magnitudes from the Sloan Digital Sky Survey catalogues, in order to estimate the zeropoints in each filter. These zeropoints were applied to calibrate the magnitudes for all the point sources detected. Our final object lists have {\it g} and {\it i} magnitudes for all three galaxies, with additional {\it r} magnitudes for NGC 2768. The object magnitudes are corrected for Galactic extinction using \citet{Schlegel1998} (see Table \ref{data}). All magnitudes discussed hereafter are extinction corrected.

\begin{figure}
\centering
 \includegraphics[scale=.49]{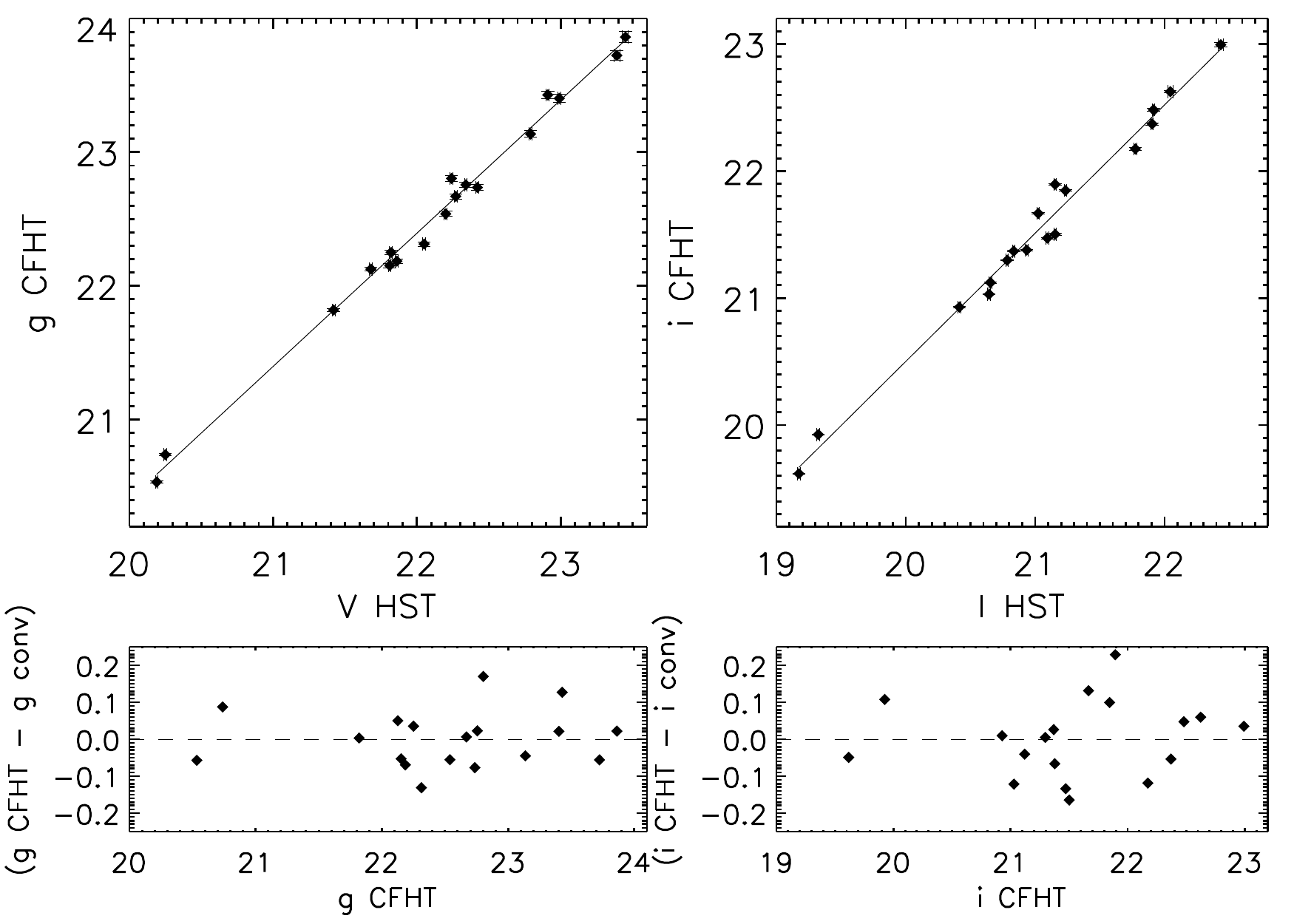} 
  \caption{Transformation of NGC 1023 GC magnitudes from HST to CFHT photometric system. The top panels show the linear fits  between HST magnitudes and CFHT magnitudes for the common GCs in {\it  g} (left panel) and {\it i} (right panel) filters. The bottom panels show the difference between measured (from CFHT) and converted magnitudes versus the measured magnitudes in the {\it g} (left panel) and {\it i} (right panel) filters.} 
\label{conv}
\end{figure}

\subsection{HST/WFPC2 GC catalogue for NGC 1023}
\citet{Larsen2000} have published a list of 221 GCs in NGC 1023 observed with HST in the {\it V} and {\it I} filters. Their selection was primarily based on sizes, colour (i.e. 0.75 < ({\it V}$-${\it I}) < 1.40) and magnitudes (i.e. 20 < {\it V} < 25). For uniformity between the catalogues, we converted the {\it V} and {\it I} magnitudes  into CFHT {\it g} and {\it i} magnitudes. \citet*{Jordi2006} transformation equations require three band magnitudes whereas the HST/WFPC2 data contain only {\it V} and {\it I} magnitudes.  In order to convert the magnitudes, we selected a set of bright objects (in the colour range 0.85 < ({\it V}$-${\it I}) < 1.35) in common between the two data sets and the magnitudes are fitted with a linear bisector relation of the form :
\begin {eqnarray}
g _{conv} &=&  [(0.996 \pm 0.021) \times V _{HST}] +   (0.473 \pm 0.175) \\
i _{conv} &=& [(1.009 \pm 0.031) \times  I  _{HST}] + (0.304 \pm 0.113)
\end{eqnarray} 
where g$_{conv}$ and i$_{conv}$ are CFHT filter equivalent magnitudes for the HST {\it V} and {\it I} magnitudes.  The top panels in Figure \ref{conv} show the magnitude conversion between the HST and the CFHT photometric systems. The bottom panels in Figure \ref{conv}  display the deviation between the measured (g$_{CFHT}$ and i$_{CFHT}$) and converted (g$_{conv}$ and i$_{conv}$) magnitudes.  The root mean square deviation of converted magnitudes (using equations 1 and 2) from the corresponding measured CFHT magnitudes are  0.07 and 0.12 magnitudes with no obvious systematic trend. This conversion is used to transform the HST photometric system to the CFHT system for the  GCs of \citet{Larsen2000}.  We also checked the colour transformation between the two photometric systems and found no systematic trend. 

\subsection{Globular cluster selection}
{\bf NGC 720:}  The GC  selection for NGC 720 is carried out on object size, magnitude and colour of individual objects. Initially however, the source position matching between the Subaru {\it g} and {\it i} band images removes spurious detections (e.g. cosmic rays) on the individual images.  To determine the object size, we measure the flux in two apertures.  Objects with surplus amount of light beyond the extraction  aperture radius are removed from the GC list. As GCs appear as point sources at the distance of NGC 720,  the probable GCs have a minimum magnitude difference between the extraction aperture and the adjacent aperture. A further selection of objects is carried out in the {\it i} band, i.e. 20.6 $\le${\it  i} $\le$ 24 (at the distance of 23.4 Mpc, objects brighter than  {\it  i} = 20.6 include ultra compact dwarfs \citep{Brodie2011}, while objects fainter than {\it  i} = 24 have magnitude errors greater than 0.15). Final selection of NGC 720 GCs is based on the  ({\it g$-$i}) colour of individual objects, i.e. 0.6 $\le$ ({\it g$-$i}) $\le$ 1.3.  In the SLUGGS survey, we have a list of spectroscopically (velocity) confirmed GCs for each of the survey galaxies. Hence we are able to check the reliability of GC selection for all the three sample galaxies.

{\bf NGC 1023:} The data for NGC 1023 include CFHT {\it g} and {\it i} band photometry and a catalogue of 221  GCs from HST \citep{Larsen2000}. The  GC system of NGC 1023 is identified based on the same selection criteria followed for NGC 720 Suprime-Cam data. Matching of object positions between the observed {\it g} and {\it i} band images cleared false detections from the list. The {\it i} band magnitude selection for NGC 1023 GCs is 18.9 $\le$ i $\le$ 23.0 based both on the distance to NGC 1023 and on the error in the measured {\it i} band magnitude. A final selection is made in colour by selecting sources in  the same colour range as used by \citet{Larsen2000}, i.e. 0.65 $\le$ ({\it g$-$i}) $\le$ 1.3. 

{\bf NGC 2768:} The data for NGC 2768 include {\it g}, {\it r} and {\it i} band Subaru imaging. False detections are primarily eliminated from the object list by matching the source position with 0.1 arcsec accuracy between the three bands. Point source objects are chosen based on the magnitude difference between the extraction and the adjacent aperture. As the data set for NGC 2768 consists of three band data, an additional selection based on two colour space is introduced (i.e. ({\it g$-$i}) versus ({\it r$-$i})). We adopted a similar GC selection process in the colour-colour diagram as used by \citet{Spitler2008} and \citet{Blom2012}. It is evident from earlier studies, viz., figure 6 in \citet{Blom2012} and figure 3 in \citet{Pota2013}, that the GCs populate a particular region in the colour-colour diagram. These GCs along with neighbouring objects showing a 2-sigma deviation from the selected region are chosen as final GC candidates. The {\it i} band magnitude cut for NGC 2768 is 20.4 $\le$ {\it i} $\le$ 24.0.  A second set of data for NGC 2768 comes from HST/ACS covering the central 2.1 arcmin region. The GCs from the HST/ACS imaging are selected in the same colour-colour diagram mentioned above for the Subaru imaging. 

\section{Analysis of  GC systems} 

\begin{figure}
\includegraphics[scale=.45] {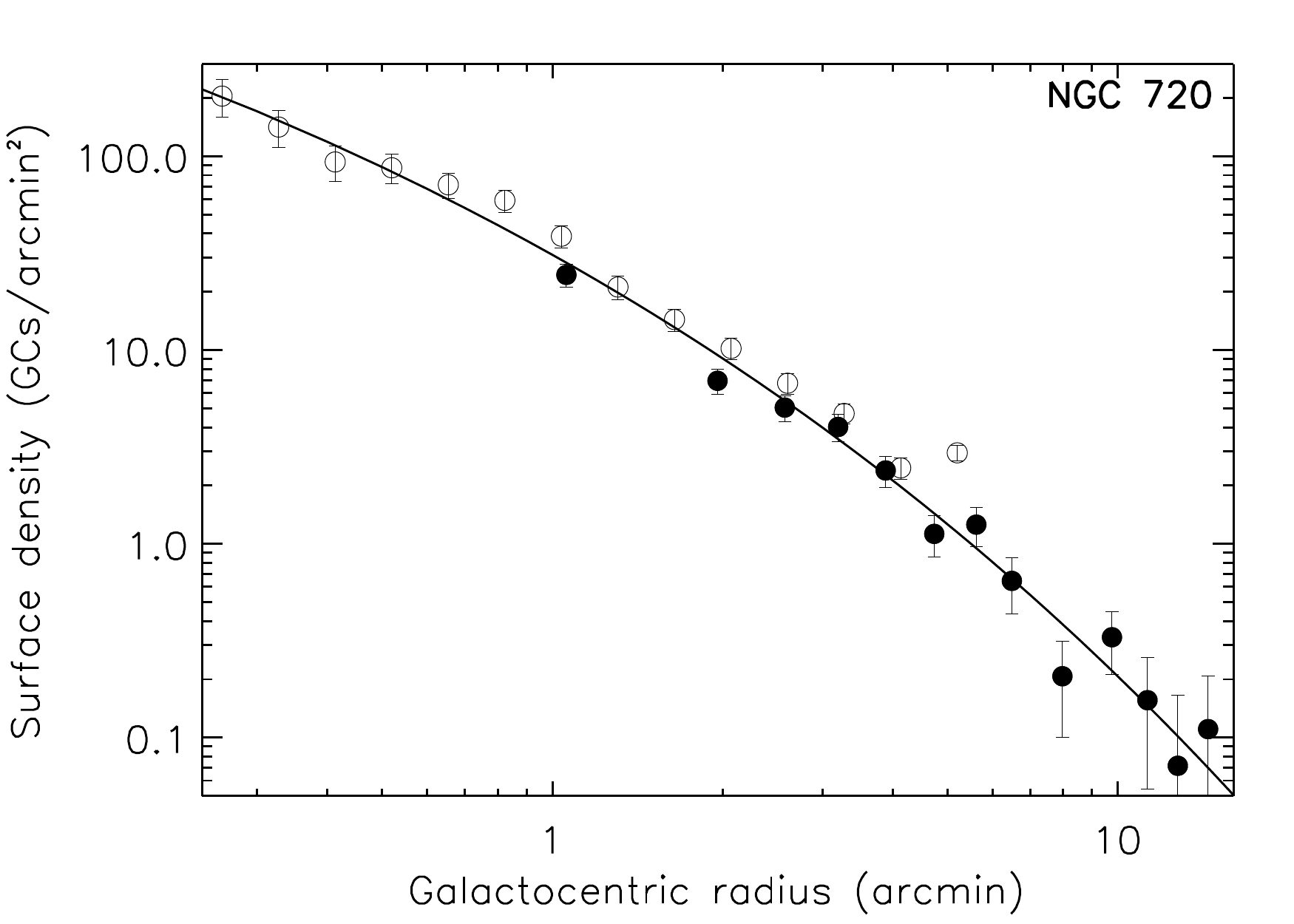}  
  \caption{Surface density profile for the GC system of NGC 720.   The plot displays the Gemini (open circles) and Subaru (filled circles) data. The GCs selected within the turnover magnitude limit, {\it i} = 23.7, are employed to derive the radial surface density values. The surface density reaches the background level around 9.8 $\pm$ 0.8 arcmin ($\sim$ 15 R$_e$) with 0.98 objects per arcmin$^{2}$.  The solid line is the fitted S\'{e}rsic profile for the GC surface density.}
\label{720surfden}
\end{figure}

\subsection{Surface density profiles}
The one dimensional radial distribution of a GC system is revealed by its surface density profile. The surface density for each radial bin is estimated by fixing a similar number of globular  clusters per circular bin and dividing by the effective covered area. The area coverage in each annuli is corrected for two factors: the presence of saturated stars and the annular area outside the image. The errors associated with the surface density distribution are given by Poisson statistics.

A combination of a S\'{e}rsic profile and a background parameter is fitted to the GC surface density distribution. The fitted profile can be written as :
\begin{equation}
N(R) = N_e ~exp \left[-b_n \left(\frac {R}{R_e}\right)^\frac{1}{n} - 1\right] + bg
\label{sersic}
\end{equation}
where {\it N$_e$} is the density of the GCs at the effective radius {\it R$_e$}, {\it n} is S\'{e}rsic index or the shape parameter for the profile, {\it b$_n$} is  given by the term 1.9992n - 0.3271 and {\it bg} represents the background parameter.  The background values obtained for the three GC systems are then subtracted from the respective radial density distribution which is shown in all density distribution plots.

\subsubsection{NGC 720}
 Figure \ref{720surfden} displays the surface density profile for NGC 720 using the Suprime-Cam and GMOS data, fitted with a S\'{e}rsic profile. The radial coverage of GMOS data reaches out to 5.6 arcmin and overlaps with the Suprime-Cam data which is detected out to a radius of $\sim$ 18 arcmin.  The GCs brighter than the turnover magnitude ({\it i} = 23.7) are selected to retrieve the radial surface density distribution.  A constant value of 0.98 objects per arcmin$^{2}$ is reached at a galactocentric radius of 9.8 $\pm$ 0.8 arcmin suggesting that the background is obtained. At a distance of 23.4 Mpc, the GCs extend to at least 68 $\pm$ 6 kpc from the centre of the galaxy.  The parameter values for the fitted profile are reported in Table \ref{surfden}. As seen from Figure \ref{720surfden}, the data sets from the Gemini and Subaru telescopes are generally consistent with each other without applying any manual adjustment.  \citet{Kissler1996} have studied the radial density distribution of NGC 720 GCs using the 2.2-meter telescope at the European Southern Observatory. They estimated the GC system reaches the background at a galactocentric distance of 2.67 arcmin. This appears to be an underestimation of the true extent by a factor of $\sim$3. This likely demonstrates our high-quality wide-field imaging and its ability to remove contamination.

\begin{figure}
 \includegraphics[scale=.45]{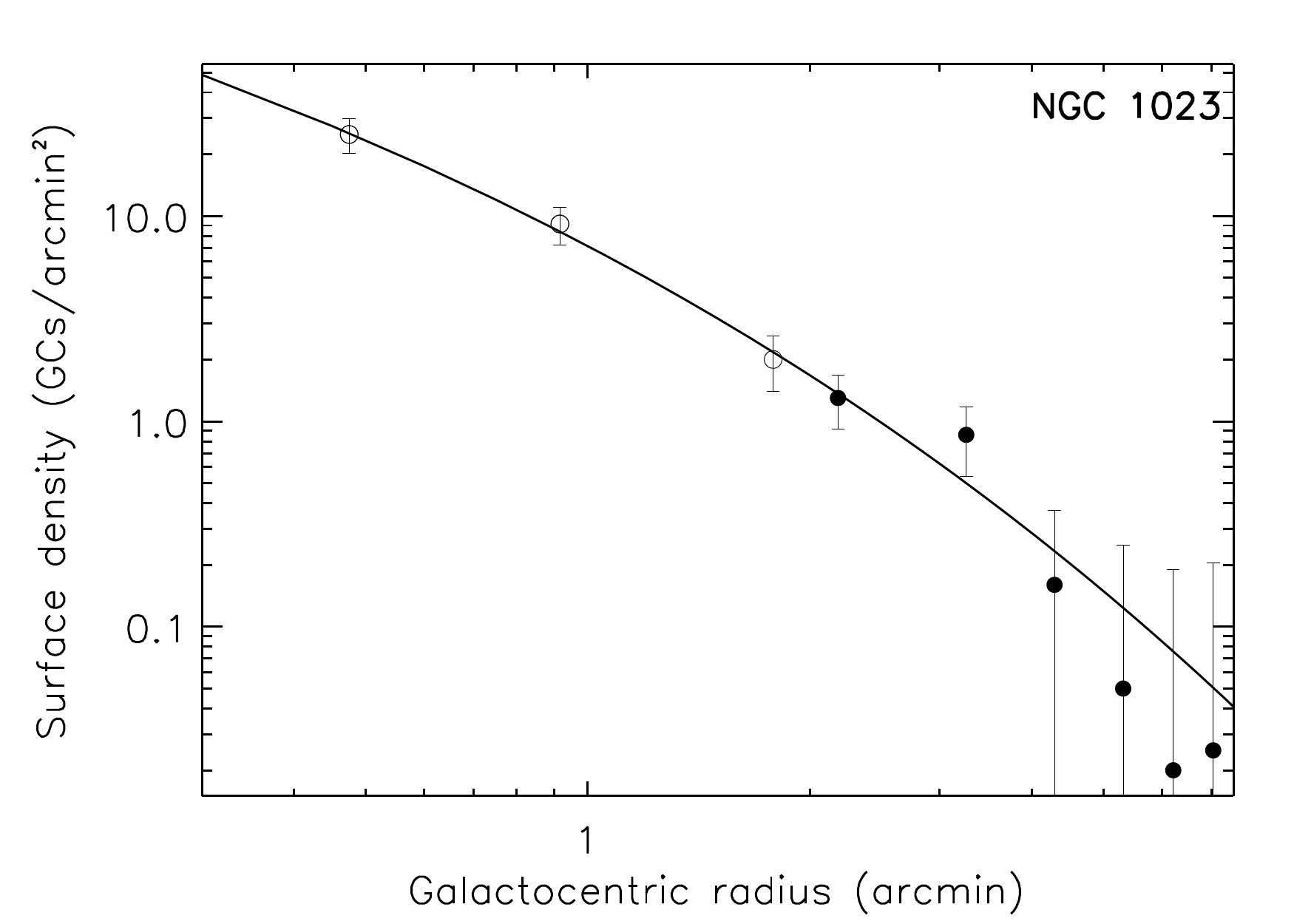} 
  \caption{Surface density profile for the GC system of NGC 1023.  The plot shows HST  (open circles) and CFHT (filled circles) data.  The limiting magnitude for the two data sets is the turnover magnitude, i.e {\it i} = 22.0. The surface density of the GC system reaches the background level around 6.2 $\pm$ 0.5 arcmin ($\sim$ 8 R$_e$) with 1.27 objects per arcmin$^{2}$. A S\'{e}rsic profile  is fitted and is shown with a solid line.}
\label{1023surfden}
\end{figure}

\subsubsection{NGC 1023}
We created a radial surface density plot for NGC 1023 using the GCs from the HST at the very centre and the CFHT for the outer regions. \citet{Larsen2000} identified a third set of GCs called red extended GCs or faint fuzzies. For the calculation of surface density, the faint fuzzies are excluded (i.e. objects with V > 22.8) as the turnover magnitude limit is {\it i} = 22.0. The area corrections are applied to account for the detector shape of HST and for  saturated stars  in the CFHT image. Figure \ref{1023surfden} shows a plot of surface density for the NGC 1023 GCs using HST  and CFHT data.  The GC surface density for NGC 1023 is fitted with equation \ref{sersic} and  fitted parameters are given in Table \ref{surfden}. The HST observations are limited to 2.2 arcmin radius and the CFHT observations extend to 15 arcmin from the centre of the galaxy. At a galactocentric radius of 6.2 $\pm$ 0.5 arcmin the GC surface density flattens to a constant value of $\sim$ 1.27 objects per arcmin$^{2}$. From the centre of NGC 1023, the GCs reach an extent of 20 $\pm$ 2 kpc.   The HST and CFHT data have not been adjusted in surface density and are consistent with each other in the region of overlap (when the two data are cut at the turnover  magnitude).  {This overlap between HST and ground based telescope is a representation of data quality.} \citet{Young2012} investigated the GC system of  NGC 1023 using the 3.5-meter WIYN telescope.  The radial extent of GC system was estimated by them to be 6.3 $\pm$ 0.8 arcmin.  Thus \citet{Young2012} and ourselves are in agreement on the radial extent of NGC 1023 GC system. 

\begin{table}
\centering
\caption{Fitted parameters for the surface density of NGC 720, NGC 1023 and NGC 2768 GC systems. The last column in the table presents the extent of the GC system in each galaxy.}
\begin{tabular}{ccccc}
\hline
Name  & R$_e$ & n & bg  & GCS ext.\\ 
NGC & (arcmin) & & (arcmin$^{-2}$) & (arcmin) \\ 
\hline\hline
720 & 1.97$\pm$0.34 &4.16$\pm$1.21 & 0.98$\pm$0.06 &9.8$\pm$0.8\\
1023& 1.00$\pm$0.35 &3.15$\pm$2.85 &1.27$\pm$0.12 &6.2$\pm$0.5\\ 
2768 & 1.66$\pm$0.23& 3.09$\pm$0.68& 0.61$\pm$0.04 &9.9$\pm$0.5 \\ 
\hline
\end{tabular}
\label{surfden}
\end{table}

\begin{figure}
 \includegraphics[scale=.45]{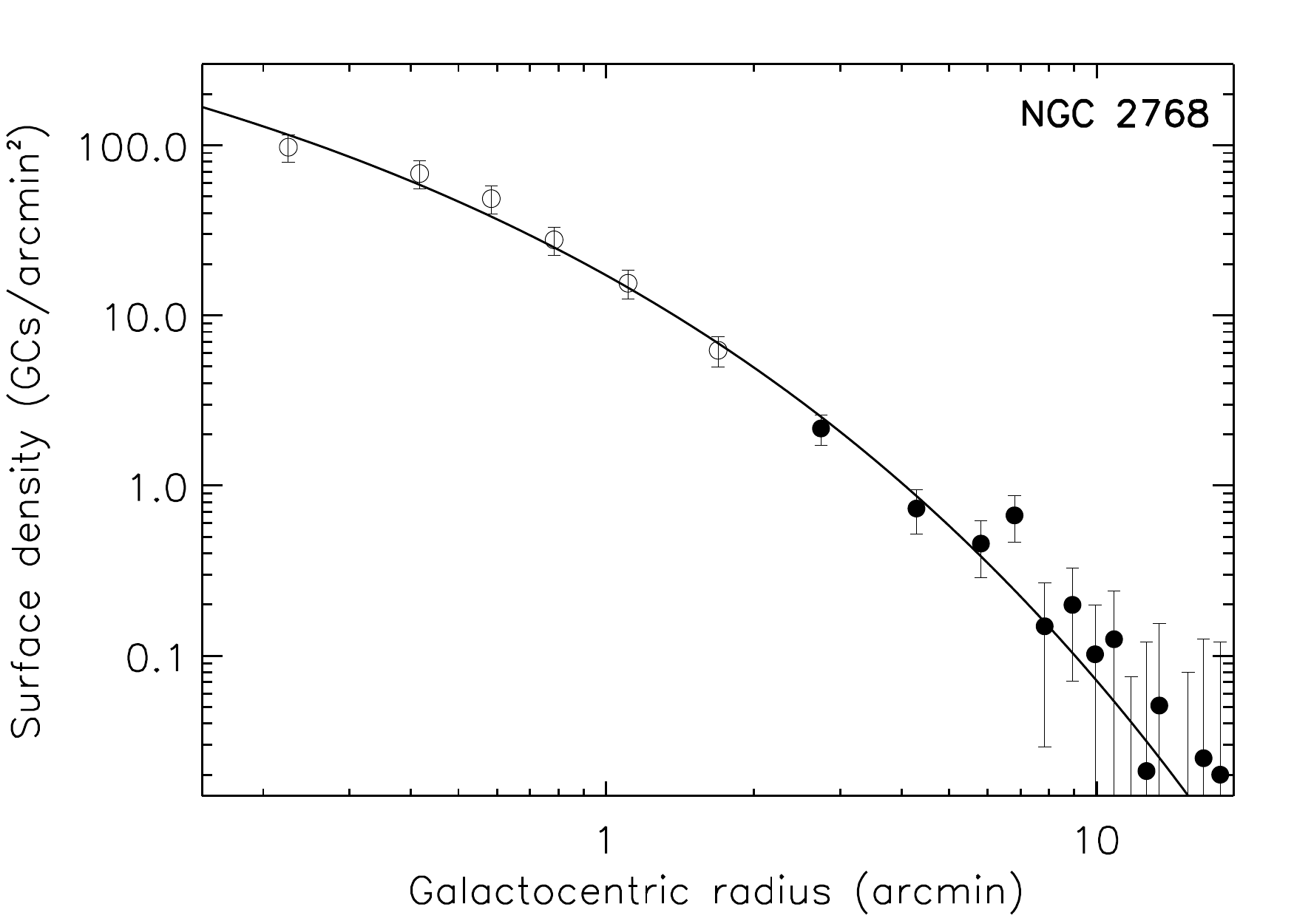}
  \caption{Surface density profile for the GC system of NGC 2768.  The plot shows HST  (open circles) and Subaru (filled circles) data.  The GCs within the turnover magnitude limit, {\it i} = 23.3, are selected for the density distribution. NGC 2768 GCs reach the background at a galactocentric distance of 9.9 $\pm$ 0.5 arcmin ($\sim$ 10 R$_e$) with 0.61 objects per arcmin$^{2}$. The solid line represents the S\'{e}rsic profile fitted on the GC density distribution.}
\label{2768surfden}
\end{figure}

\subsubsection{NGC 2768}
Figure \ref{2768surfden} displays the radial distribution of the GC system of NGC 2768. The data points in the inner 2.1 arcmin radius of the galaxy were obtained from the HST data and the area beyond that was covered by the Subaru  data. The data points shown in the Figure \ref{2768surfden}  are generated from the GCs with {\it i} $<$  23.3 (i.e. the turnover magnitude).  The HST data points are corrected for the detector shape. The presence of saturated stars in the inner annular radii and the area outside the detector were taken into account in the area calculation for the Subaru data points. The GC system of NGC 2768 reaches a background value of 0.61 objects per arcmin$^{2}$ at a galactocentric distance of 9.9 $\pm$ 0.5 arcmin. The surface density distribution of the GCs is fitted with a S\'{e}rsic profile and is shown in Figure \ref{2768surfden}.  The extent of the GC system of NGC 2768 is at least 63 $\pm$ 3 kpc. Since both data sets are cut at the turnover magnitude, the good overlap between HST and Subaru data sets confirms the magnitude completeness of the Subaru data. We are unable to find any previous work which has studied the GC extent for this galaxy.

\begin{figure*}
\centering
\includegraphics[scale=.5]{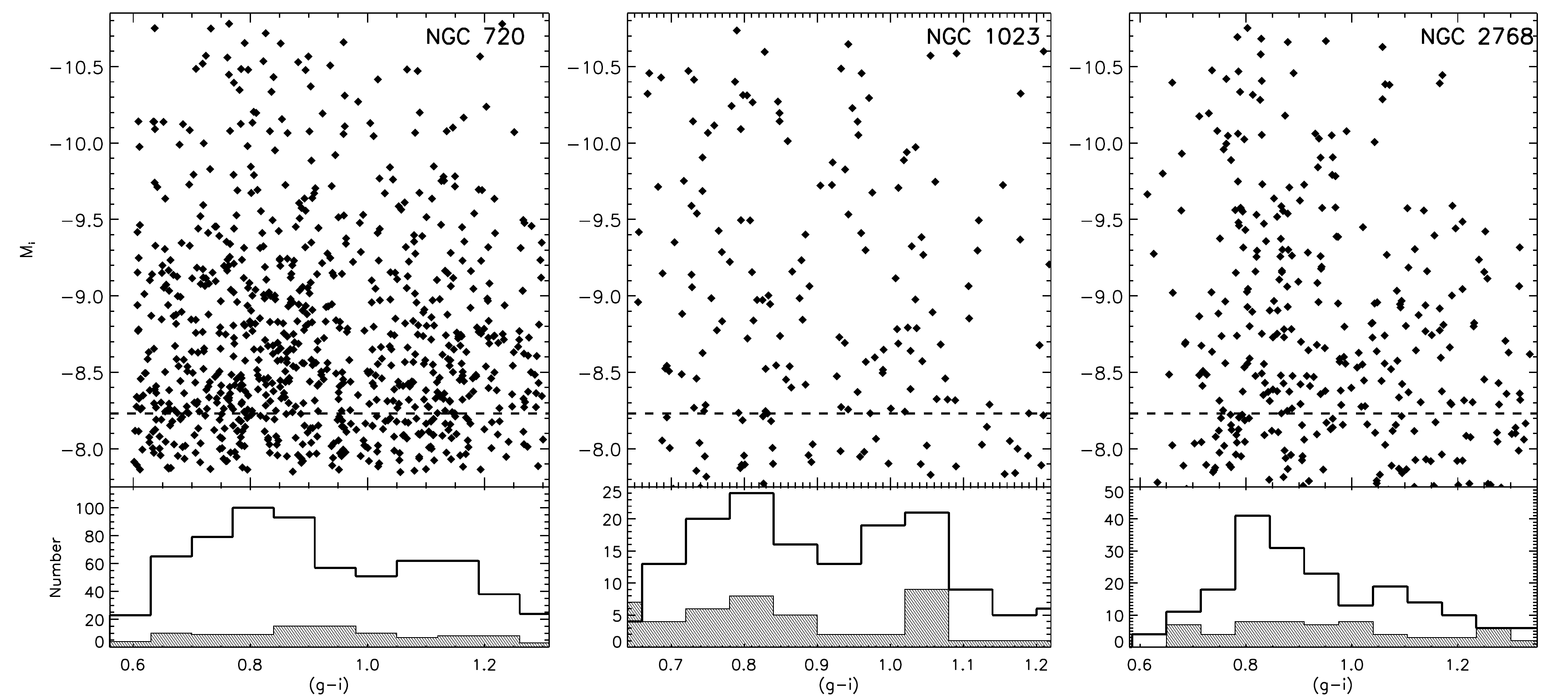}
\caption{Colour magnitude diagrams for the selected GC candidates using wide-field data. GC candidates shown in the Figure include objects brighter than M$_i$ = $-$7.75 mag within the measured GC system extent.  The turnover magnitude in the {\it i}-filter is M$_i$ = $-$8.23 mag, shown as a dashed line in all three top panels.  The {\it top left} panel shows the GC candidates of NGC 720 observed using  Subaru/Suprime-Cam. The open histogram is plotted in the {\it bottom left} panel representing the Subaru data with objects detected above the turnover magnitude. The shaded area represents the estimated background for the Subaru data. The {\it top middle} panel shows the GC candidates of NGC 1023 observed using CFHT/MegaCam. The {\it bottom middle} panel displays the histogram of GC candidates from the  CFHT (open) and the background contamination (shaded). The {\it top right} panel shows the GC candidates of NGC 2768 using Subaru/Suprime-Cam. The histogram of the GC candidates (open) for NGC 2768 is shown along with the background (shaded) in the {\it bottom right} panel.}
\label{cmd}
\end{figure*}

\subsection{Colour magnitude diagrams}

The top panels in Figure \ref{cmd} show the colour magnitude diagrams (CMDs) of GC candidates for the sample galaxies, based on the selection discussed in Section 2.4.  The CMDs display all the detected objects brighter than M$_i$ = $-$7.75 mag (0.5 mag fainter than the turnover magnitude) for the respective galaxies. The bottom panels display the {(\it g$-$i}) colour histograms of the same GC candidates along with the background contamination for the respective galaxies. In this figure, we have displayed only the data from the wide-field imaging and not from the space-based data. Also the histograms represent only the GC candidates detected above the turnover magnitude. In order to estimate the colour distribution of background objects within the GC extent, we have made use of the objects detected outside the GC system extent. First the colour distribution of the objects outside the GC extent is analysed and correct for the relevant areal coverage. Then this colour distribution (shown in lower panels of Figure \ref{cmd}) is subtracted from the corresponding GC system colours to obtain the uncontaminated GC colour distribution. The colour distribution of the background objects generally shows a broad colour range and does not strongly affect the GC subpopulation peaks.

\begin{figure*}
\includegraphics[scale=.5] {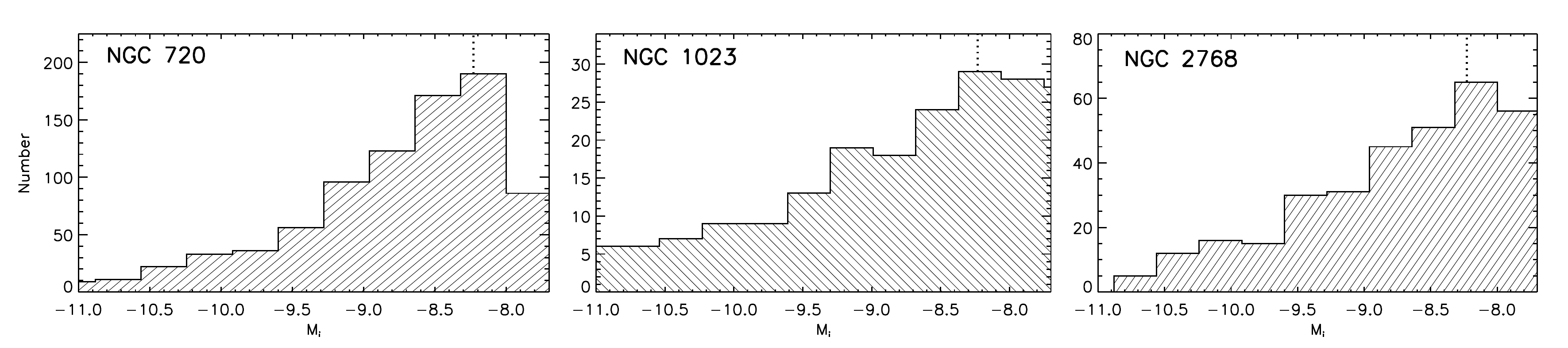}  
  \caption{Globular cluster luminosity function in {\it i} band filter. The histograms represent the globular cluster luminosity function of the GC systems detected for the individual galaxies. The histograms only include the GCs detected till the GC system extent estimated from the radial surface density distributions.  The dotted line at M$_i$ = $-$8.23 represents the turnover magnitude in {\it i} band filter.}
\label{GCLF}
\end{figure*}

All CMDs have displayed objects detected above the magnitude M$_i$ = $-$7.75 mag. The top left panel shows the CMD for NGC 720 GC candidates detected within a  galactocentric radius of 9.8 arcmin (see Section 3.1), observed using the Subaru/Suprime-Cam telescope. The colour histogram of detected GC candidates above the turnover magnitude along with the background is displayed in the bottom left panel. The CMD for the NGC 1023 GC candidates is plotted in the top middle panel, detected from the CFHT/MegaCam data. As the surface density of GC candidates reaches the background at  6.2 arcmin from the centre, the CMD is plotted with the objects within that radius only. The bottom middle panel displays the colour histogram for the GC candidates and the background. The top right panel in Figure \ref{cmd} displays the CMD for the NGC 2768 GC candidates. The diagram exhibits the GCs detected using Subaru/Suprime-Cam data. Only the GC candidates detected within a galactocentric radius of 9.9 arcmin are included in the plot and the respective colour histogram for GC candidates along with background is shown in the bottom right panel. The globular cluster luminosity function (GCLF) for the detected GCs is plotted in Figure \ref{GCLF} for the three galaxies.

\begin{figure*}
\centering
\includegraphics[scale=.5]{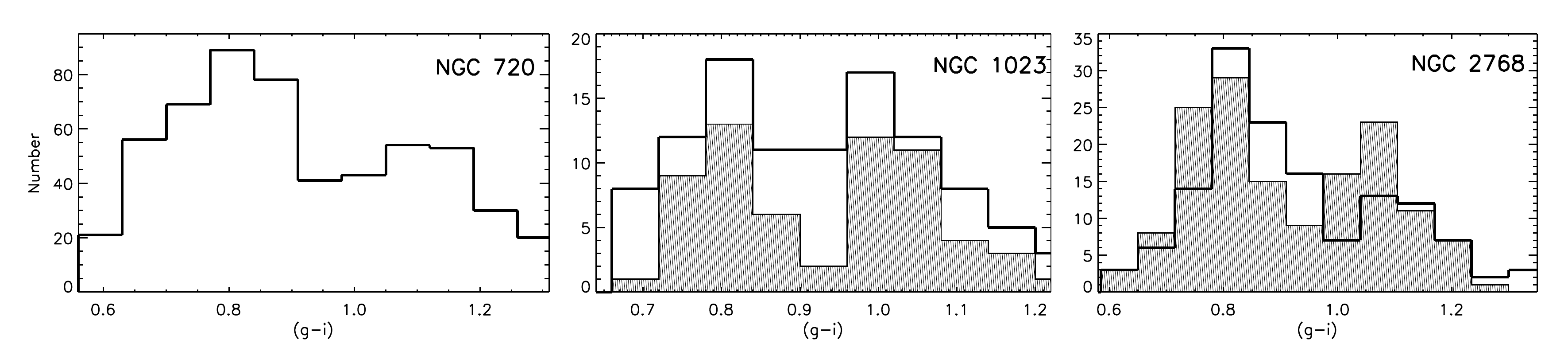}
\caption{Colour histograms of GCs after the correction for background contamination. The estimated background contamination for the respective GC system is subtracted from the total GCs and the corrected GCs are represented in histograms.  The {\it left} panel shows the final GCs of NGC 720 detected using Subaru/Suprime-Cam. The histogram shows a clear bimodal colour distribution for NGC 720. The {\it middle} panel shows the GCs of NGC 1023 observed using HST/WFPC2 (shaded area) and CFHT/MegaCam (open area) data. The {\it right} panel shows the GCs of NGC 2768 using HST/ACS (shaded area) and Subaru/Suprime-Cam (open area). The background subtraction has improved the colour histograms shown in Figure \ref{cmd} and now the peaks for the blue and red subpopulations are more distinctly seen.}
\label{histogram}
\end{figure*}

\subsection{GC bimodality}
\subsubsection{Colour histograms}
Figure \ref{histogram} illustrates the colour histograms of GCs corrected for the background contamination. The background contaminations for each GC system (shown in bottom panels of Figure \ref{cmd}), after area correction, is subtracted and the final GC colour distribution is shown. The final list of detected GCs above the turnover magnitude after background contamination correction includes 554 (Subaru) for NGC 720, 62 (HST) and 105 (CFHT) for NGC 1023, and 147 (HST) and 139 (Subaru) for NGC 2768. 

The left panel of Figure \ref{histogram} shows the NGC 720 GC colour histogram using Subaru/Suprime-Cam data. The galaxy shows a clear distinction between the blue and red subpopulations with more blue than red GCs.  The blue and red GC subpopulations of NGC 720 peak in colour around {(\it g}$-${\it i}) = 0.8 and 1.1 respectively. The middle panel shows the colour histogram for NGC 1023 GCs using HST/WFPC2 and CFHT/MegaCam data. The colour  distribution shows a bimodal nature with two peaks around {(\it g}$-${\it i}) = 0.8 and 1.05.  The right panel represents the colour histogram of NGC 2768 GCs detected using HST/ACS and Subaru/Suprime-Cam data. Both data sets show a bimodal colour distribution. The blue and red subpopulations peak in colour at {(\it g}$-${\it i}) = 0.8 and 1.1 respectively.

The CMDs and colour histograms for the three sample galaxies strengthen the bimodal distribution of GCs for the galaxies. \citet{Kissler1996} studied the GC system of NGC 720, but did not detect bimodality. \citet{Larsen2000} confirm the bimodal distribution for NGC 1023 GCs using the HST/WFPC2 data. Later, \citet{Young2012} reconfirmed the presence of multiple subpopulations in NGC 1023 using WIYN data. NGC 2768 was the only galaxy detected with a clear bimodal colour distribution in a survey of 29 S0 galaxies by \citet{Kundu2001}.

\subsubsection{Gaussian mixture modeling}
Gaussian mixture modeling (GMM) is an algorithm to statistically quantify whether a distribution is unimodal or multimodal \citep{Muratov2010}. The well known Kaye\textquoteright s Mixture Model (KMM, \citealt{Ashman1994}) algorithm is among the general class of algorithms of GMM. Based on three statistics, the GMM signifies the presence of a multimodal distribution over unimodal. They are: 1. confidence level from the parametric bootstrap method (low values indicate a multi-modal distribution), 2. separation (D) of the means relative to their widths (D > 2 implies a multi-modal distribution) and 3. kurtosis of the input distribution (negative kurtosis for multi-modal distributions).

{\bf NGC 720:} The GMM algorithm fit to the NGC 720 GC data gives a bimodal colour  distribution with two peaks at ({\it g$-$i}) = 0.793 $\pm$ 0.010 and 1.125 $\pm$ 0.012. The widths for the blue and red GCs are 0.104 and 0.090 respectively. The GMM algorithm partitions the  total GC system into 64 percent blue and 36 percent red GC subpopulations. The parametric bootstrap method rules out the unimodal distribution with a confidence level better than 0.01 percent (implying that a multimodal distribution is supported with $>$99.9 percent probability) and D = 3.42 $\pm$ 0.16 for the NGC 720 GCs.

{\bf NGC 1023:} Using GMM on the HST data, the GC system of NGC 1023 has D = 3.55 $\pm$ 0.53 supporting multi-modality.  The peaks of the blue and red subpopulations are ({\it g$-$i}) = 0.785 $\pm$ 0.015 and 1.017 $\pm$ 0.022 respectively. The estimated widths for the subpopulations are 0.033 and 0.086. The total GC system consists of 38 percent blue and 62 percent red subpopulations. The heteroscedastic fit for the GCs of NGC 1023 from CFHT data gives a blue peak at ({\it g$-$i}) = 0.799 $\pm$ 0.020 and a red peak at 1.038 $\pm$ 0.022. GMM algorithm divides the total GCs into 43 and 57 percent blue and red GCs respectively. The blue and red peaks have a width of 0.069 and  0.091 respectively. GMM provides similar peak values for the subpopulations from the two data sets.  \citet{Larsen2000} give the peak values of two subpopulations from the KMM test, i.e.  ({\it V$-$I}) = 1.02 and 1.25, which are in reasonable agreement with the values derived from GMM  i.e. ({\it V$-$I}) = 0.99 $\pm$ 0.01 and 1.26 $\pm$ 0.02. 

{\bf NGC 2768:} The GMM algorithm gives a multimodal colour distribution for the NGC 2768 GC system from the HST data. The blue and red subpopulations peak in colour around ({\it g$-$i}) =  0.821 $\pm$ 0.017 and 1.101 $\pm$ 0.025 respectively. GMM provides the widths of the two subpopulations as 0.085 and 0.109.  The value of D statistic is greater than 2.89, supporting  two well separated  subpopulations for the NGC 2768 GC system. We then applied the GMM algorithm to the GC colours from the Subaru imaging. The heteroscedastic split in GCs peak at ({\it g$-$i}) = 0.819 $\pm$ 0.015 and  1.076 $\pm$ 0.017 with respective widths of 0.075 and 0.079 for the two subpopulations. The separation between two subpopulations is 3.65, supports bimodal distribution. The total GC system is divided into 65 percent blue and 35 percent red subpopulations. 

\subsubsection{Colour - metallicity transformation}
Usher et al. (2012) give the colour - metallicity relation derived from an analysis of 903 GCs. The relation for GCs with ({\it g$-$i}) $>$ 0.77 is of the form: 
\begin{equation}
[Z/H] = [(3.49\pm0.12) \times (g-i)]+(-4.03\pm0.11).
\label{meta1}
\end{equation}
We have converted the peak colours for the GC subpopulations of the three galaxies into metallicity, and listed them in Table \ref{meta2}. The peak metallicity for the blue and red subpopulations agrees with the GC colour/metallicity - galaxy luminosity relation \citep{Peng2006, Faifer2011}.

\begin{figure*}
\includegraphics[scale=.5] {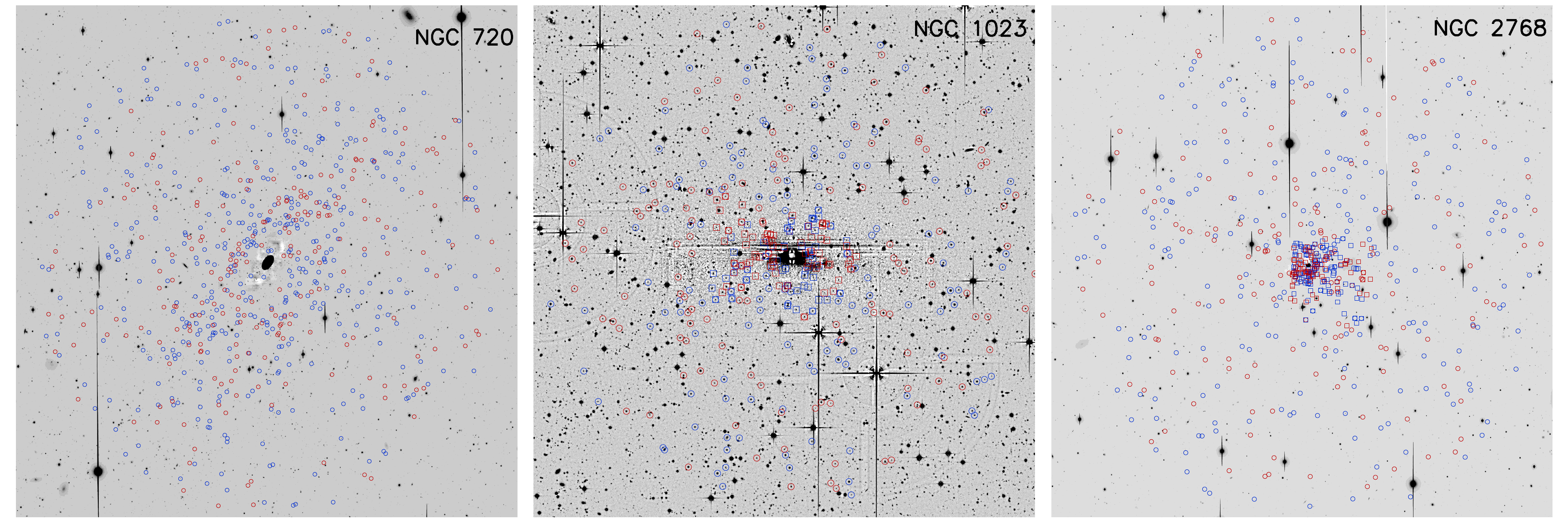}  
  \caption{Two dimensional sky images of three galaxies: NGC 720, NGC 1023 and NGC 2768. The galaxy stellar light is subtracted from the individual images with North up and East on the left. Each galaxy image covers an area of 10, 6.3 and 10 square arcmin centered on the galaxy, respectively, for NGC 720, NGC 1023 and NGC 2768. The blue and red open circles represent the positions of the blue and red GC candidates detected from the ground based telescopes, whereas the blue and red open squares represent the positions of the blue and red GC candidates detected from the HST. }
\label{sky}
\end{figure*}

\begin{table}
\centering
\caption{The peak values of colour for the blue and red GC subpopulations derived from GMM. The colour  - metallicity relation given by equation \ref{meta1} is used to derive the corresponding metallicity shown below. For NGC 1023 and NGC 2768, the peak colour and metalicity values from both data are recorded. }
\begin{tabular}{ccccc}
\hline
\multicolumn{1}{c}{Galaxy}&\multicolumn{2}{c}{Blue GCs}&\multicolumn{2}{c}{Red GCs}\\
NGC & ({\it g$-$i}) & [Z/H] & ({\it g$-$i}) & [Z/H] \\
\hline\hline
720 & 0.793$\pm$0.010 &-1.26$\pm$0.07&1.125$\pm$0.012 & -0.10$\pm$0.08\\
\multirow{2}{*}{1023}&0.785$\pm$0.015&-1.29$\pm$0.10&1.017$\pm$0.022&-0.48$\pm$0.15\\
&0.799$\pm$0.020&-1.24$\pm$0.14&1.038$\pm$0.022&-0.41$\pm$0.15\\
\multirow{2}{*}{2768}&0.821$\pm$0.017&-1.16$\pm$0.12&1.101$\pm$0.025 &-0.19$\pm$0.17\\
&0.819$\pm$0.015&-1.17$\pm$0.10&1.076$\pm$0.017 &-0.27$\pm$0.12\\
\hline\hline
\end{tabular}
\label{meta2}
\end{table}

\begin{figure}
\centering
 \includegraphics[scale=.5]{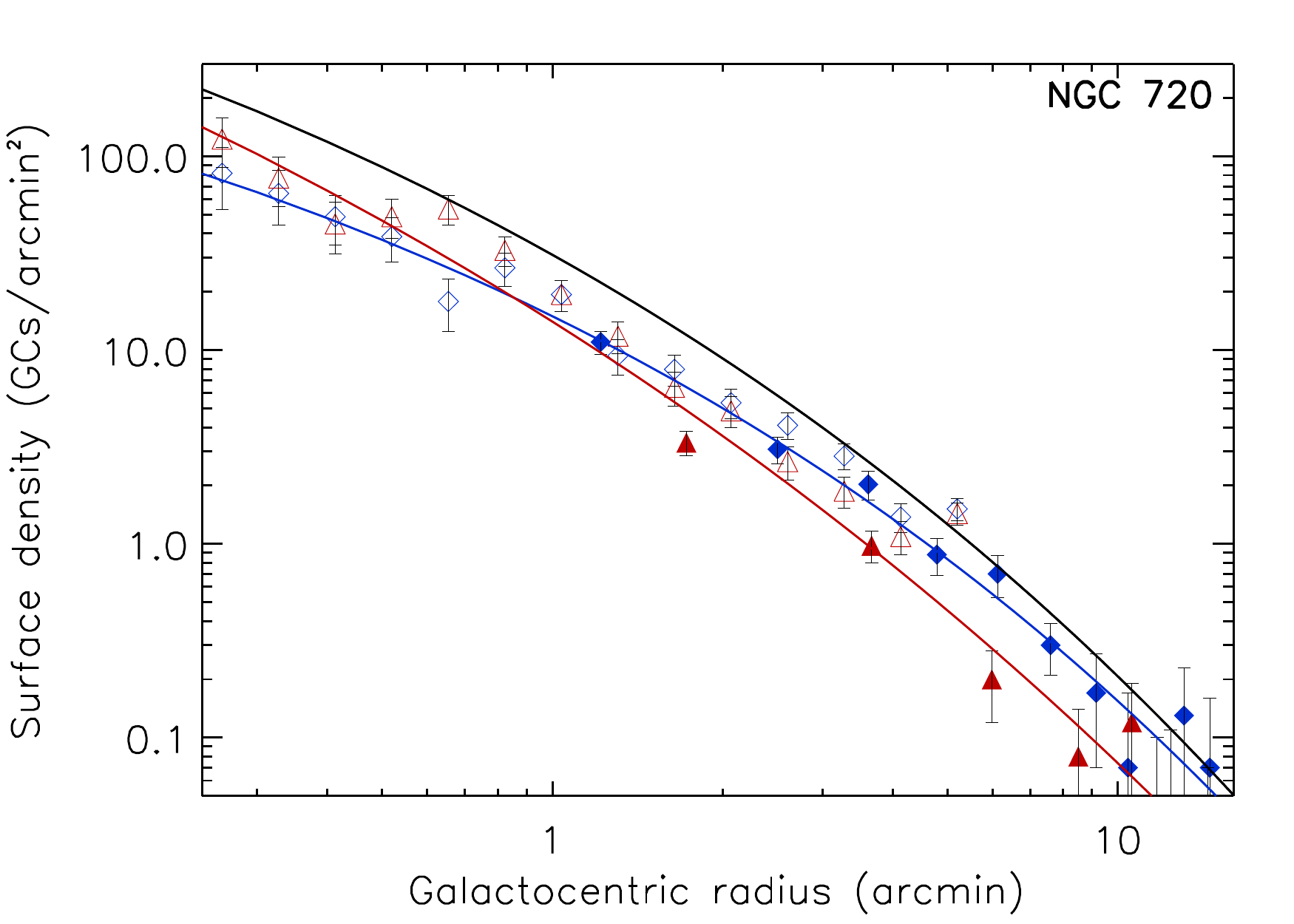} 
  \caption{GC subpopulations of NGC 720. Surface densities for the blue (diamonds) and red (triangles) GCs of NGC 720 are shown. The open and the filled symbols represent the Gemini and the Subaru data respectively. A S\'{e}rsic profile is fitted to the three GC distributions  and is displayed in respective colour solid lines along with the total system in a black solid line. }
\label{720subpop}
\end{figure}

\begin{figure}
\centering
 \includegraphics[scale=.5]{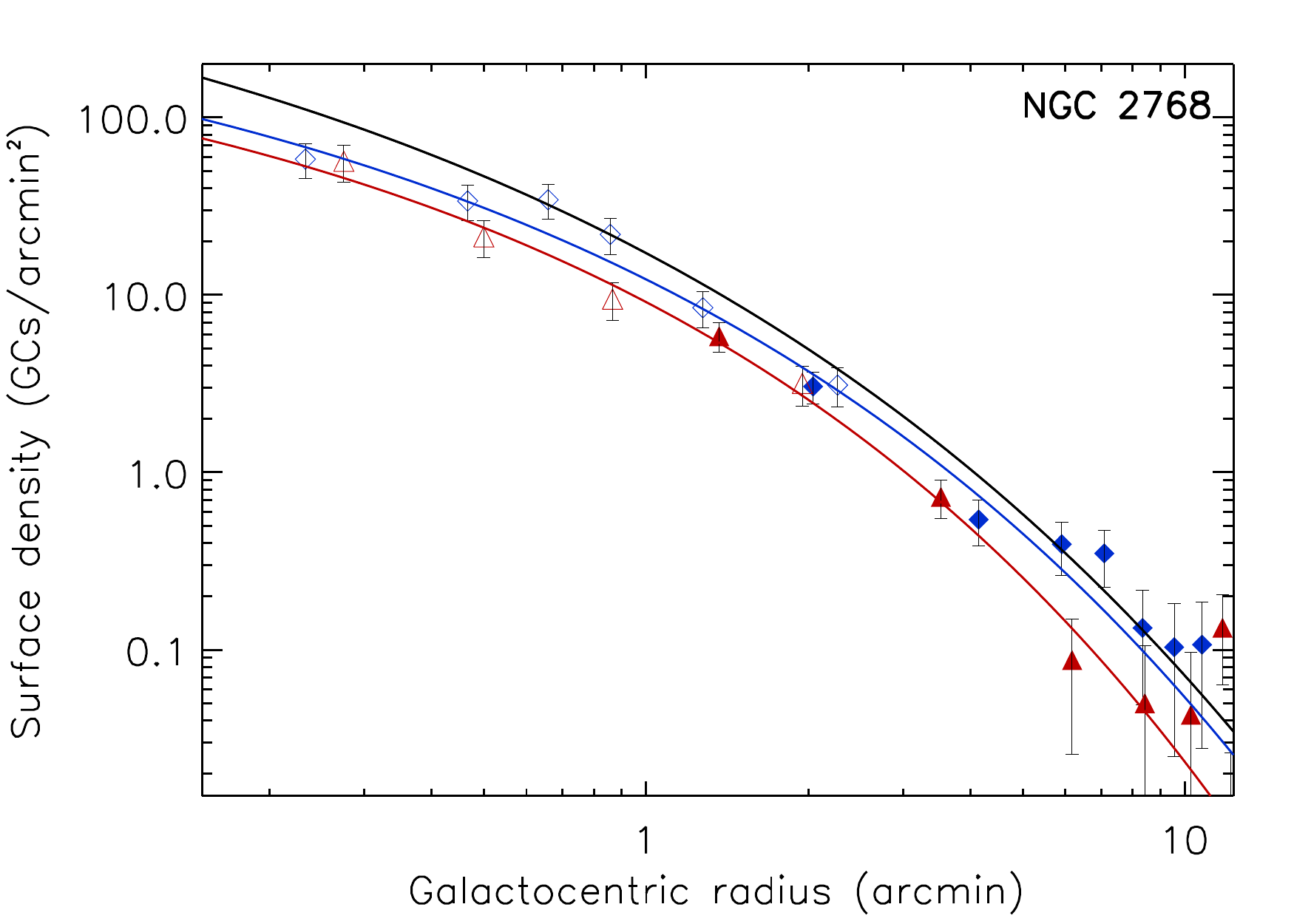}
  \caption{GC subpopulations of NGC 2768. The data sets include HST (open symbols) in the inner 2.1 arcmin radius and Subaru (filled symbols)  to 20 arcmin.  The radial density distribution for the blue (diamonds) and red (triangles) GCs are shown. The solid lines are the S\'{e}rsic profiles for the two subpopulations and the total system in black solid line. }
\label{2768subpop}
\end{figure}

\subsection{GC subpopulations}
With our high quality photometric data, we are able to separate the GC subpopulations and investigate their properties.  Figure \ref{sky} shows the two dimensional images of the three galaxies after the subtraction of galaxy stellar light. The positions of the blue and red GCs are displayed on each galaxy image. Only the GCs detected within the turnover magnitude are used in the study of GC subpopulations. First the surface density distribution of GC subpopulations with galactocentric radius is analysed. For this, the GC system of NGC 720 is classified into blue and red subpopulations dividing at the colour ({\it g$-$i}) = 0.98 (the colour at which the Gaussian distributions for the two subpopulations cross in the GMM fit).  The subpopulations are separately binned in galactocentric radius and the surface density values are calculated. Figure \ref{720subpop} displays the estimated values of background subtracted surface density for the blue and red GCs along with the total system. The Gemini and Subaru data are merged together to obtain the distribution from a galactocentric radius of 0.18 to 18 arcmin. The surface densities are fitted with a S\'{e}rsic profile (see Equation \ref{sersic}).  The fitted parameters for the blue and red GCs are recorded in Table \ref{subpop}. The blue subpopulation has a density enhancement over the red subpopulations over the whole range of radius except in the central 0.9 arcmin. The effective radius for the blue subpopulation is larger than for the red subpopulation.

Due to the small number of detected GCs within the turnover magnitude, we are unable to fit the distribution of GC subpopulations of NGC 1023.   

For NGC 2768, the GCs are classified into blue and red subpopulations at ({\it g$-$i}) = 0.96 (from the GMM fit).  The background subtracted surface density values for the blue and red subpopulations are plotted in Figure \ref{2768subpop}. Both the HST and Subaru data are incorporated in the figure. The radial density distributions for blue and red subpopulations are fitted with a S\'{e}rsic profile. Table \ref{subpop} tabulates the fitted parameters for the blue and red GC density distributions. The blue  and  red GCs have similar density profiles, with the more extended blue subpopulation. 

\begin{table}
\centering
\caption{Fitted parameters for the surface density of blue and red GC subpopulations of NGC 720 and NGC 2768. We are not able to fit the GC subpopulations of NGC 1023. }
\begin{tabular}{ccccc}
\hline
 NGC& GCs & R$_e$ & n & bg \\ 
 & &(arcmin) & & (arcmin$^{-2}$) \\ 
\hline\hline
\multirow{2}{*}{720}&Blue & 3.93$\pm$2.30 &4.78$\pm$2.30 & 0.63$\pm$0.06\\ 
       &Red& 1.33$\pm$0.31 &5.55$\pm$2.53 & 0.39$\pm$0.04 \\ 
\hline
\multirow{2}{*}{2768} &Blue&1.83$\pm$0.27 & 2.78$\pm$0.64 & 0.33$\pm$0.03\\
         &Red&1.50$\pm$0.23&2.53$\pm$0.79&0.25$\pm$0.05\\
\hline
\end{tabular}
\label{subpop}
\end{table}

\subsection{Radial colour distribution}
\begin{figure}
\centering
 \includegraphics[scale=.5]{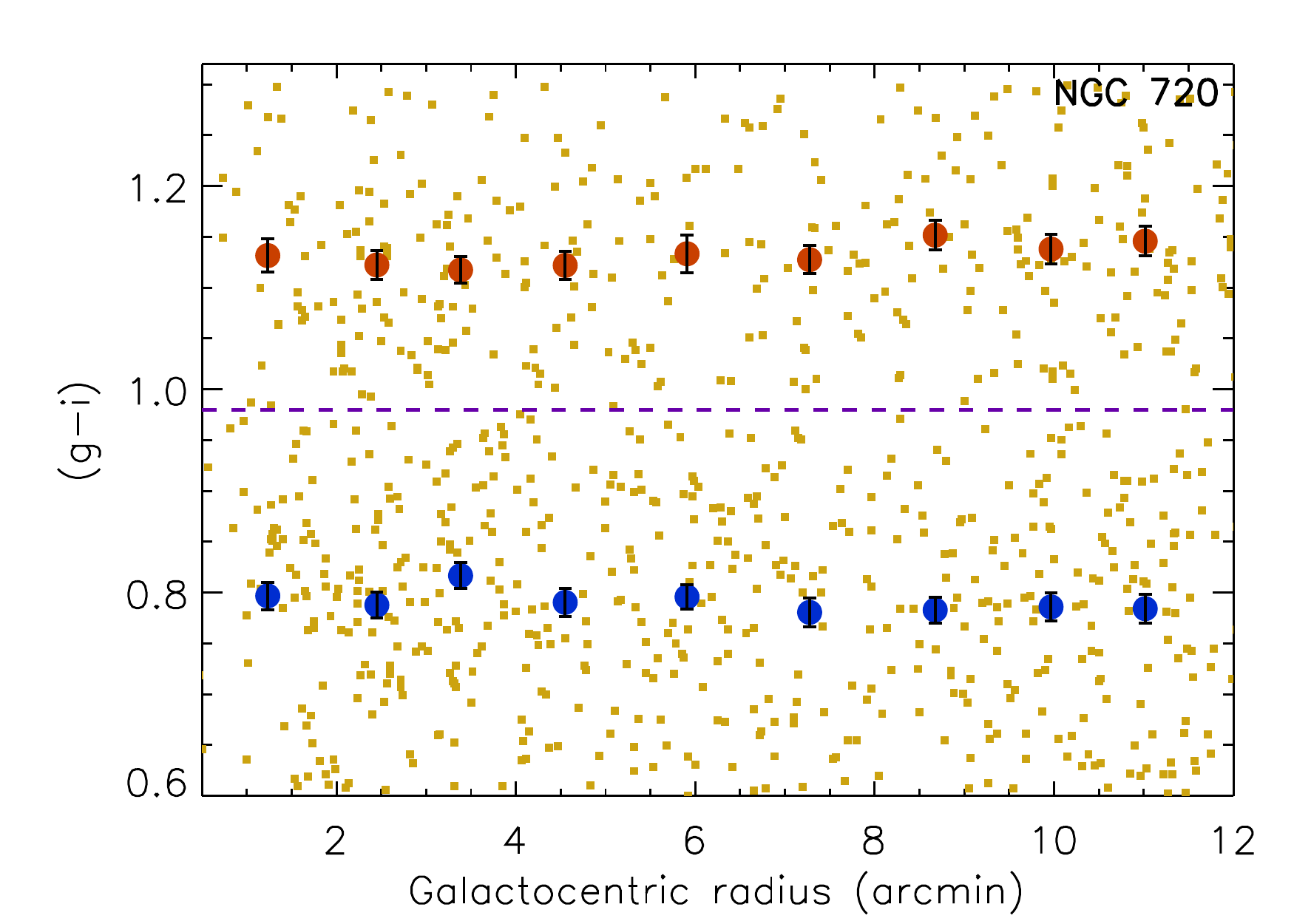} 
 \caption{Colour distribution of NGC 720 GC system with galactocentric radius. The individual GCs from the Subaru data are represented by yellow squares. The mean colours over  particular bins in radius are shown as filled circles for Subaru data.  The separation for blue and red GCs is shown with a dashed line at ({\it g$-$i}) = 0.98.  }
\label{720colour}
\end{figure}

The blue and red subpopulations of NGC 720 are separated at a colour of ({\it g$-$i}) = 0.98.  The average colour in each radial bin is estimated separately for blue and red subpopulations.    Neither the red nor the blue subpopulations from the Subaru data reveal a colour gradient. The average colour values for the two subpopulations with galactocentric radius are displayed in Figure \ref{720colour}. 

\begin{figure}
\centering
 \includegraphics[scale=.5]{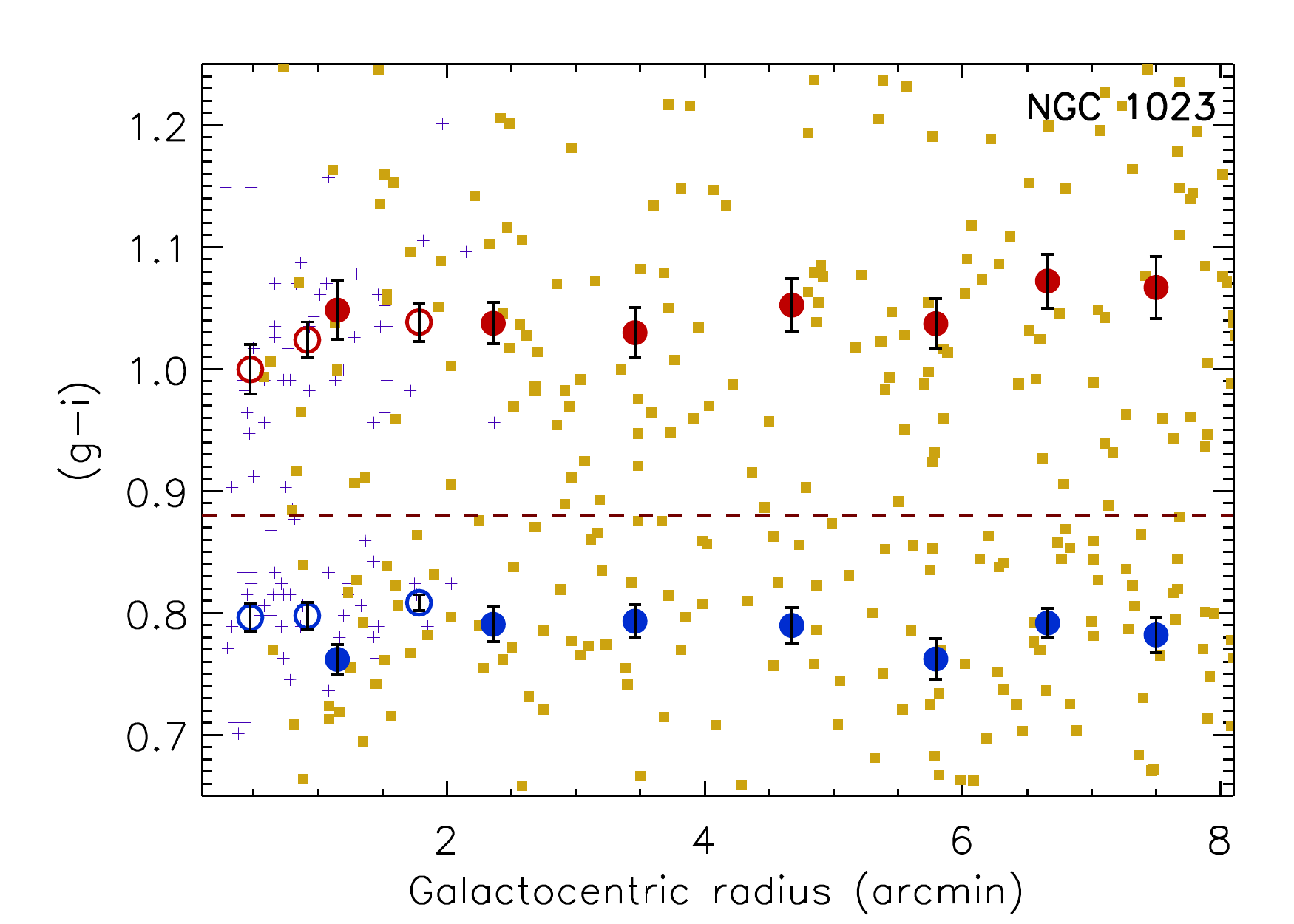} 
 \caption{Colour distribution of NGC 1023 GC system with galactocentric radius. The plot shows the average colour for the blue and red subpopulations using HST (open circles) and CFHT  (filled circles) data. The individual GCs are represented by plus signs (HST) and squares (CFHT).   The separation between the blue and the red GCs is shown with a dashed line at ({\it g$-$i}) = 0.88. The blue GCs show a constant colour with galactocentric radius, while the red GCs show a positive slope (0.028 $\pm$ 0.009 mag per arcmin) in the inner region and a constant colour for larger radii. }
\label{1023colour}
\end{figure}

The separation between the two subpopulations for NGC 1023 GCs is ({\it g$-$i}) = 0.88 (from the GMM fit) for the HST and the CFHT data.  The averaged colour values in each radial bin for the HST  and the CFHT data sets are plotted in Figure \ref{1023colour}. The individual GCs from the HST and CFHT are also plotted in the figure. A positive colour gradient is visible for the HST red subpopulation (slope = 0.028 $\pm$ 0.009 mag per arcmin). 

\begin{figure}
\centering
 \includegraphics[scale=.5]{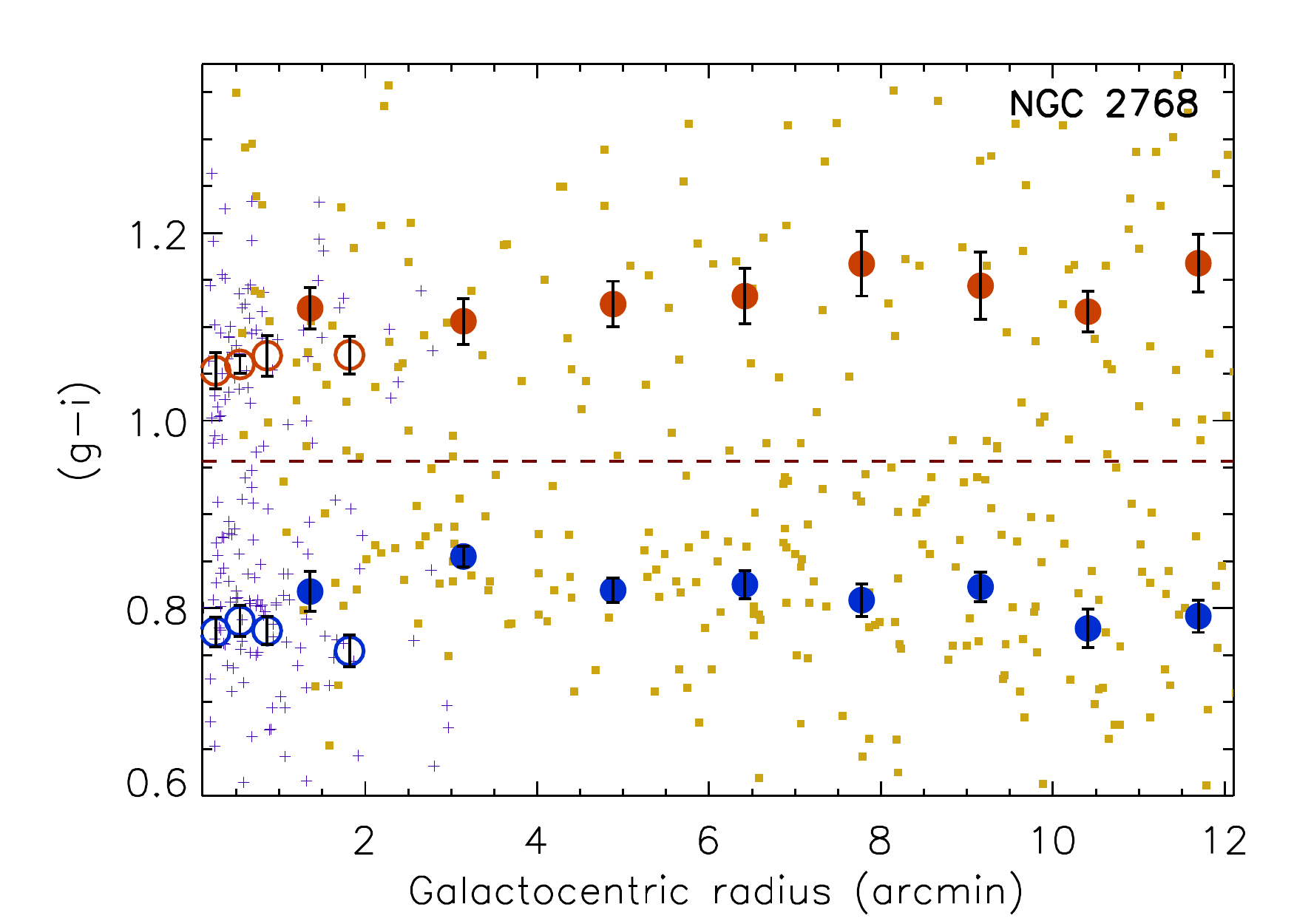} 
 \caption{Colour distribution of NGC 2768 GC system with galactocentric radius.  The HST (open circles) and the Subaru (filled circles) data are incorporated in this figure. The average colour values in radial bins for the blue and the red subpopulations are represented by blue and red circles respectively. The individual GCs from the HST (plus signs) and the Subaru (squares) are also displayed in the figure. The separation for blue and red GCs is shown with a dashed line at ({\it g$-$i}) = 0.96. The blue GCs selected from the HST data show a slight negative gradient with a slope of 0.007 $\pm$ 0.002 mag per arcmin.}
\label{2768colour}
\end{figure}

Figure \ref{2768colour} shows the radial colour distribution for the blue and red GCs of NGC 2768 to a galactocentric distance of 12 arcmin from the centre. The GCs are categorised  into blue and red subpopulations at ({\it g$-$i}) = 0.96. Figure \ref{2768colour} displays the individual GCs from the HST and the Subaru data.  The radial colour distribution from the Subaru data does not show any statistically significant radial trend, which might be caused by the contamination from the ground based data. But the inner blue GCs from the HST data show a slight negative slope (0.007 $\pm$ 0.002 mag per arcmin).

The radial colour distribution is an important tool to study different GC formation scenarios. In the cases of NGC 1407 and M87, both GC subpopulations show a negative colour gradient, supporting an in-situ dissipative formation scenario for the GCs. Beyond a transition radius, the GCs do not show a colour gradient. The GCs exterior to the transition region may be formed by ongoing accretion/mergers. The data used for the NGC 1407 study \citep{Forbes2011} came from three band imaging with subarcsecond seeing using the Subaru telescope. The colour gradient observed for the M87 GCs \citep{Harris2009b} was taken with multi-band filters using the CFHT and the seeing for the observation was 0.8 arcsec. 

\begin{figure}
\centering
\includegraphics[scale=.5]{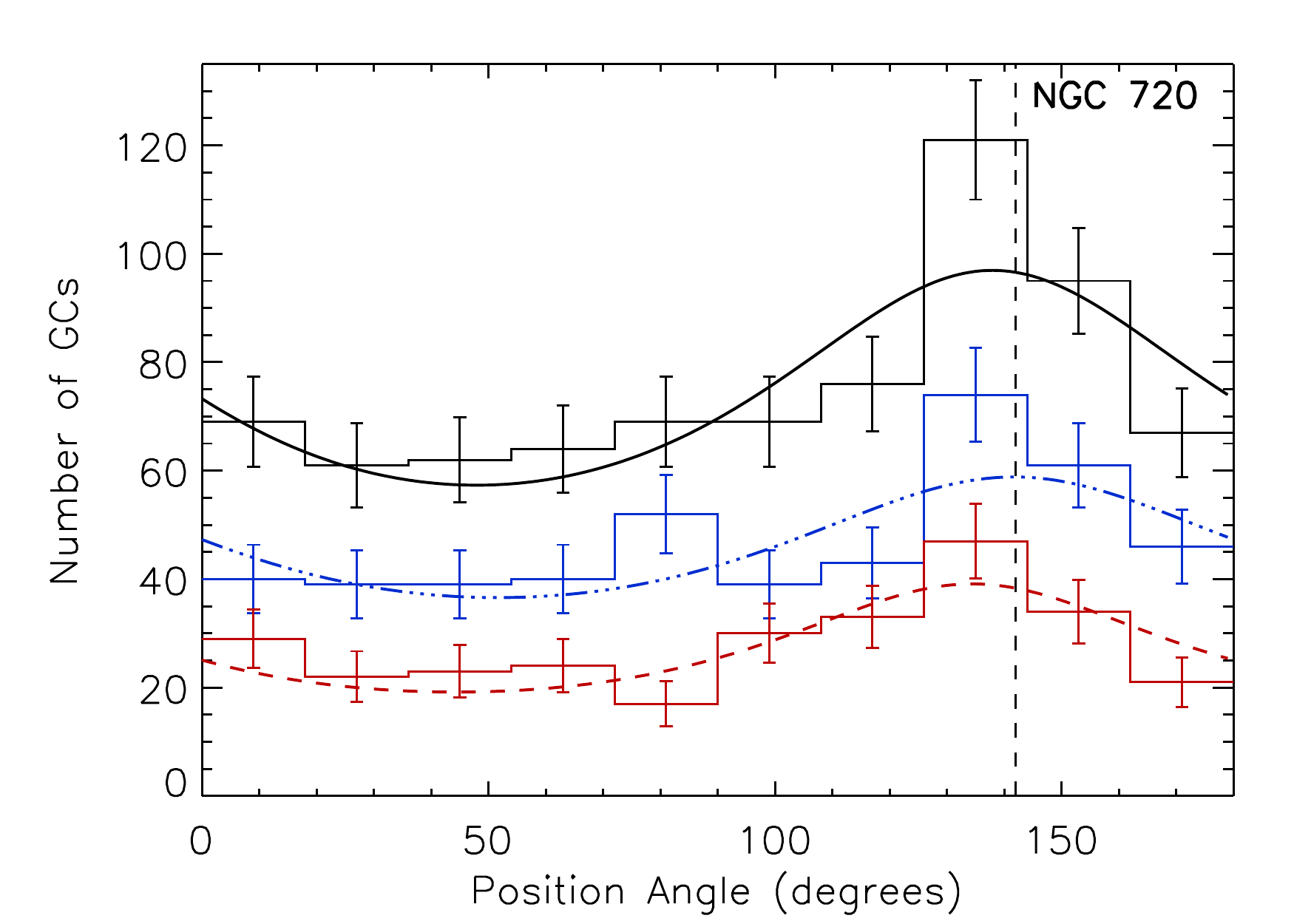}  
\caption{Azimuthal distribution of NGC 720 GCs. The histograms in black, blue and red represent the azimuthal distribution of total, blue and red GCs of NGC 720 respectively.  The distribution is fitted with the profile given by equation \ref{azimuth} which is also plotted in the figure as solid (total system), dotted (blue subpopulation) and dashed (red subpopulation) lines. The host galaxy starlight (dashed vertical line) is aligned at a position angle of 142 $\pm$ 5 degrees which matches with the total system, the blue and red subpopulations of GCs. } 
\label{azi_720}
\end{figure}

\subsection{Azimuthal distribution}
We study the azimuthal distribution of  the GC systems and  their blue and red subpopulations.  The position angles of individual GCs ({\it $\theta$}) are estimated from the Right Ascension and Declination from the centre of the galaxy keeping 0 degree for North and measuring counter-clockwise.  We binned the GCs in wedges of 18 degrees and fitted  a profile \citep{McLaughlin1994} of the form:
\begin{eqnarray}\nonumber
 \sigma(R,\theta) &=& kR^{-\alpha} \left[cos^2(\theta - \text{PA}) + \right. \\
                               && \left. (1-\epsilon^2)^{-2} sin^2(\theta - \text{PA})\right]^{-\alpha/2} + bg
\label{azimuth}
\end{eqnarray}
where {\it $\sigma(R,\theta)$} is the azimuthal distribution of GCs at radius, {\it R} and angle {\it $\theta$}, {\it $\alpha$ }is the power law index fitted to the surface density of GCs, {\it bg} is the background estimated from the S\'{e}rsic fits (see  Section 3.1) and {\it k} is the normalization constant. The profile is iterated with the position angle of the GC system (PA) and the ellipticity ($\epsilon$) as free parameters. 

\begin{figure}
\centering
\includegraphics[scale=.5]{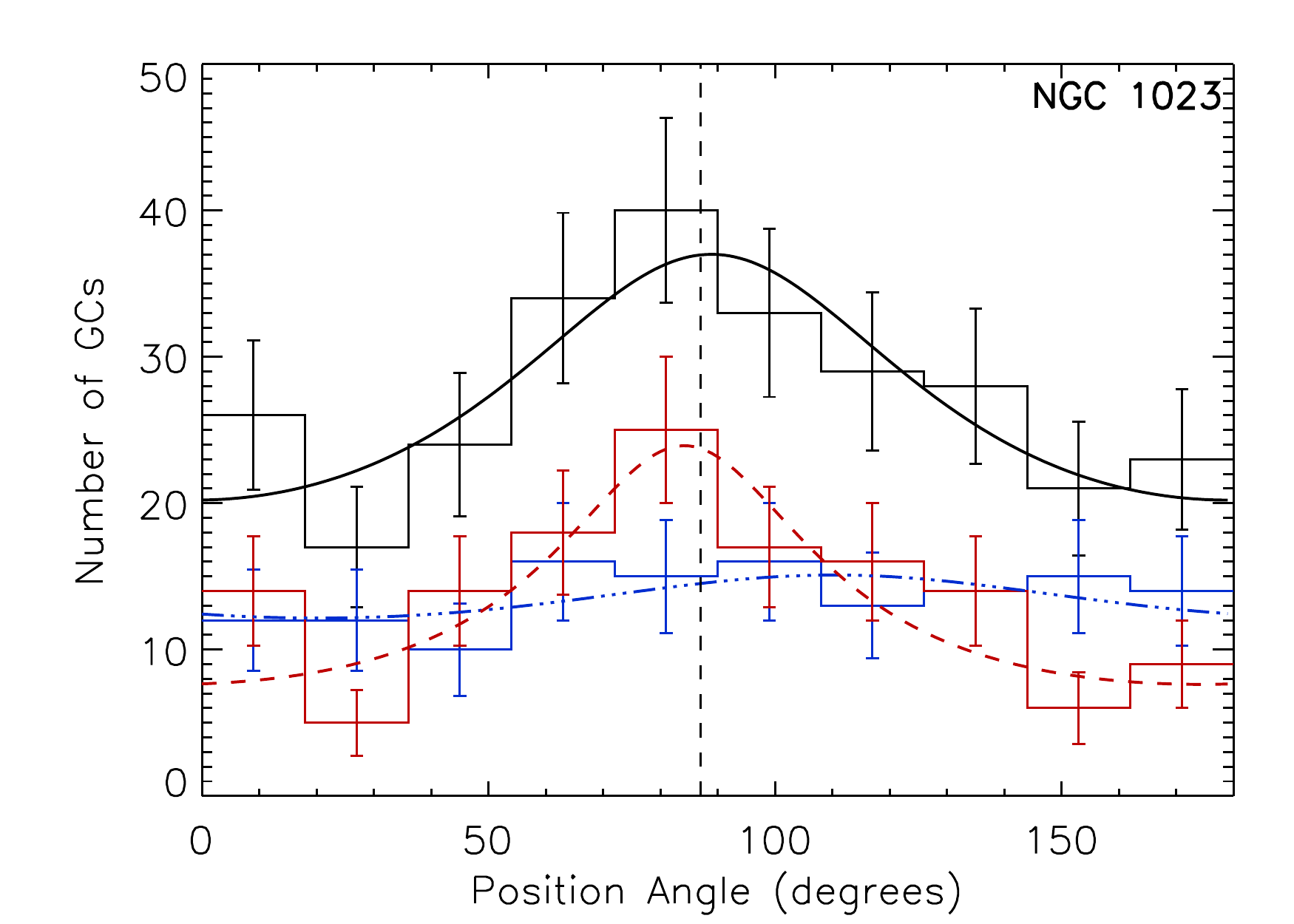}
\caption{Azimuthal distribution of NGC 1023 GCs. The density of total system of GCs and blue and red GC subpopulations are shown in black, blue and red histograms. The fitted lines (same patterns and colours as given in Figure \ref{azi_720}) represent the profile given by equation \ref{azimuth} for NGC 1023 GCs. The dashed vertical line represents the position angle of the galaxy light (PA = 87 degrees). The total system and red subpopulation of NGC 1023 GCs are arranged in elliptical rings along the position angle of the galaxy light. In contrast the blue subpopulation shows a nearly flat azimuthal distribution (indicating a more circular distribution).}
\label{azi_1023}
\end{figure}

\begin{table*}
\centering
\caption{Position angle and ellipticity for the GC systems of NGC 720, NGC 1023 and NGC 2768. The values for the GCs are determined by fitting equation \ref{azimuth} to the histograms of azimuthal distribution. The table displays the values of the parameters for the total system, blue and red GCs along with the host galaxy stellar properties  obtained from HyperLeda \citep{Paturel2003}.}
\begin{tabular*}{0.98\textwidth}{cccc|cccc|cccc}
\hline
\multicolumn{4}{c}{NGC 720} &\multicolumn{4}{c}{NGC 1023} &\multicolumn{4}{c}{NGC 2768}\\
\hline
 & Type & PA & $\epsilon$  & & Type & PA & $\epsilon$  & & Type & PA & $\epsilon$ \\
 &           & (degrees) &   & &            &(degrees) & &   &          &(degrees) & \\
 \hline\hline
 Galaxy & Stars & 142$\pm$5& 0.47$\pm$0.05 &  Galaxy & Stars & 87$\pm$5& 0.58$\pm$0.05 &  Galaxy & Stars & 93$\pm$3 & 0.60$\pm$0.03 \\
 GCs &Total& 138$\pm$6 & 0.28$\pm$0.06 &   GCs &Total&89$\pm$7 &  0.35$\pm$0.09 &  GCs &Total&89$\pm$2 & 0.59$\pm$0.03\\
 GCs & Blue  & 142$\pm$8 & 0.26$\pm$0.06 &  GCs & Blue & 110$\pm$32 & 0.15$\pm$0.15 & GCs & Blue & 90$\pm$3 & 0.57$\pm$0.04\\
 GCs & Red    & 134$\pm$6 & 0.37$\pm$0.08 &  GCs  & Red & 84$\pm$6 &0.57$\pm$0.08 &  GCs  &Red & 87$\pm$3 & 0.60$\pm$0.05\\
 \hline  
\end{tabular*}
\label{value_azi}
\end{table*}

\subsubsection{NGC 720}
For NGC 720, the position angle of the galaxy light is 142 degrees and the number of GCs in the azimuthal distribution peaks around 138 degrees for the total  GC population (see Figure \ref{azi_720}). The ellipticity  value determined for the total GC system is 0.28 $\pm$ 0.06, while the galaxy light has an ellipticity of 0.47 $\pm$ 0.05. The GC system of NGC 720 matches with the galaxy light in position angle but not in ellipticity.  The azimuthal distribution is also determined for the blue and red subpopulations and  recorded in Table \ref{value_azi}.   Both the blue and red subpopulations are aligned along the position angle of the galaxy light. Also the ellipticity of the red subpopulation is in good agreement with the galaxy stellar light. \citet{Kissler1996} studied the  shape of  the GC system and the host galaxy. They estimated the position angle and ellipticity for the GC system as 147 $\pm$ 10 degrees and 0.5 $\pm$ 0.1, whereas the starlight had 142 $\pm$ 3 degrees and 0.45 $\pm$ 0.05 respectively. We conclude that our findings about the position angle and ellipticity of the GC system of NGC 720 matches well with \citet{Kissler1996}. They have also found that the position angle (115 $\pm$ 15 degrees) and ellipticity (0.2 - 0.3, \citealt{Buote1994}) of the X-ray gas in NGC 720  differs from those shown by both the host galaxy stars and the total GC system.  We note that the ellipticity of the X-ray gas and the blue subpopulation are in reasonable agreement. Although the  ellipticities are consistent, we note that the PAs are not. This consistency in ellipticities implies that both the X-ray gas and blue subpopulation might have a common dynamical behaviour and hence strengthens the connection between the blue subpopulation and galaxy haloes \citep{Forbes2012a}. 

\subsubsection{NGC 1023}
The azimuthal distribution for the NGC 1023 GCs are shown in Figure \ref{azi_1023}.  The profiles obtained from equation \ref{azimuth} are fitted to the different GC subpopulations and displayed in Figure \ref{azi_1023}. The photometric position angle for the galaxy NGC 1023 is 87 degrees and the best fitted profile for the total and red GCs peaks at a similar values within errors.  The red GCs of NGC 1023 are aligned along the position angle of the galaxy light with ellipticity, $\epsilon$ = 0.57 $\pm$ 0.08. The best fitted profile generated by equation \ref{azimuth} for the blue GCs shows a flat distribution. The profile peaks at 110 $\pm$ 32 degrees and represents a nearly circular distribution for the blue subpopulation of NGC 1023.

\subsubsection{NGC 2768}
\begin{figure}
\centering
\includegraphics[scale=.5]{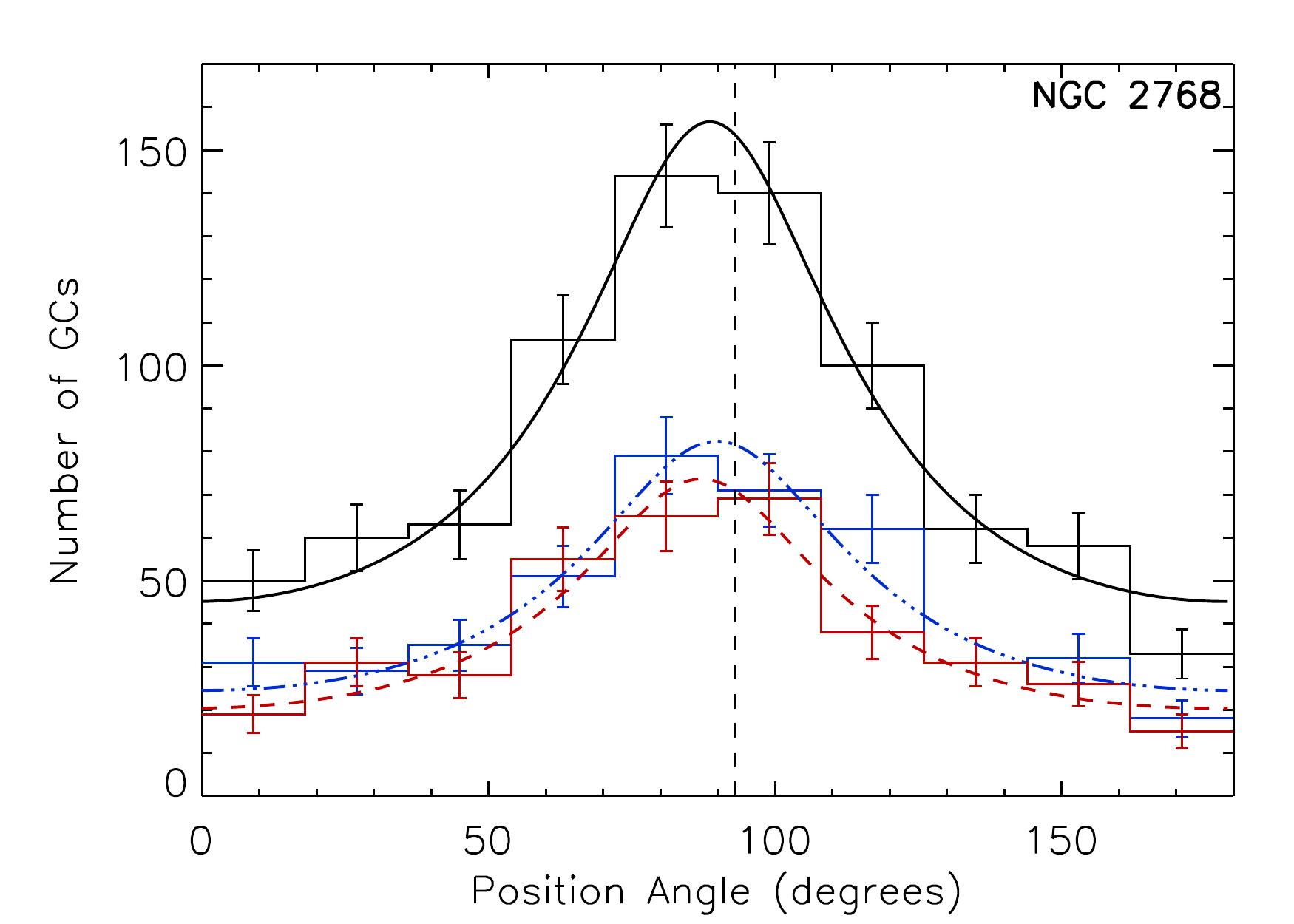}
\caption{Azimuthal distribution of NGC 2768 GCs. The histograms in black, blue and red represent the azimuthal distribution of total, blue and red GCs of NGC 2768. The fitted lines (same patterns and colours as given in Figure \ref{azi_720}) represent the profile given by equation \ref{azimuth}  for NGC 2768 GCs. The position angle of the galaxy stellar light (PA = 93 degrees) is represented by the dashed vertical line. The total system, red and blue GCs of NGC 2768 have an ellipticity value of 0.58 $\pm$ 0.06. The position angles of the GC system and subpopulations match with the galaxy light of NGC 2768.  }
\label{azi_2768}
\end{figure}

Figure \ref{azi_2768} displays the azimuthal distribution of the total system, blue and red  subpopulations of NGC 2768 GCs. The distributions are fitted with sinusoidal profiles given by the equation \ref{azimuth} and are shown in the figure. Table \ref{value_azi} displays the position angle and ellipticity values estimated from the fitted profiles.  Both the blue and red GC subpopulations are distributed with $\epsilon$ $\sim$ 0.58 along the position angle of galaxy light (PA = 93 degrees). In addition, the estimated values for the total GC system match well with both the subpopulations. 

\

\subsection{Specific frequency}
Two key properties of a GC system that can be estimated accurately using wide-field imaging data are the total number of GCs and the specific frequency.  The specific frequency (S$_N$) of a GC system is  the total number of GCs in a galaxy per unit host galaxy luminosity.  In order to compare the GC systems of galaxies, the value of S$_N$ is a useful parameter. The value of S$_N$ may be dependent on galaxy morphology, mass, luminosity and environment.  For elliptical and lenticular galaxies, the value of S$_N$ ranges from 2 to 6 \citep{Elmegreen1999, Harris1991} depending on the host galaxy mass/luminosity.  The value of S$_N$ is defined  by the relation of \citet{Harris1981}: 
\begin{equation}
\text{S}_N = \text{N}_\text{GC} ~10^{0.4(M_V^T + 15)}.
\label{sn}
\end{equation}
  
The parameter N$_\text{GC}$ (the total number of GCs) is estimated from the surface density distribution of GC systems.  To determine the total number of GCs,  the area under the S\'{e}rsic profile fitted to the radial density distribution of GCs (from the centre out to the radius at which it reaches the background) is integrated and then doubled (by assuming a symmetric GC luminosity function, only GCs within the turnover magnitude have been counted).  M$_V^T$ in equation \ref{sn} represents the total absolute magnitude in the V band.
 
 For NGC 720, NGC 1023 and NGC 2768, the total number of GCs is estimated to be 1489 $\pm$ 96,  548 $\pm$ 59 and 744 $\pm$ 68 respectively. The total visual magnitude for the respective galaxies is M$_V^T$ = $-$21.68 $\pm$ 0.05, $-$21.07 $\pm$ 0.06 and $-$21.91 $\pm$ 0.1 mag. Hence the specific frequency of GCs in NGC 720, NGC 1023 and NGC 2768 is calculated to be 3.2 $\pm$ 0.2, 1.8 $\pm$ 0.2 and 1.3 $\pm$ 0.1. 
 
  \citet{Kissler1996} estimated the total number of GCs for NGC 720 to be 660 $\pm$ 190. They derived a specific frequency of 2.2 $\pm$ 0.9. The GC extent used to derive these properties is 2.67 arcmin, but the extent from our study is 9.8 $\pm$ 0.8 arcmin. The difference in the estimation of GC extent is responsible for the difference in N$_\text{GC}$ and hence S$_N$. For NGC 1023, \citet{Young2012} estimated  N$_\text{GC}$ = 490 $\pm$ 30 and S$_N$ = 1.7 $\pm$ 0.3 for the GC system of NGC 1023.  With the estimation of a similar extent for GC system of NGC 1023, we have derived N$_\text{GC}$ =  548 $\pm$ 59 and S$_N$ = 1.8 $\pm$ 0.2. Both the estimations are in good agreement with each other for NGC 1023. \citet{Kundu2001} studied the GC system of NGC 2768 using HST/WFPC2 data and calculated the total number of GCs in their field of view as 343 with a local S$_N$ of 1.2 $\pm$ 0.4 using M$_V^{FOV}$ = $-$21.2. The estimated N$_\text{GC}$ using our wider field of view is double the number determined from the smaller field of view of WFPC2. We note that NGC 2768 is found to have a lower S$_N$ value compared with S0 galaxies of similar luminosity \citep{Brodie2006}.
    
\section{Global relations of GC systems}
 In this section, we explore five global scaling relations between the GC systems and their host galaxy.  Along with the above discussed three galaxies, we include 33 literature studied galaxies plus four (NGC 821, NGC 1407,  NGC 4278 and NGC 4365) galaxies from earlier SLUGGS studies. We have  carried out a selection of galaxies based on their available GC system properties and used the same selection criteria as adopted in \citet{Spitler2008}, followed from \citet{Rhode2005}.  The main criteria followed for the selection of literature galaxies are: the GC system must have been observed in two filters with an estimate of total GC number, the fraction of blue to red GCs must have been given or can be calculated and the uncertainties in the estimated parameters should be < 40 percent. In our sample of 40 galaxies selected for this scaling relation study, three lack an estimate of GC system extent and the other two lack  the ratio of blue to red GCs, but all have a reliable total GC number estimate. 

\subsection{GC system extent versus galaxy stellar mass}
\citet{Rhode2007} and \citet*{Rhode2010}  have given a relation between the radial extent of a GC system and the host galaxy stellar mass for 11 galaxies. The extent of a GC system is defined as the radial distance at which the  GC surface density distribution reaches the background. The host galaxy mass is estimated from the absolute visual magnitude making use of mass to light ratios given by \citet{Zepf1993}. The mass to light ratios applied for the different Hubble types are as follows : {\it M/L} = 10 for elliptical galaxies, 7.6 for S0 galaxies, 6.1 for Sa - Sb galaxies and 5 for Sc galaxies. Before discussing the GC extent versus galaxy stellar mass relation, we discuss the possible sources of error.

The galaxy stellar mass is derived from the galaxy V band magnitude, distance and mass to light ratio. Measurement of the total magnitude involves a typical error of 0.05 to 0.2 mag. Another large error comes from the mass to light ratio for different galaxy morphologies.  For a given Hubble type, the mass to light ratio for a sample of galaxies is not constant. For example,  NGC 1316 is included as an elliptical galaxy, and assumed to have a value of  {\it M/L$_V$} = 10 \citep{Zepf1993}. However, \citet{Shaya1996} found a lower mass to light ratio of 2.2 for the galaxy. A possible explanation for the lower value is the presence of an intermediate age stellar population \citep{Shaya1996, Kuntschner2000}. Estimation of mass to light ratios for individual galaxies is a difficult process. Here we use the \citet{Zepf1993} values, but note the potentially large source of error. 

\begin{figure}
\centering
\includegraphics[scale=.47]{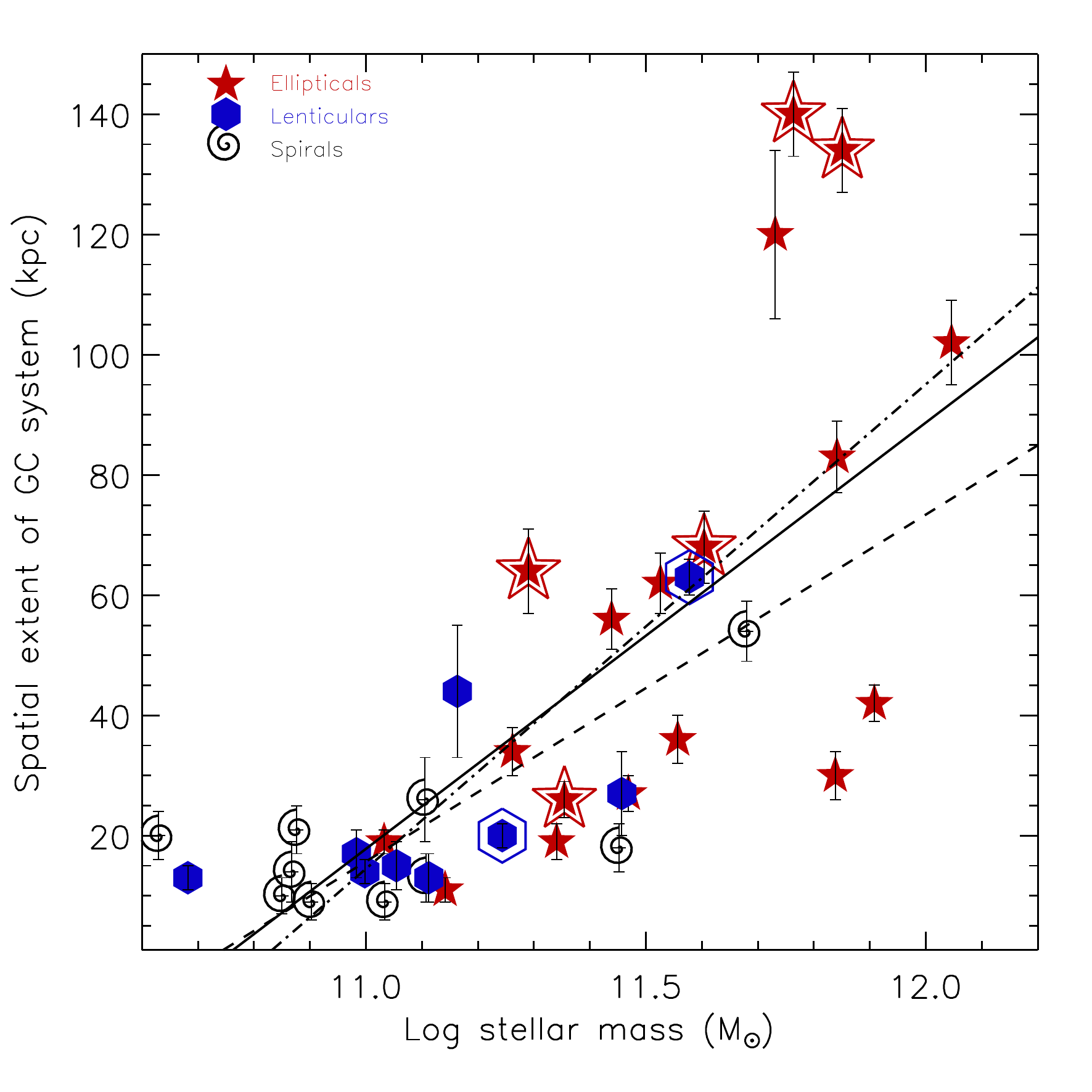}
\caption{Radial extent of GC system versus log galaxy mass. The galaxies studied by the  SLUGGS survey are represented with double star and double hexagon symbols, while others represent galaxies studied using wide-field photometry from the literature. Elliptical, lenticular and spiral galaxies  are drawn with star, hexagon and spiral symbols respectively. The linear fits  given by equations 7 and 10 are shown by a straight line and a dot-dash line respectively. The dashed line depicts the linear relation given by equation \ref{linRh07} from \citet{Rhode2007}.  }
\label{GCext1}
\end{figure} 

Errors in the GC system extent include the galaxy distance errors, the bin size errors involved in GC surface density distribution and issues due to image quality.  The main component determining a precise GC spatial extent is the imaging quality. In order to completely observe the extent of a GC system, wide-field imaging data from a large aperture telescope must be used.  Imaging data needs to be observed in good seeing conditions which reduces the contamination in point source identification.  Also GC selection from multi-filtered imaging data reduces the contamination. For example, NGC 720 has a GC system 3 times larger (this work) than the literature estimate \citep{Kissler1996}, with the use of better quality and wider field data. The amount of contamination in a three filter imaging data can be as low as $\sim$5 percent \citep{Romanowsky2009}. Hence accurate estimations of GC system extent using wide-field imaging data is needed to reduce errors. 

 We have expanded the \citet{Rhode2007,Rhode2010} studies of GC system extent versus host galaxy mass (for 11 galaxies) by including another 26 galaxies: three from this work, four from the earlier SLUGGS studies and nineteen from other literature studies (as the GC system extent is not estimated for the other three galaxies). Table \ref{GCext} tabulates the distance, total visual magnitude, estimated galaxy mass and the GC extent for the sample of 37 galaxies. The extent of GC systems against the host galaxy mass is plotted and displayed in Figure \ref{GCext1}.  As the galaxy mass increases, it is evident from the figure that the extent of GC systems grows. Or in other words, more massive galaxies accommodate more extended GC systems.

 The best fitted linear and second order polynomial (not shown in the Figure \ref{GCext1}) are:
 \begin{eqnarray}
  y &=& [(70.9 \pm 11.2) \times x ] - (762 \pm 127) \\
  y &=& [(40.9 \pm 4.3)  \times x^{2}] - [(879 \pm 97)  \times x ]   \\
        && +(4726 \pm 546) \nonumber
\label{linGCeq2b}
\end{eqnarray}
respectively, where \textit{x} is the log stellar mass in M$_\odot$ and \textit{y} is the spatial extent of GC system in kpc. Figure \ref{GCext1} also shows the linear fit from \citet{Rhode2007}:  
\begin{equation}
 y = [(57.7 \pm 3.7) \times x] - (619 \pm 41).  
\label{linRh07}
\end{equation}
The slope of the linear fit  has changed with the addition of more galaxies and is steeper than in \citet{Rhode2007}.  The second order polynomial fit given by Equation 8 also provides a reasonable match to the data. 
 
In order to better understand the relation between the GC system extent and host galaxy mass, we analysed the host galaxy's morphology. The total sample is divided into different Hubble types and shown with different symbols in the figure (see the caption of Figure \ref{GCext1}). Spiral galaxies are positioned at the bottom left side of the figure. Since the extent of a GC system for spiral galaxies in the sample  is found to be independent of the host galaxy mass, we  did a separate analysis excluding them.  In the total sample of galaxies, we have seventeen elliptical galaxies and ten lenticular galaxies. Although most of the early-type galaxies agree well with the fitted linear relation (within error bars), some are displaced from the fit (i.e. NGC 4365, NGC 1407, NGC 4374).  Another linear fit is carried out only for the 27 early-type galaxies and is shown in Figure \ref{GCext1}, which is given by:  
\begin{equation}
y = [(80.5 \pm 15.7)  \times x ] - (872 \pm 179). \\
\label{linGC2}
\end{equation}
 
It is evident from Figure \ref{GCext1} that the spatial extent of GC systems is larger for more luminous early-type galaxies. 

With this limited sample of galaxies, we conclude that the spatial extent of GC systems is  proportional to the host galaxy stellar mass. This result is in agreement  with \citet{Rhode2007}, but our linear fit is steeper than \citet{Rhode2007} (since the majority of their sample was spiral galaxies), when more galaxies are included. The main errors affecting the relation are the image quality and the assumed constant mass to light ratios for  galaxies of individual Hubble type.  With our sample of galaxies, we also infer that the extent of a GC system is only weakly dependent on galaxy stellar mass for late-type galaxies.

\subsection{GC extent versus galaxy effective radius}

Given the errors associated with determining galaxy stellar mass, we now examine galaxy effective radius. The effective radius (R$_e$) is defined as the galaxy radius comprising half of the total luminosity.  We exclude the late-type galaxies from this analysis because the effective radius for late-type galaxies includes the bulge plus extended disc components, but only the bulge component for early-type galaxies. This is done for the sake of uniformity.

\begin{figure}
\centering
\includegraphics[scale=.47]{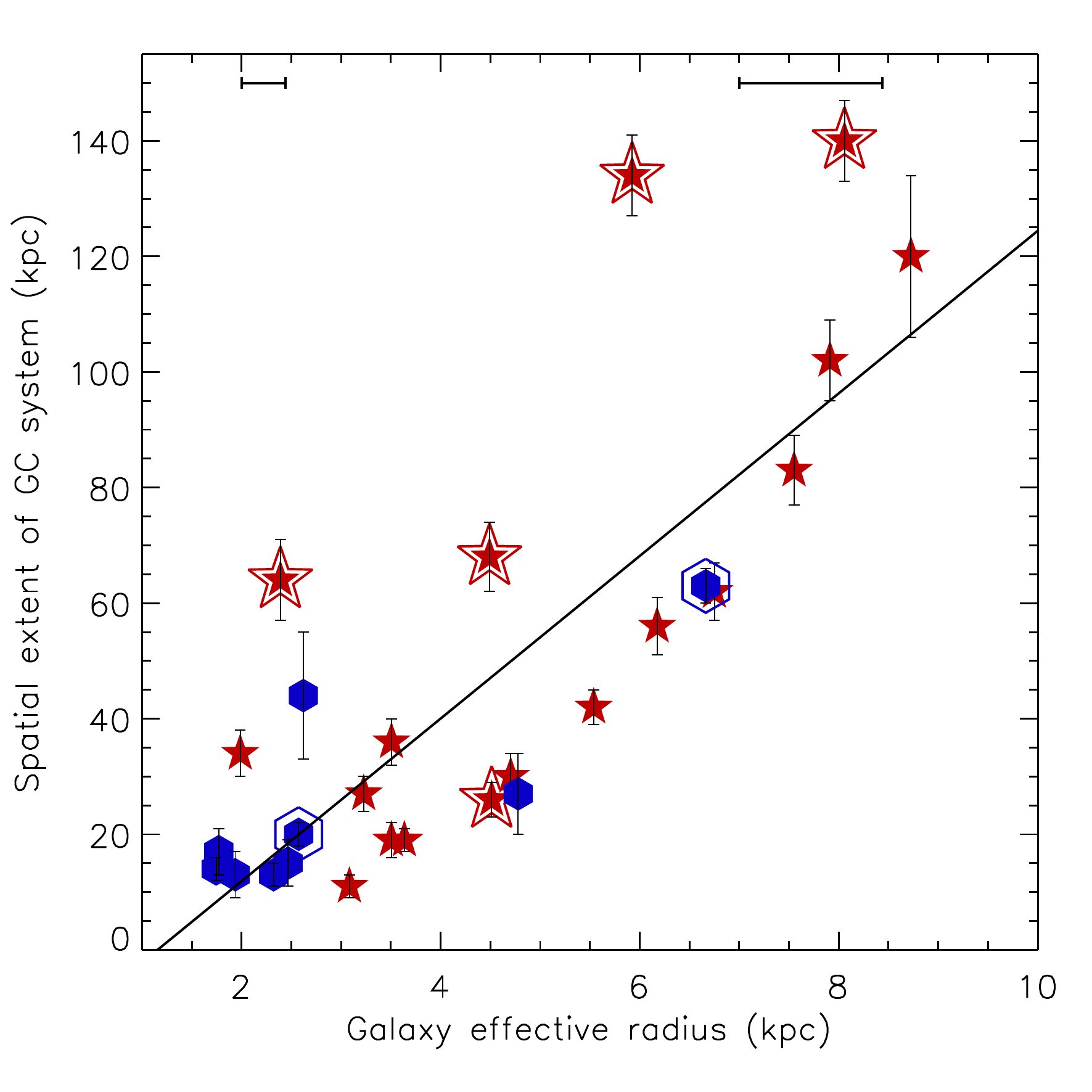}
\caption{Radial extent of GC system versus galaxy effective radius for early-type galaxies. The symbols  shown in the figure are same as in the Figure \ref{GCext1}. The linear fit given by equation \ref{linRe} is drawn with a solid line. The typical 20 percent error at R$_e$ = 2 and 7 kpc  are shown at the top of the figure. }
\label{GCextRad}
\end{figure}
 
The effective radii for early-type galaxies are taken from \citet{Faber1989}, \citet{Bender1992} and \citet{Cappellari2011}.  The effective radius for NGC 1387 is taken from \citet{de1991} and we have multiple measurements for other galaxies. \citet{Faber1989}  and \citet{Bender1992} estimated  effective radii from de Vaucouleurs fits to the surface brightness profiles. \citet{Cappellari2011} derived the effective radii combining the RC3 and 2MASS determinations, both measurements are based on growth curves. Estimation of the effective radius includes a large error of $\sim$ 20 percent \citep{Cappellari2011}. This error has a greater effect on larger sized galaxies (as shown in Figure \ref{GCextRad}). The priority for effective radius values used here are  \citet{Cappellari2011}, then \citet{Faber1989} and finally \citet{Bender1992}.  The effective radii for the sample galaxies are also recorded in Table \ref{GCext}. Figure \ref{GCextRad} shows the GC system extent versus effective radius for early-type galaxies. As evident from the figure, the GC system extent is larger for greater effective radii. A linear fit is carried out for the sample of 27 galaxies and is represented with a straight line in Figure \ref{GCextRad}. The fitted linear relation between GC  system size and galaxy size is given by: 
 \begin{equation}
y = [(14.1 \pm 2.1) \times x ] - (16.2 \pm 10.1) \\
\label{linRe}
\end{equation}
where {\it x} represents the galaxy effective radius and {\it y} represents the spatial extent of GC system. For a sample of 27 early-type galaxies the extent of a GC system is $\sim$14 times the effective radius of the host galaxy. An advantage of this relation is that it is independent of an assumed mass to light ratio as needed in Section 4.1.  Hence, Figure \ref{GCextRad} provides a better understanding between GC system extent and their host galaxies.

\subsection{GC system effective radius versus galaxy effective radius}

\begin{figure}
\centering
\includegraphics[scale=.5]{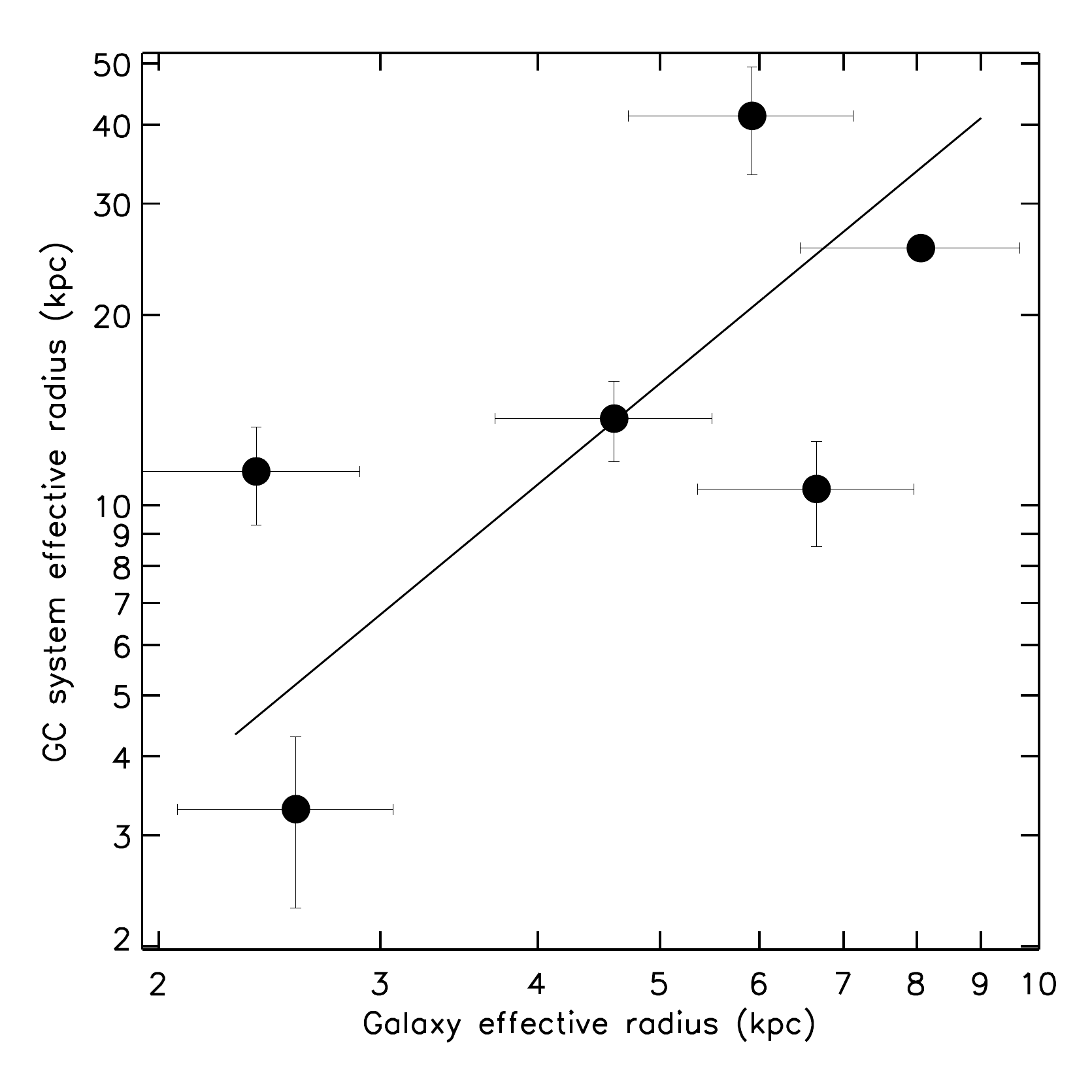}
\caption{GC system effective radius versus galaxy effective radius. The plot displays  six galaxies from the SLUGGS survey. The GC system effective radius is derived from the S\'{e}rsic profile fitted to the radial surface density distribution of GCs and the galaxy effective radius is discussed in Section 4.2. The straight line shown in the figure is given by fitting the data with the bootstrap technique. }
\label{GCReGalaXRe}
\end{figure}

Although we can confirm the correlation of GC spatial extent with galaxy mass as found by \citet{Rhode2007}, we also find evidence that the measurement of the spatial extent is strongly dependent on the quality of the data used.  Thus the \citet{Rhode2007} correlation should be considered more as a general trend than a quantitative relation. A better quantity to use is the effective radius of the GC system, although this has only been measured for a handful of GC systems to date. 

Here we plot GC system effective radius versus galaxy effective radius. The effective radius of the GC system is derived from a S\'{e}rsic profile fitted to the radial GC surface density profile. Most literature studies have used a power law or de Vaucouleurs law (a S\'{e}rsic fit with n fixed to 4) to analyse the GC radial density distribution.  Figure \ref{GCReGalaXRe} shows the plot for six SLUGGS galaxies available with both parameters (recorded in Table \ref{GCGalRe}). We have linearly fitted the data with the bootstrap technique and found:
\begin{equation}
y = [(5.2 \pm 3.7) \times x ]   - ( 8.5 \pm 6.5) \\
\label{linReGal}
\end{equation}
where {\it x} and  {\it y} represent galaxy and GC system effective radius respectively.

The GC system spatial extent (shown in Figure \ref{GCext1} \& \ref{GCextRad}) includes errors mainly from quality of data used. But the GC system effective radius is a more reliable parameter as it is derived from a S\'{e}rsic profile. Hence, we suggest GC system effective radius versus galaxy effective radius as a better version of Figure \ref{GCext1}. 

\begin{table}
\centering
\caption{Effective radius of GC systems from a S\'{e}rsic fit and their host galaxy. The reference for GC system and galaxy effective radii are given in the last column respectively.}
\begin{tabular}{crcc}
\hline
\multicolumn{1}{c}{Galaxy}&\multicolumn{2}{c}{Effective radius}&\multicolumn{1}{c}{Ref.}\\
\hline
NGC   & GC system (kpc) & Stellar light (kpc)&\\ 
\hline\hline
720    &   13.7$\pm$2.2   &   4.60$\pm$0.9   &1,   5 \\
1023  &    3.3$\pm$0.9  &    2.57$\pm$0.5   & 1,   6\\
1407  &    25.5$\pm$1.4  &    8.06$\pm$1.6 &2,   5\\  
2768  &   10.6$\pm$1.8  &    6.66$\pm$1.3  &1,   6\\ 
4278  &   11.3$\pm$1.5  &    2.39$\pm$0.5  &3,   6\\ 
4365  &    41.3$\pm$8.1  &    5.92$\pm$1.2 &4,   6\\ 
\hline
\end{tabular}
\newline
1. This work  2. \citet{Spitler2012}  3. \citet{Usher2013}  4. \citet{Blom2012}  5. \citet{Faber1989} 
\newline6. \citet{Cappellari2011} 
\label{GCGalRe}
\end{table}

\subsection{Ratio of blue to red GC number as a function of host galaxy density}
\citet{Tonini2013} has performed a series of Monte Carlo simulations to study the assembly history of galaxies and the formation of associated GC systems. One prediction is that galaxies in higher density environments are expected to have a higher minor merger/accretion frequency and hence to contain a higher number of accreted blue GCs. According to Tonini's prediction, the ratio of blue to red GCs should be larger for galaxies in higher density environments. 

To quantify the density of environment around a galaxy, we have employed the local density parameter. We use the local environment density as a proxy for the merger history in comparison with \citet{Tonini2013}. \citet{Tully1988} has estimated the local density parameter $\rho$ (in Mpc$ ^{-3} $) for 2367 galaxies in the Nearby Galaxies Catalog. He defined it as the number of galaxies per Mpc$^{-3}$ found around a galaxy within a smoothing length $\sigma$. The density parameter is given by: 
\begin{equation}
\rho = \sum_i C ~\text{exp}~[ -r_i^2/2\sigma^2]
\end{equation}  
where \textit{C} = 1/$(2 \pi \sigma ^2)^{3/2}$ = 0.0635/$\sigma^3$ is a normalization coefficient, {\it r$_i$} is the projected distance towards the {\it i}$^{th}$ galaxy from the central galaxy and the distribution around each galaxy is fitted with a Gaussian profile of half width $\sigma $. The density parameter $\rho$ is the sum over all galaxies excluding the central galaxy. The definition of $\rho$ given above does not take into account the incompleteness of the catalogue at large distances. Hence, the $\rho$ values calculated by \citet{Tully1988} have a large uncertainty factor. The environmental measure should ideally give an indication of the merger/interaction history for individual galaxies. As such a measure is unavailable, we use $\rho$ as a proxy.

\begin{figure}
\centering
\includegraphics[scale=.5]{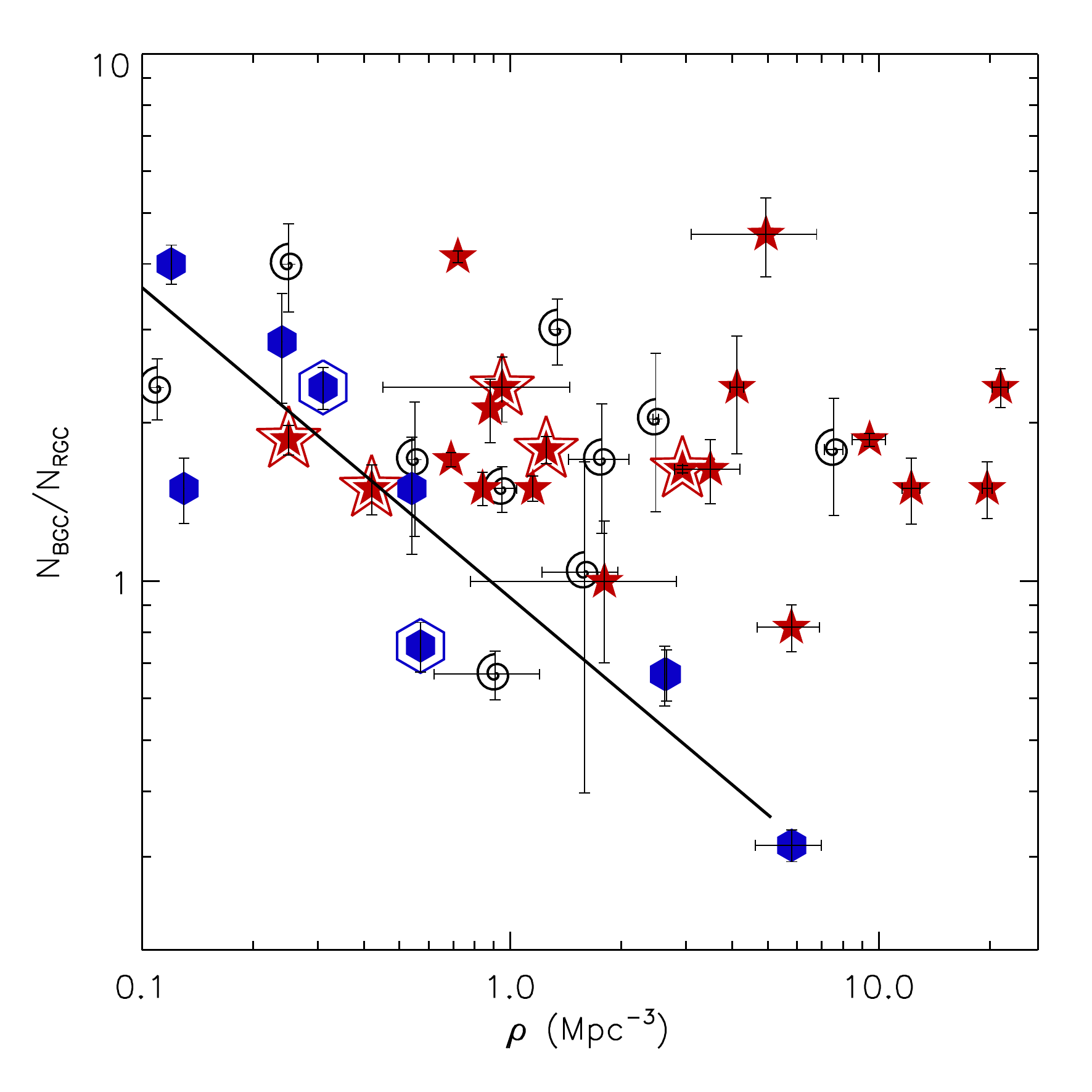}
\caption{Ratio of  blue to red GCs versus density of environment. Spirals, lenticular galaxies and elliptical galaxies are reperesented by spirals, hexagons and stars respectively. The double symbol points are the galaxies from SLUGGS survey and others from the literature.  We did not find any correlation for the spiral and elliptical galaxies. But we found that the ratio of blue to red GCs decreases with the density of environment  for lenticular galaxies (the fitted linear relation is shown as a straight line). Note the presence of two overlapping galaxies, NGC 4754 and  NGC 4762, around the coordinates (2.6, 0.7).}
\label{GCnumRho}
\end{figure}

Using our total sample of galaxies, we searched for confirmation of Tonini's prediction. The galaxies with reliable GC number ratios are selected based on the criteria mentioned in Section 4.1 and are tabulated in Table \ref{GCext} along with $\rho$. Figure \ref{GCnumRho} shows the ratio of blue to red GCs versus the local density parameter $\rho$. Galaxies of different morphological types are shown with different symbols in the figure. It is visible from the figure that there is no strong correlation between the density of environment and the blue to red GC number ratio for elliptical and spiral galaxies. However,  we find an anti-correlation for the lenticular galaxies: the ratio of blue to red GCs  decreases with increasing local density. With the bootstrapping technique, a best fit linear relation to the data points of lenticular galaxies is:
\begin{equation}
 y =  [ (-0.59 \pm 0.07) \times x] + (-0.031 \pm 0.052)
\end{equation}  
where {\it x} and {\it y} represent log($\rho$) and log(N$_\text{BGC}$/N$_\text{RGC}$) respectively. This negative slope implies that there is a higher relative number of red GCs for lenticular galaxies in denser environments. We note that the correlation still holds if the galaxy with the lowest blue to red ratio (NGC 1387) is removed from the sample.

\citet{Cho2012}  studied ten early-type galaxies in low density environments using HST/ACS data. They  compared the properties with cluster galaxies from the ACS Virgo Cluster Survey (ACSVCS, \citealt{Cote2004}). They found that the mean colour of GCs is bluer and also the relative fraction of red GCs is lower for field galaxies than for the cluster galaxies from the ACSVCS. From these trends, they inferred that the galaxy environment has only a weak effect on the formation and mean metallicities of GCs, while the host galaxy luminosity/mass plays a major role. They also suggested a possible explanation for the environmental dependence whereby the GC formation in dense environments is affected by neighbouring galaxies through interaction/accretion which produces a large variation in the GC system properties for galaxies in high density environments.

 \citet{Spitler2008} investigated the relationships of T$_{blue}$ (the number of blue GCs normalised to the host galaxy stellar mass) with host galaxy stellar mass (M$_*$) and local density $\rho$. They studied a sample of 25, mostly elliptical, galaxies with only two lenticular galaxies. Both T$_{blue}$ versus M$_*$ and T$_{blue}$ versus $\rho$, showed positive trends implying a lower T$_{blue}$ value for lower mass galaxies and lower density environments. This supports the idea that the T parameter  has a dependence on either mass and density or possibly both. Trying to disentangle the dependance, they noticed a slight positive trend in a residual plot of T$_{blue}$ versus M$_*$ after removing the dependence of T$_{blue}$ with $\rho$. They argued that GC formation efficiency is highly dependent on host galaxy stellar mass, but much less so on environmental density. 

In our sample, the relative fraction of red GCs in lenticular galaxies  increases with the environmental density (Figure \ref{GCnumRho}), while the same trend is not shown by elliptical or spiral galaxies. \citet{Cho2012} detected an increase in the relative fraction of red GCs with the environmental density. The majority of their galaxies  are also lenticular galaxies, after combining with the ACSVCS data. Hence our result matches with \citet{Cho2012}. Thus from our study, we also confirm the dependence of GC formation on the galaxy environment, at least for lenticular galaxies, as seen in \citet{Cho2012}.  However, \citet{Spitler2008} found the GC formation is dependent on host galaxy mass, and only weakly on environmental density. Their result was based on a sample mostly of elliptical galaxies and does not show any environmental dependence. Similarly the elliptical galaxies in our sample do not show any dependence on environment.  Spiral galaxies in our sample also exhibit an independence of blue to red ratio from their environments.

We find a relatively higher fraction of red GCs in lenticular galaxies residing in high density environments.   Among the various galaxy interactions which can cause variations in GC numbers, as discussed in \citet{Forbes1997}, is tidal stripping which removes the outer halo or blue GCs. For example, NGC 1387 is an S0 galaxy in our sample with the lowest relative fraction of blue to red GCs (N$_\text{BGC}$/N$_\text{RGC}$ = 0.32). The lack of blue GCs could be caused by a tidal interaction between NGC 1387 and NGC 1399. \citet{Bassino2006a, Bassino2006b} observed a low number of blue GCs around NGC 1387 and an overabundance in the direction near to NGC 1399. They proposed this as a case of tidal stripping through which NGC 1399 has stripped away the outer halo of NGC 1387, creating a deficit of blue GCs compared to the red GCs.  Using numerical simulations, \citet{Bekki2003} confirmed an asymmetry in the distribution of blue GCs around NGC 1399 and also suggested the influence of a tidal interaction with the nearby galaxies. We propose that the tidal stripping might be the cause for the observed trend by lenticular galaxies.

\subsection{GC ellipticity  versus galaxy stellar light ellipticity}

\begin{figure}
\centering
\includegraphics[scale=.47]{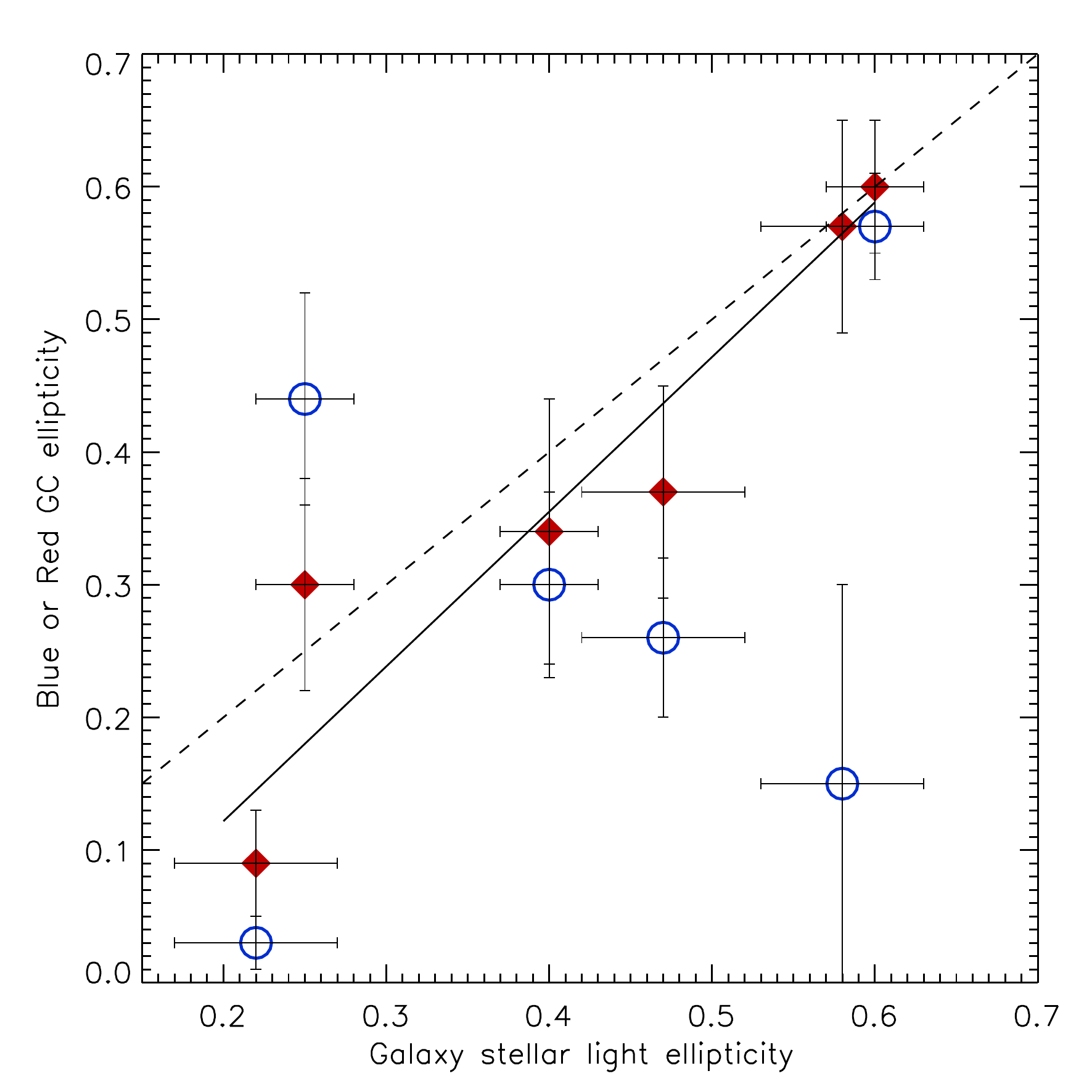}
\caption{GC ellipticity versus galaxy stellar light ellipticity. The plot shows the relation between ellipticities of  blue GCs or red GCs versus galaxy stellar light for six galaxies, recorded in Table \ref{ellipticity}. The blue and the red GCs are represented in blue open circles and red filled diamonds respectively. A linear fit to the red GCs is drawn as a solid line and a one-to-one relation is shown as a dashed line. }
\label{GCellipt}
\end{figure}

Figure \ref{GCellipt} shows the relation between ellipticity for GC subpopulations and for galaxy stellar light. The ellipticity values of galaxy stellar light are derived by fitting ellipses on the radial light distribution and the GC subpopulations are estimated from the azimuthal distribution of GCs. Most literature studies have examined the azimuthal distribution of the total GC system and not for individual GC subpopulations. Hence we have accessible values for GC subpopulation ellipticity for only a handful of galaxies. Table \ref{ellipticity} displays the ellipticity values for blue and red GCs and the galaxy stellar light for six available galaxies. 

We observe a positive correlation between the ellipticity for red GCs and galaxy stellar light (Figure \ref{GCellipt}). But the distribution of blue GCs shows no trend. Using the bootstrap technique, we are able to fit a linear relation to the red GCs:
\begin{equation}
y= [(1.1 \pm 0.2) \times x] + (-0.1 \pm 0.1)
\end{equation}
where {\it x} and {\it y} are ellipticity for galaxy stellar light and red GCs. This linear relation between the galaxy stellar light and red GCs is also consistent with the one-to-one relation, which is also drawn in Figure \ref{GCellipt}. This suggests that both the red GC subpopulation and the underlying stellar populations share a common evolutionary history (see also \citealt{Forbes2012a}). This supports the GC formation scenarios which predict the red GC subpopulation have originated along with the majority of galaxy stars. These scenarios suggested that the blue GC subpopulation formed before the red GCs.   

Recently, \citet{Park2013} studied the ellipticities of blue and red GC subpopulations and the host galaxy stellar light in 23 early-type galaxies using HST/ACS Virgo Cluster Survey. They found a tight correlation between the  ellipticities of the red GC subpopulation and the galaxy stellar light, while a less tight trend for the blue GC subpopulation. Thus their findings support our results from a smaller sample with wider-field data.



\begin{table}
\centering
\caption{Ellipticity values for GC subpopulations and their respective galaxy stellar light for the six galaxies. The reference for GCs and galaxy stellar light are given in the last column.}
\begin{tabular}{ccccc}
\hline
\multicolumn{1}{c}{Galaxy}&\multicolumn{3}{c}{Ellipticity}&\multicolumn{1}{c}{Ref.}\\
\hline
NGC   & Blue GCs & Red GCs  & Stellar light&\\ 
\hline\hline
720 & 0.26$\pm$0.06 & 0.37$\pm$0.08 & 0.47$\pm$0.05 & 1, 5\\ 
1023&0.15$\pm$0.15 & 0.57$\pm$0.08 & 0.58$\pm$0.05 &1, 5\\ 
2768 &0.57$\pm$0.04&  0.60$\pm$0.05 & 0.60$\pm$0.03 &1, 5 \\ 
4365 &  0.44$\pm$0.08 & 0.30$\pm$0.08 & 0.25$\pm$0.03&2, 2\\ 
4486 &  0.30$\pm$0.07 & 0.34$\pm$0.10  & 0.40$\pm$0.05 &3, 3\\ 
4649&   0.03$\pm$0.02 & 0.09$\pm$0.04  & 0.22$\pm$0.05 & 4, 4\\ 
\hline
\end{tabular}
\newline
1. This work  2. \citet{Blom2012}   3. \citet{Strader2011}  \newline 4. \citet{Lee2008} 5. \citet{Paturel2003}
\label{ellipticity}
\end{table}

\section{Conclusions}
We have carried out a detailed study of GC systems in three early-type galaxies: NGC 720, NGC 1023 and NGC 2768. The galaxies were observed in multi-band wide-field images using the 8-meter Subaru Telescope, the 3.6-meter Canada France Hawaii Telescope and the 2.4-meter Hubble Space Telescope. The main conclusions are discussed below. 
\begin{enumerate}
\item The spatial extent of the GC systems of NGC 720, NGC 1023 and NGC 2768 are estimated as 68 $\pm$ 6, 20 $\pm$ 2 and 63 $\pm$ 3 kpc respectively. The spatial extent matches well with the literature for NGC 1023 and we provide a first estimate of the GC system extent for NGC 720 and NGC 2768.
\item The radial surface densities of GCs are fitted with S\'{e}rsic  profiles. From the S\'{e}rsic fits, we estimated the effective radii for the GC systems of NGC 720, NGC 1023 and NGC 2768 are 13.7 $\pm$ 2.2, 3.3 $\pm$ 0.9 and 10.6 $\pm$ 1.8 kpc respectively. 
\item Colour magnitude diagrams show bimodal colour distributions of GCs in all three galaxies with greater than 99.99 percent probability in all three galaxies. 
\item The total number of GCs are estimated as 1489 $\pm$ 96, 548 $\pm$ 59 and 744 $\pm$ 68 for NGC 720, NGC 1023 and NGC 2768 respectively. The S$_N$ values for the corresponding galaxies are 3.2 $\pm$ 0.2, 1.8 $\pm$ 0.2 and 1.3 $\pm$ 0.1. 
\item The peak colour of the blue and red globular cluster subpopulation agrees with the globular cluster colour - host galaxy luminosity relation \citep{Peng2006,Faifer2011}. This strengthens the fact that more massive galaxies have more metal enrichment. 
\item  The position angle of the host galaxy matches with both the blue and red subpopulation in all three galaxies. Ellipticity values of the host galaxies match better with the red subpopulation than the blue subpopulation for all three galaxies. 
\end{enumerate}

We  discuss five global relationships between the  host galaxy and the GC system. We found that the spatial extent of a GC system is dependent on the host galaxy stellar mass/luminosity and the effective radius of the galaxy. Knowing the host galaxy luminosity, or the size of the galaxy, can therefore provide an estimation of the extent of the GC system. The extent of a GC system is determined to be $\sim$ 14 times the effective radius of the host galaxy. The spatial extent of GC systems in elliptical and lenticular galaxies show a strong dependence on host galaxy stellar mass, but not for spiral galaxies. 

We have analysed the relation between ellipticities for blue and red GC subpopulations and galaxy stellar light for a sample of six galaxies. The ellipticity for the red GC subpopulation appears to be correlated with the galaxy stellar light ellipticity for this sample. We support the view that the red GCs and the galaxy stellar light have a coeval formation.  This result from a small sample of 6 galaxies is supported by \citet{Park2013}.

We have also found that the relative fraction of blue to red GCs decreases with galaxy environment density for lenticular galaxies. This result is in general agreement with the observations of \citet{Cho2012} and in disagreement with the predictions of \citet{Tonini2013}. We did not observe any specific trend for elliptical (supporting \citealt{Spitler2008}) and spiral galaxies with galaxy environment density.  An interaction between galaxies, which can decrease the blue GC number in cluster environments, is tidal stripping. Through tidal effects, the outer halo (containing the blue GCs) of the small galaxy may stripped away giving a lower fraction of blue to red GCs \citep{Forbes1997, Bassino2006a, Bassino2006b} for lenticular galaxies in cluster environments. 

\section*{Acknowledgments}
We thank the anonymous referee for his/her careful reading of the manuscript and the valuable feedbacks. We thank the members of SAGES group for the support and enlightening discussions. We thank N. Pastorello for the careful reading of the manuscript. We also acknowledge the useful comments from C. Tonini, C. Blom, C. Usher, V. Pota and  N. Pastorello. With great pleasure we thank F. Faifer for providing us with the Gemini/GMOS data. SSK thanks the Swinburne University for the SUPRA fellowship. DAF thanks the ARC for support via DP-130100388. This research is supported in part by National Science Foundation grants AST-0909237 and AST-1211995. This paper was based in part on data collected at Subaru Telescope, which is operated by the National Astronomical Observatory of Japan. This paper uses data products produced by the OIR Telescope Data Center, supported by the Smithsonian Astrophysical Observatory.  Based on observations made with the NASA/ESA Hubble Space Telescope, and obtained from the Hubble Legacy Archive, which is a collaboration between the Space Telescope Science Institute (STScI/NASA), the Space Telescope European Coordinating Facility (ST-ECF/ESA) and the Canadian Astronomy Data Centre (CADC/NRC/CSA). Based on observations obtained with MegaPrime/MegaCam, a joint project of CFHT and CEA/DAPNIA, at the Canada-France-Hawaii Telescope (CFHT) which is operated by the National Research Council (NRC) of Canada, the Institut National des Science de l'Univers of the Centre National de la Recherche Scientifique (CNRS) of France, and the University of Hawaii.  This research has made use of the NASA/IPAC Extragalactic Data base (NED) operated by the Jet Propulsion Laboratory, California Institute of Technology, under contract with the National Aeronautics and Space Administration.  We acknowledge the usage of the HyperLeda database (http://leda.univ-lyon1.fr).

\begin{table*}
\caption{Properties of our galaxy sample. The top part of the table includes data for galaxies in the SLUGGS survey and the bottom part of the table lists other literature galaxies. Morphological type is taken from NED. The distances are obtained from \citet{Cappellari2011} if available, otherwise from NED.  The total visual magnitude for the galaxies is taken from \citet{de1991} and hence we derive the absolute magnitude, M$_V^T$. The distance, absolute magnitude and the mass to light ratio (given by \citealt{Zepf1993}) are incorporated to determine galaxy stellar mass (M$_\star$). GC numbers (N$_\text{GC}$) is taken from different references as recorded in the footnote.  N$_\text{BGC}$/N$_\text{RGC}$ represents the ratio of blue to red GCs. The reference corresponding to galaxy effective radii are also mentioned in the footnote. The local density parameter is taken from Tully (1988). }
\begin{tabular}{rlrrrlrcccr}
\hline
  \multicolumn{1}{r}{NGC} &
  \multicolumn{1}{c}{Type}&
  \multicolumn{1}{c}{D} &
  \multicolumn{1}{r}{V$_T^0$} &
  \multicolumn{1}{c|}{M$_V^T$} &
  \multicolumn{1}{c}{log(M$_\star$)} &
  \multicolumn{1}{r}{GCExt} &
  \multicolumn{1}{c}{N$_\text{GC}$} &
  \multicolumn{1}{c}{N$_\text{BGC}$/N$_\text{RGC}$} &  
  \multicolumn{1}{c}{R$_\text{e}$} & 
 \multicolumn{1}{c}{$\rho$} \\
 & & (Mpc) & (mag) &(mag)&(M$_\odot$) & (kpc)&  &&(kpc) &  (Mpc$^{-3}$)\\
\hline\hline
      720 &     E5   &    23.44  &    10.17  &     -21.68     &          11.604    &         68 $\pm$  6  &     1584 $\pm$ 190$^a$     & 1.85   &  4.60$^x$   &       0.25 \\
       821 &    E6   &    23.40  &    10.79  &     -21.06     &          11.354    &         26 $\pm$  3  &      320 $\pm$  45$^b$      &  2.33    &4.51$^z$   &       0.95 \\
      1023 &    S0   &    11.10  &     9.15  &     -21.07     &          11.243    &         20 $\pm$  2  &      572 $\pm$  94$^a$      &   0.75   &2.57$^z$   &       0.57 \\
      1407 &    E0   &    23.11  &     9.74  &     -22.08     &          11.764    &        140 $\pm$  7  &     6400 $\pm$ 700$^c$      &  1.50   & 8.06$^x$   &       0.42 \\
      2768 &    S0   &    21.80  &     9.78  &     -21.91     &          11.578    &         63 $\pm$  3  &      714 $\pm$ 162$^a$      &    2.33  &6.66$^z$   &       0.31 \\
      4278 &    E1   &    15.60  &    10.07  &     -20.90     &          11.290    &         64 $\pm$  7  &     1700 $\pm$ 100$^d$      &  1.78   & 2.39$^z$   &       1.25 \\
      4365 &    E3   &    23.30  &     9.54  &     -22.30     &          11.851    &        134 $\pm$  7  &     6450 $\pm$ 110$^e$      &   1.63   &5.92$^z$   &       2.93 \\
\hline
       891 &    Sb   &     8.36  &     8.82  &     -20.79     &          11.034    &          9 $\pm$  3  &       70 $\pm$  20$^f$     &   1.70   &4.14$^z$   &       0.55 \\
      1052 &    Sb   &    19.60  &    10.44  &     -21.02     &          11.341    &         19 $\pm$  3  &      400 $\pm$ 120$^g$      & 1.00  &  3.50$^x$   &       1.80 \\
      1055 &    Sb   &    16.30  &    10.09  &     -20.97     &          11.106    &         26 $\pm$  7  &      210 $\pm$  40$^h$      &  4.00    &5.34$^z$   &       0.25 \\
      1316 &   E&    20.14  &     8.53  &     -22.99     &          11.526    &         62 $\pm$  5  &      636 $\pm$  35$^i$      &   1.50   &6.75$^z$   &       1.15 \\
      1379 &    E0   &    17.71  &    10.99  &     -20.25     &          11.032    &         19 $\pm$  2  &      225 $\pm$  23$^j$      &   0.82  & 3.64$^x$   &       5.79 \\
      1387 &    S0   &    17.24  &    10.72  &     -20.46     &          10.998    &         14 $\pm$  2  &      390 $\pm$  27$^j$    &    0.32  &1.75$^t$   &       5.80 \\
      1427 &    E3   &    19.35  &    10.91  &     -20.52     &          11.141    &         11 $\pm$  2  &      470 $\pm$  80$^k$      &  4.56   & 3.08$^x$   &       4.94 \\
      2683 &    Sb   &     7.70  &     8.97  &     -20.46     &          10.902    &          9 $\pm$  3  &      120 $\pm$  40$^l$      &    2.03  &2.10$^z$   &       2.48 \\
      3258 &    E1   &    32.10 &    11.30 &      -21.23  &           11.425     &        -                    &       6000 $\pm$ 150$^u$  &   3.2   &   4.26$^x$  &   0.72\\
      3268 &    E2   &    34.80  &    11.30 &     -21.41   &            11.495   &         -                    &      4750  $\pm$ 150$^u$  &  1.6 &    6.08$^x$   &    0.69\\
      3379 &    E1   &    10.30  &     9.24  &     -20.82     &          11.262    &         34 $\pm$  4  &      270 $\pm$  68$^m$      &  2.33   & 1.99$^z$   &       4.12 \\
      3384 &    S0   &    11.30  &     9.84  &     -20.43     &          10.983    &         17 $\pm$  4  &      120 $\pm$  30$^n$      &  1.50    &1.77$^z$   &       0.54 \\
      3556 &    Sb   &     7.10  &     9.26  &     -20.00     &          10.629    &         20 $\pm$  4  &      290 $\pm$  80$^l$      &   1.70   &3.00$^z$   &       1.77 \\
      3585 &    E6   &    18.30  &     9.75  &     -21.56     &          11.557    &         36 $\pm$  4  &      550 $\pm$  55$^o$      &  -    &3.51$^x$   &       0.12 \\
      4013 &    Sb   &    15.10  &    10.52  &     -20.37     &          10.867    &         14 $\pm$  5  &      140 $\pm$  20$^f$      &   3.00   &3.42$^z$   &       1.34 \\
      4157 &    Sb   &    14.70  &    10.44  &     -20.40     &          10.876    &         21 $\pm$  4  &       80 $\pm$  20$^l$      &   1.78   &2.71$^z$   &       7.55 \\
      4261 &    E2   &     30.80  &    10.39 &    -22.05    &            11.753   &         -                       &    1242 $\pm$ 90$^v$    & 1.50     &  5.67$^z$  &   0.84\\   
      4374 &    E1   &    18.50  &     9.07  &     -22.27     &          11.838    &         30 $\pm$  4  &     1775 $\pm$ 150$^p$      &   2.33  & 4.70$^z$   &      21.38 \\
      4406 &    E3   &    16.70  &     8.84  &     -22.27     &          11.841    &         83 $\pm$  6  &     2900 $\pm$ 415$^m$      &    1.50  &7.55$^z$   &      12.25 \\
      4472 &    E2   &    17.10  &     8.38  &     -22.78     &          12.046    &        102 $\pm$  7  &     5900 $\pm$ 721$^q$      &   1.50   &7.91$^z$   &      19.68 \\
      4594 &    Sa   &     9.80  &     7.55  &     -22.41     &          11.775    &         54 $\pm$  5  &     1900 $\pm$ 189$^m$      &  1.50   & 7.28$^y$   &       0.95 \\
      4636 &    E0   &    14.30  &     9.51  &     -21.27     &          11.439    &         56 $\pm$  5  &     4200 $\pm$ 120$^r$     &   1.86  & 6.17$^z$   &       9.44 \\
      4649 &    E2   &     17.30 &        8.75 &   -22.44    &            11.910   &         42 $\pm$  3 &       3600  $\pm$ 500$^t$           &    1.67 & 5.54$^z$   &       3.49\\
      4754 &    S0   &    16.10  &    10.43  &     -20.60     &          11.054    &         15 $\pm$  4  &      115 $\pm$  15$^n$      &  0.67   & 2.47$^z$   &       2.62 \\
      4762 &    S0   &    22.60  &    10.16  &     -21.61     &          11.457    &         27 $\pm$  7  &      270 $\pm$  30$^n$      &   0.67   &4.78$^z$   &       2.65 \\
      5812 &    E0   &    27.95  &    10.89  &     -21.34     &          11.469    &         27 $\pm$  3  &      400 $\pm$  40$^o$       &   -   &3.23$^x$   &       0.19 \\
      5813 &    E1   &    31.30  &    10.48  &     -22.00     &          11.731    &        120 $\pm$ 14  &     2900 $\pm$ 400$^n$      & 2.13   &  8.73$^z$   &       0.88 \\ 
      5866 &    S0   &    14.90  &     9.99  &     -20.88     &          11.163    &         44 $\pm$ 11  &      340 $\pm$  80$^n$      &  2.85    &2.62$^z$   &       0.24 \\
      7331 &    Sb   &    13.10  &     8.75  &     -21.84     &          11.452    &         18 $\pm$  4  &      210 $\pm$ 130$^l$      &   1.04   &3.91$^z$   &       1.59 \\
      7332 &    S0   &    23.00  &    11.06  &     -20.75     &          11.112    &         13 $\pm$  4  &      175 $\pm$  15$^h$     &  4.00   & 1.94$^z$   &       0.12 \\
      7339 &    Sbc  &    22.40  &    11.42  &     -20.33     &          10.850    &         10 $\pm$  3  &       75 $\pm$  10$^h$      &   2.33  & 2.66$^z$   &       0.11 \\
      7457 &    S0   &    13.20  &    10.93  &     -19.67     &          10.682    &         13 $\pm$  2  &      210 $\pm$  30$^s$      &   1.50   &2.32$^z$   &       0.13 \\
      7814 &    Sb   &    17.17  &    10.20  &     -20.97     &          11.107    &         13 $\pm$  4  &      190 $\pm$  20$^q$      &    0.67  &3.39$^z$   &       0.91 \\
\hline
\end{tabular}
\\
 References : {\it a}. This work, {\it b}. \citet{Spitler2008}, {\it c}. \citet{Forbes2011}, {\it d}. \citet{Usher2013}, {\it e}. \citet{Blom2012}, {\it f}. \citet{Rhode2010}, {\it g}. \citet{Forbes2001}, {\it h}. \citet{Young2012}, {\it i}. \citet{Richtler2012}, {\it j}. \citet{Bassino2006b}, {\it k}. \citet{Forte2001}, {\it l}. \citet{Rhode2007}, {\it m}. \citet{Rhode2004}, {\it n}. \citet{Hargis2012}, {\it o}.  \citet{Lane2013}, {\it p}. \citet{Gomez2004}, {\it q}. \citet{Rhode2003}, {\it r}. \citet{Dirsch2005}, {\it s}. \citet{Hargis2011}, {\it t}. \citet{Lee2008}, {\it u}. \citet{Bassino2008}, {\it v}. \citet{Bonfini2012},  {\it x}. \citet{Faber1989},  {\it y}. \citet{Bender1992}, {\it z}.  \citet{Cappellari2011}.
\label{GCext}
\end{table*}

\bibliographystyle{mn2e}

\bibliography{reference}

\begin{thebibliography}{93}
\expandafter\ifx\csname natexlab\endcsname\relax\def\natexlab#1{#1}\fi

\bibitem[{{Ashman} {et~al}\mbox{.}(1994){Ashman}, {Bird}, \&
  {Zepf}}]{Ashman1994}
{Ashman} K.~M., {Bird} C.~M., {Zepf} S.~E., 1994, \aj, 108, 2348

\bibitem[{{Ashman} \& {Zepf}(1992)}]{Ashman1992}
{Ashman} K.~M., {Zepf} S.~E., 1992, \apj, 384, 50

\bibitem[{{Bassino} {et~al}\mbox{.}(2006{\natexlab{a}}){Bassino}, {Faifer},
  {Forte}, {Dirsch}, {Richtler}, {Geisler}, \& {Schuberth}}]{Bassino2006a}
{Bassino} L.~P., {Faifer} F.~R., {Forte} J.~C., {Dirsch} B., {Richtler} T.,
  {Geisler} D., {Schuberth} Y., 2006{\natexlab{a}}, \aap, 451, 789

\bibitem[{{Bassino} {et~al}\mbox{.}(2006{\natexlab{b}}){Bassino}, {Richtler},
  \& {Dirsch}}]{Bassino2006b}
{Bassino} L.~P., {Richtler} T., {Dirsch} B., 2006{\natexlab{b}}, \mnras, 367,
  156

\bibitem[{{Bassino} {et~al}\mbox{.}(2008){Bassino}, {Richtler}, \&
  {Dirsch}}]{Bassino2008}
{Bassino} L.~P., {Richtler} T., {Dirsch} B., 2008, \mnras, 386, 1145

\bibitem[{{Bekki} {et~al}\mbox{.}(2003){Bekki}, {Forbes}, {Beasley}, \&
  {Couch}}]{Bekki2003}
{Bekki} K., {Forbes} D.~A., {Beasley} M.~A., {Couch} W.~J., 2003, \mnras, 344,
  1334

\bibitem[{{Bender} {et~al}\mbox{.}(1992){Bender}, {Burstein}, \&
  {Faber}}]{Bender1992}
{Bender} R., {Burstein} D., {Faber} S.~M., 1992, \apj, 399, 462

\bibitem[{{Bertin}(2006)}]{Bertin2006}
{Bertin} E., 2006, in Astronomical Society of the Pacific Conference Series,
  Vol. 351, Astronomical Data Analysis Software and Systems XV, {Gabriel} C.,
  {Arviset} C., {Ponz} D., {Enrique} S., eds., p. 112

\bibitem[{{Bertin} \& {Arnouts}(1996)}]{Bertin1996}
{Bertin} E., {Arnouts} S., 1996, \aaps, 117, 393

\bibitem[{{Bertin} {et~al}\mbox{.}(2002){Bertin}, {Mellier}, {Radovich},
  {Missonnier}, {Didelon}, \& {Morin}}]{Bertin2002}
{Bertin} E., {Mellier} Y., {Radovich} M., {Missonnier} G., {Didelon} P.,
  {Morin} B., 2002, in Astronomical Society of the Pacific Conference Series,
  Vol. 281, Astronomical Data Analysis Software and Systems XI, {Bohlender}
  D.~A., {Durand} D., {Handley} T.~H., eds., p. 228

\bibitem[{{Blom} {et~al}\mbox{.}(2012){Blom}, {Spitler}, \&
  {Forbes}}]{Blom2012}
{Blom} C., {Spitler} L.~R., {Forbes} D.~A., 2012, \mnras, 420, 37

\bibitem[{{Bonfini} {et~al}\mbox{.}(2012){Bonfini}, {Zezas}, {Birkinshaw},
  {Worrall}, {Fabbiano}, {O'Sullivan}, {Trinchieri}, \& {Wolter}}]{Bonfini2012}
{Bonfini} P., {Zezas} A., {Birkinshaw} M., {Worrall} D.~M., {Fabbiano} G.,
  {O'Sullivan} E., {Trinchieri} G., {Wolter} A., 2012, \mnras, 421, 2872

\bibitem[{{Boulade} {et~al}\mbox{.}(2003){Boulade}, {Charlot}, {Abbon}, {Aune},
  {Borgeaud}, {Carton}, {Carty}, {Da Costa}, {Deschamps}, {Desforge},
  {Eppell{\'e}}, {Gallais}, {Gosset}, {Granelli}, {Gros}, {de Kat}, {Loiseau},
  {Ritou}, {Rouss{\'e}}, {Starzynski}, {Vignal}, \& {Vigroux}}]{Boulade2003}
{Boulade} O. {et~al.}, 2003, in Society of Photo-Optical Instrumentation
  Engineers (SPIE) Conference Series, Vol. 4841, Society of Photo-Optical
  Instrumentation Engineers (SPIE) Conference Series, {Iye} M., {Moorwood}
  A.~F.~M., eds., pp. 72--81

\bibitem[{{Brodie} \& {Huchra}(1991)}]{Brodie1991}
{Brodie} J.~P., {Huchra} J.~P., 1991, \apj, 379, 157

\bibitem[{{Brodie} {et~al}\mbox{.}(2011){Brodie}, {Romanowsky}, {Strader}, \&
  {Forbes}}]{Brodie2011}
{Brodie} J.~P., {Romanowsky} A.~J., {Strader} J., {Forbes} D.~A., 2011, \aj,
  142, 199

\bibitem[{{Brodie} \& {Strader}(2006)}]{Brodie2006}
{Brodie} J.~P., {Strader} J., 2006, \araa, 44, 193

\bibitem[{{Brodie} {et~al}\mbox{.}(2012){Brodie}, {Usher}, {Conroy}, {Strader},
  {Arnold}, {Forbes}, \& {Romanowsky}}]{Brodie2012}
{Brodie} J.~P., {Usher} C., {Conroy} C., {Strader} J., {Arnold} J.~A., {Forbes}
  D.~A., {Romanowsky} A.~J., 2012, \apjl, 759, L33

\bibitem[{{Buote} \& {Canizares}(1994)}]{Buote1994}
{Buote} D.~A., {Canizares} C.~R., 1994, \apj, 427, 86

\bibitem[{{Buote} \& {Canizares}(1996)}]{Buote1996}
{Buote} D.~A., {Canizares} C.~R., 1996, \apj, 468, 184

\bibitem[{{Buote} \& {Canizares}(1997)}]{Buote1997}
{Buote} D.~A., {Canizares} C.~R., 1997, \apj, 474, 650

\bibitem[{{Buote} {et~al}\mbox{.}(2002){Buote}, {Jeltema}, {Canizares}, \&
  {Garmire}}]{Buote2002}
{Buote} D.~A., {Jeltema} T.~E., {Canizares} C.~R., {Garmire} G.~P., 2002, \apj,
  577, 183

\bibitem[{{Capaccioli} {et~al}\mbox{.}(1986){Capaccioli}, {Lorenz}, \&
  {Afanasjev}}]{Capaccioli1986}
{Capaccioli} M., {Lorenz} H., {Afanasjev} V.~L., 1986, \aap, 169, 54

\bibitem[{{Cappellari} {et~al}\mbox{.}(2011){Cappellari}, {Emsellem},
  {Krajnovi{\'c}}, {McDermid}, {Scott}, {Verdoes Kleijn}, {Young}, {Alatalo},
  {Bacon}, {Blitz}, {Bois}, {Bournaud}, {Bureau}, {Davies}, {Davis}, {de
  Zeeuw}, {Duc}, {Khochfar}, {Kuntschner}, {Lablanche}, {Morganti}, {Naab},
  {Oosterloo}, {Sarzi}, {Serra}, \& {Weijmans}}]{Cappellari2011}
{Cappellari} M. {et~al.}, 2011, \mnras, 413, 813

\bibitem[{{Cho} {et~al}\mbox{.}(2012){Cho}, {Sharples}, {Blakeslee}, {Zepf},
  {Kundu}, {Kim}, \& {Yoon}}]{Cho2012}
{Cho} J., {Sharples} R.~M., {Blakeslee} J.~P., {Zepf} S.~E., {Kundu} A., {Kim}
  H.-S., {Yoon} S.-J., 2012, \mnras, 422, 3591

\bibitem[{{Cortesi} {et~al}\mbox{.}(2011){Cortesi}, {Merrifield}, {Arnaboldi},
  {Gerhard}, {Martinez-Valpuesta}, {Saha}, {Coccato}, {Bamford}, {Napolitano},
  {Das}, {Douglas}, {Romanowsky}, {Kuijken}, {Capaccioli}, \&
  {Freeman}}]{Cortesi2011}
{Cortesi} A. {et~al.}, 2011, \mnras, 414, 642

\bibitem[{{C{\^o}t{\'e}} {et~al}\mbox{.}(2004){C{\^o}t{\'e}}, {Blakeslee},
  {Ferrarese}, {Jord{\'a}n}, {Mei}, {Merritt}, {Milosavljevi{\'c}}, {Peng},
  {Tonry}, \& {West}}]{Cote2004}
{C{\^o}t{\'e}} P., {Blakeslee} J.~P., {Ferrarese} L., {Jord{\'a}n} A., {Mei}
  S., {Merritt} D., {Milosavljevi{\'c}} M., {Peng} E.~W., {Tonry} J.~L., {West}
  M.~J., 2004, \apjs, 153, 223

\bibitem[{{C{\^o}t{\'e}} {et~al}\mbox{.}(1998){C{\^o}t{\'e}}, {Marzke}, \&
  {West}}]{Cote1998}
{C{\^o}t{\'e}} P., {Marzke} R.~O., {West} M.~J., 1998, \apj, 501, 554

\bibitem[{{C{\^o}t{\'e}} {et~al}\mbox{.}(2000){C{\^o}t{\'e}}, {Marzke}, {West},
  \& {Minniti}}]{Cote2000}
{C{\^o}t{\'e}} P., {Marzke} R.~O., {West} M.~J., {Minniti} D., 2000, \apj, 533,
  869

\bibitem[{{C{\^o}t{\'e}} {et~al}\mbox{.}(2002){C{\^o}t{\'e}}, {West}, \&
  {Marzke}}]{Cote2002}
{C{\^o}t{\'e}} P., {West} M.~J., {Marzke} R.~O., 2002, \apj, 567, 853

\bibitem[{{Crocker} {et~al}\mbox{.}(2008){Crocker}, {Bureau}, {Young}, \&
  {Combes}}]{Crocker2008}
{Crocker} A.~F., {Bureau} M., {Young} L.~M., {Combes} F., 2008, \mnras, 386,
  1811

\bibitem[{{de Vaucouleurs} {et~al}\mbox{.}(1991){de Vaucouleurs}, {de
  Vaucouleurs}, {Corwin}, {Buta}, {Paturel}, \& {Fouqu{\'e}}}]{de1991}
{de Vaucouleurs} G., {de Vaucouleurs} A., {Corwin}, Jr. H.~G., {Buta} R.~J.,
  {Paturel} G., {Fouqu{\'e}} P., 1991, {Third Reference Catalogue of Bright
  Galaxies. Volume I: Explanations and references. Volume II: Data for galaxies
  between 0$^{h}$ and 12$^{h}$. Volume III: Data for galaxies between 12$^{h}$
  and 24$^{h}$.}

\bibitem[{{Dirsch} {et~al}\mbox{.}(2005){Dirsch}, {Schuberth}, \&
  {Richtler}}]{Dirsch2005}
{Dirsch} B., {Schuberth} Y., {Richtler} T., 2005, \aap, 433, 43

\bibitem[{{Elmegreen}(1999)}]{Elmegreen1999}
{Elmegreen} B.~G., 1999, \apss, 269, 469

\bibitem[{{Faber} {et~al}\mbox{.}(1989){Faber}, {Wegner}, {Burstein}, {Davies},
  {Dressler}, {Lynden-Bell}, \& {Terlevich}}]{Faber1989}
{Faber} S.~M., {Wegner} G., {Burstein} D., {Davies} R.~L., {Dressler} A.,
  {Lynden-Bell} D., {Terlevich} R.~J., 1989, \apjs, 69, 763

\bibitem[{{Faifer} {et~al}\mbox{.}(2011){Faifer}, {Forte}, {Norris}, {Bridges},
  {Forbes}, {Zepf}, {Beasley}, {Gebhardt}, {Hanes}, \& {Sharples}}]{Faifer2011}
{Faifer} F.~R., {Forte} J.~C., {Norris} M.~A., {Bridges} T., {Forbes} D.~A.,
  {Zepf} S.~E., {Beasley} M., {Gebhardt} K., {Hanes} D.~A., {Sharples} R.~M.,
  2011, \mnras, 416, 155

\bibitem[{{Forbes} {et~al}\mbox{.}(1997){Forbes}, {Brodie}, \&
  {Grillmair}}]{Forbes1997}
{Forbes} D.~A., {Brodie} J.~P., {Grillmair} C.~J., 1997, \aj, 113, 1652

\bibitem[{{Forbes} {et~al}\mbox{.}(2012{\natexlab{a}}){Forbes}, {Cortesi},
  {Pota}, {Foster}, {Romanowsky}, {Merrifield}, {Brodie}, {Strader}, {Coccato},
  \& {Napolitano}}]{Forbes2012b}
{Forbes} D.~A., {Cortesi} A., {Pota} V., {Foster} C., {Romanowsky} A.~J.,
  {Merrifield} M.~R., {Brodie} J.~P., {Strader} J., {Coccato} L., {Napolitano}
  N., 2012{\natexlab{a}}, \mnras, 426, 975

\bibitem[{{Forbes} {et~al}\mbox{.}(1996){Forbes}, {Franx}, {Illingworth}, \&
  {Carollo}}]{Forbes1996}
{Forbes} D.~A., {Franx} M., {Illingworth} G.~D., {Carollo} C.~M., 1996, \apj,
  467, 126

\bibitem[{{Forbes} {et~al}\mbox{.}(2001){Forbes}, {Georgakakis}, \&
  {Brodie}}]{Forbes2001}
{Forbes} D.~A., {Georgakakis} A.~E., {Brodie} J.~P., 2001, \mnras, 325, 1431

\bibitem[{{Forbes} {et~al}\mbox{.}(2012{\natexlab{b}}){Forbes}, {Ponman}, \&
  {O'Sullivan}}]{Forbes2012a}
{Forbes} D.~A., {Ponman} T., {O'Sullivan} E., 2012{\natexlab{b}}, \mnras, 425,
  66

\bibitem[{{Forbes} {et~al}\mbox{.}(2011){Forbes}, {Spitler}, {Strader},
  {Romanowsky}, {Brodie}, \& {Foster}}]{Forbes2011}
{Forbes} D.~A., {Spitler} L.~R., {Strader} J., {Romanowsky} A.~J., {Brodie}
  J.~P., {Foster} C., 2011, \mnras, 413, 2943

\bibitem[{{Forte} {et~al}\mbox{.}(2001){Forte}, {Geisler}, {Ostrov}, {Piatti},
  \& {Gieren}}]{Forte2001}
{Forte} J.~C., {Geisler} D., {Ostrov} P.~G., {Piatti} A.~E., {Gieren} W., 2001,
  \aj, 121, 1992

\bibitem[{{Forte} {et~al}\mbox{.}(2012){Forte}, {Vega}, \&
  {Faifer}}]{Forte2012}
{Forte} J.~C., {Vega} E.~I., {Faifer} F., 2012, \mnras, 421, 635

\bibitem[{{G{\'o}mez} \& {Richtler}(2004)}]{Gomez2004}
{G{\'o}mez} M., {Richtler} T., 2004, \aap, 415, 499

\bibitem[{{G{\'o}mez} {et~al}\mbox{.}(2001){G{\'o}mez}, {Richtler}, {Infante},
  \& {Drenkhahn}}]{Gomez2001}
{G{\'o}mez} M., {Richtler} T., {Infante} L., {Drenkhahn} G., 2001, \aap, 371,
  875

\bibitem[{{Gwyn}(2008)}]{Gwyn2008}
{Gwyn} S.~D.~J., 2008, \pasp, 120, 212

\bibitem[{{Hargis} \& {Rhode}(2012)}]{Hargis2012}
{Hargis} J.~R., {Rhode} K.~L., 2012, \aj, 144, 164

\bibitem[{{Hargis} {et~al}\mbox{.}(2011){Hargis}, {Rhode}, {Strader}, \&
  {Brodie}}]{Hargis2011}
{Hargis} J.~R., {Rhode} K.~L., {Strader} J., {Brodie} J.~P., 2011, \apj, 738,
  113

\bibitem[{{Harris}(1991)}]{Harris1991}
{Harris} W.~E., 1991, \araa, 29, 543

\bibitem[{{Harris}(2009{\natexlab{a}})}]{Harris2009a}
{Harris} W.~E., 2009{\natexlab{a}}, \apj, 699, 254

\bibitem[{{Harris}(2009{\natexlab{b}})}]{Harris2009b}
{Harris} W.~E., 2009{\natexlab{b}}, \apj, 703, 939

\bibitem[{{Harris} \& {van den Bergh}(1981)}]{Harris1981}
{Harris} W.~E., {van den Bergh} S., 1981, \aj, 86, 1627

\bibitem[{{Hook} {et~al}\mbox{.}(2004){Hook}, {J{\o}rgensen},
  {Allington-Smith}, {Davies}, {Metcalfe}, {Murowinski}, \&
  {Crampton}}]{Hook2004}
{Hook} I.~M., {J{\o}rgensen} I., {Allington-Smith} J.~R., {Davies} R.~L.,
  {Metcalfe} N., {Murowinski} R.~G., {Crampton} D., 2004, \pasp, 116, 425

\bibitem[{{Jordi} {et~al}\mbox{.}(2006){Jordi}, {Grebel}, \&
  {Ammon}}]{Jordi2006}
{Jordi} K., {Grebel} E.~K., {Ammon} K., 2006, \aap, 460, 339

\bibitem[{{Kissler-Patig} {et~al}\mbox{.}(1996){Kissler-Patig}, {Richtler}, \&
  {Hilker}}]{Kissler1996}
{Kissler-Patig} M., {Richtler} T., {Hilker} M., 1996, \aap, 308, 704

\bibitem[{{Kron}(1980)}]{Kron1980}
{Kron} R.~G., 1980, \apjs, 43, 305

\bibitem[{{Kundu} \& {Whitmore}(2001)}]{Kundu2001}
{Kundu} A., {Whitmore} B.~C., 2001, \aj, 122, 1251

\bibitem[{{Kuntschner}(2000)}]{Kuntschner2000}
{Kuntschner} H., 2000, \mnras, 315, 184

\bibitem[{{Lane} {et~al}\mbox{.}(2013){Lane}, {Salinas}, \&
  {Richtler}}]{Lane2013}
{Lane} R.~R., {Salinas} R., {Richtler} T., 2013, \aap, 549, A148

\bibitem[{{Larsen} \& {Brodie}(2000)}]{Larsen2000}
{Larsen} S.~S., {Brodie} J.~P., 2000, \aj, 120, 2938

\bibitem[{{Lee} {et~al}\mbox{.}(2008){Lee}, {Park}, {Kim}, {Hwang}, {Kim}, \&
  {Geisler}}]{Lee2008}
{Lee} M.~G., {Park} H.~S., {Kim} E., {Hwang} H.~S., {Kim} S.~C., {Geisler} D.,
  2008, \apj, 682, 135

\bibitem[{{Liu} {et~al}\mbox{.}(2011){Liu}, {Peng}, {Jord{\'a}n}, {Ferrarese},
  {Blakeslee}, {C{\^o}t{\'e}}, \& {Mei}}]{Liu2011}
{Liu} C., {Peng} E.~W., {Jord{\'a}n} A., {Ferrarese} L., {Blakeslee} J.~P.,
  {C{\^o}t{\'e}} P., {Mei} S., 2011, \apj, 728, 116

\bibitem[{{McLaughlin} {et~al}\mbox{.}(1994){McLaughlin}, {Harris}, \&
  {Hanes}}]{McLaughlin1994}
{McLaughlin} D.~E., {Harris} W.~E., {Hanes} D.~A., 1994, \apj, 422, 486

\bibitem[{{Miyazaki} {et~al}\mbox{.}(2002){Miyazaki}, {Komiyama}, {Sekiguchi},
  {Okamura}, {Doi}, {Furusawa}, {Hamabe}, {Imi}, {Kimura}, {Nakata}, {Okada},
  {Ouchi}, {Shimasaku}, {Yagi}, \& {Yasuda}}]{Miyazaki2002}
{Miyazaki} S. {et~al.}, 2002, \pasj, 54, 833

\bibitem[{{Muratov} \& {Gnedin}(2010)}]{Muratov2010}
{Muratov} A.~L., {Gnedin} O.~Y., 2010, \apj, 718, 1266

\bibitem[{{Ouchi} {et~al}\mbox{.}(2004){Ouchi}, {Shimasaku}, {Okamura},
  {Furusawa}, {Kashikawa}, {Ota}, {Doi}, {Hamabe}, {Kimura}, {Komiyama},
  {Miyazaki}, {Miyazaki}, {Nakata}, {Sekiguchi}, {Yagi}, \&
  {Yasuda}}]{Ouchi2004}
{Ouchi} M. {et~al.}, 2004, \apj, 611, 660

\bibitem[{{Park} \& {Lee}(2013)}]{Park2013}
{Park} H.~S., {Lee} M.~G., 2013, arXiv:1307.5395

\bibitem[{{Paturel} {et~al}\mbox{.}(2003){Paturel}, {Petit}, {Prugniel},
  {Theureau}, {Rousseau}, {Brouty}, {Dubois}, \& {Cambr{\'e}sy}}]{Paturel2003}
{Paturel} G., {Petit} C., {Prugniel} P., {Theureau} G., {Rousseau} J., {Brouty}
  M., {Dubois} P., {Cambr{\'e}sy} L., 2003, \aap, 412, 45

\bibitem[{{Peng} {et~al}\mbox{.}(2006){Peng}, {Jord{\'a}n}, {C{\^o}t{\'e}},
  {Blakeslee}, {Ferrarese}, {Mei}, {West}, {Merritt}, {Milosavljevi{\'c}}, \&
  {Tonry}}]{Peng2006}
{Peng} E.~W., {Jord{\'a}n} A., {C{\^o}t{\'e}} P., {Blakeslee} J.~P.,
  {Ferrarese} L., {Mei} S., {West} M.~J., {Merritt} D., {Milosavljevi{\'c}} M.,
  {Tonry} J.~L., 2006, \apj, 639, 95

\bibitem[{{Peng} {et~al}\mbox{.}(2008){Peng}, {Jord{\'a}n}, {C{\^o}t{\'e}},
  {Takamiya}, {West}, {Blakeslee}, {Chen}, {Ferrarese}, {Mei}, {Tonry}, \&
  {West}}]{Peng2008}
{Peng} E.~W., {Jord{\'a}n} A., {C{\^o}t{\'e}} P., {Takamiya} M., {West} M.~J.,
  {Blakeslee} J.~P., {Chen} C.-W., {Ferrarese} L., {Mei} S., {Tonry} J.~L.,
  {West} A.~A., 2008, \apj, 681, 197

\bibitem[{{Pota} {et~al}\mbox{.}(2013){Pota}, {Forbes}, {Romanowsky}, {Brodie},
  {Spitler}, {Strader}, {Foster}, {Arnold}, {Benson}, {Blom}, {Hargis},
  {Rhode}, \& {Usher}}]{Pota2013}
{Pota} V. {et~al.}, 2013, \mnras, 428, 389

\bibitem[{{Rhode} {et~al}\mbox{.}(2010){Rhode}, {Windschitl}, \&
  {Young}}]{Rhode2010}
{Rhode} K.~L., {Windschitl} J.~L., {Young} M.~D., 2010, \aj, 140, 430

\bibitem[{{Rhode} \& {Zepf}(2003)}]{Rhode2003}
{Rhode} K.~L., {Zepf} S.~E., 2003, \aj, 126, 2307

\bibitem[{{Rhode} \& {Zepf}(2004)}]{Rhode2004}
{Rhode} K.~L., {Zepf} S.~E., 2004, \aj, 127, 302

\bibitem[{{Rhode} {et~al}\mbox{.}(2007){Rhode}, {Zepf}, {Kundu}, \&
  {Larner}}]{Rhode2007}
{Rhode} K.~L., {Zepf} S.~E., {Kundu} A., {Larner} A.~N., 2007, \aj, 134, 1403

\bibitem[{{Rhode} {et~al}\mbox{.}(2005){Rhode}, {Zepf}, \&
  {Santos}}]{Rhode2005}
{Rhode} K.~L., {Zepf} S.~E., {Santos} M.~R., 2005, \apjl, 630, L21

\bibitem[{{Richtler} {et~al}\mbox{.}(2012){Richtler}, {Bassino}, {Dirsch}, \&
  {Kumar}}]{Richtler2012}
{Richtler} T., {Bassino} L.~P., {Dirsch} B., {Kumar} B., 2012, \aap, 543, A131

\bibitem[{{Romanowsky} {et~al}\mbox{.}(2009){Romanowsky}, {Strader}, {Spitler},
  {Johnson}, {Brodie}, {Forbes}, \& {Ponman}}]{Romanowsky2009}
{Romanowsky} A.~J., {Strader} J., {Spitler} L.~R., {Johnson} R., {Brodie}
  J.~P., {Forbes} D.~A., {Ponman} T., 2009, \aj, 137, 4956

\bibitem[{{Sancisi} {et~al}\mbox{.}(1984){Sancisi}, {van Woerden}, {Davies}, \&
  {Hart}}]{Sancisi1984}
{Sancisi} R., {van Woerden} H., {Davies} R.~D., {Hart} L., 1984, \mnras, 210,
  497

\bibitem[{{Sandage} \& {Bedke}(1994)}]{Sandage1994}
{Sandage} A., {Bedke} J., 1994, {The Carnegie Atlas of Galaxies. Volumes I,
  II.}

\bibitem[{{Schlegel} {et~al}\mbox{.}(1998){Schlegel}, {Finkbeiner}, \&
  {Davis}}]{Schlegel1998}
{Schlegel} D.~J., {Finkbeiner} D.~P., {Davis} M., 1998, \apj, 500, 525

\bibitem[{{Shaya} {et~al}\mbox{.}(1996){Shaya}, {Dowling}, {Currie}, {Faber},
  {Ajhar}, {Lauer}, {Groth}, {Grillmair}, {Lynd}, \& {O'Neil}}]{Shaya1996}
{Shaya} E.~J., {Dowling} D.~M., {Currie} D.~G., {Faber} S.~M., {Ajhar} E.~A.,
  {Lauer} T.~R., {Groth} E.~J., {Grillmair} C.~J., {Lynd} R., {O'Neil}, Jr.
  E.~J., 1996, \aj, 111, 2212

\bibitem[{{Sinnott} {et~al}\mbox{.}(2010){Sinnott}, {Hou}, {Anderson},
  {Harris}, \& {Woodley}}]{Sinnott2010}
{Sinnott} B., {Hou} A., {Anderson} R., {Harris} W.~E., {Woodley} K.~A., 2010,
  \aj, 140, 2101

\bibitem[{{Spitler} {et~al}\mbox{.}(2008){Spitler}, {Forbes}, {Strader},
  {Brodie}, \& {Gallagher}}]{Spitler2008}
{Spitler} L.~R., {Forbes} D.~A., {Strader} J., {Brodie} J.~P., {Gallagher}
  J.~S., 2008, \mnras, 385, 361

\bibitem[{{Spitler} {et~al}\mbox{.}(2012){Spitler}, {Romanowsky}, {Diemand},
  {Strader}, {Forbes}, {Moore}, \& {Brodie}}]{Spitler2012}
{Spitler} L.~R., {Romanowsky} A.~J., {Diemand} J., {Strader} J., {Forbes}
  D.~A., {Moore} B., {Brodie} J.~P., 2012, \mnras, 423, 2177

\bibitem[{{Strader} {et~al}\mbox{.}(2011){Strader}, {Romanowsky}, {Brodie},
  {Spitler}, {Beasley}, {Arnold}, {Tamura}, {Sharples}, \&
  {Arimoto}}]{Strader2011}
{Strader} J., {Romanowsky} A.~J., {Brodie} J.~P., {Spitler} L.~R., {Beasley}
  M.~A., {Arnold} J.~A., {Tamura} N., {Sharples} R.~M., {Arimoto} N., 2011,
  \apjs, 197, 33

\bibitem[{{Tonini}(2013)}]{Tonini2013}
{Tonini} C., 2013, \apj, 762, 39

\bibitem[{{Tully}(1988)}]{Tully1988}
{Tully} R.~B., 1988, {Nearby galaxies catalog}

\bibitem[{{Usher} {et~al}\mbox{.}(2012){Usher}, {Forbes}, {Brodie}, {Foster},
  {Spitler}, {Arnold}, {Romanowsky}, {Strader}, \& {Pota}}]{Usher2012}
{Usher} C., {Forbes} D.~A., {Brodie} J.~P., {Foster} C., {Spitler} L.~R.,
  {Arnold} J.~A., {Romanowsky} A.~J., {Strader} J., {Pota} V., 2012, \mnras,
  426, 1475

\bibitem[{{Usher} {et~al}\mbox{.}(2013){Usher}, {Forbes}, {Spitler}, {Brodie},
  {Romanowsky}, {Strader}, \& {Woodley}}]{Usher2013}
{Usher} C., {Forbes} D.~A., {Spitler} L.~R., {Brodie} J.~P., {Romanowsky}
  A.~J., {Strader} J., {Woodley} K.~A., 2013, arXiv1308.6585

\bibitem[{{Wang} {et~al}\mbox{.}(2013){Wang}, {Peng}, {Blakeslee},
  {C{\^o}t{\'e}}, {Ferrarese}, {Jord{\'a}n}, {Mei}, \& {West}}]{Wang2013}
{Wang} Q., {Peng} E.~W., {Blakeslee} J.~P., {C{\^o}t{\'e}} P., {Ferrarese} L.,
  {Jord{\'a}n} A., {Mei} S., {West} M.~J., 2013, \apj, 769, 145

\bibitem[{{Young} {et~al}\mbox{.}(2012){Young}, {Dowell}, \&
  {Rhode}}]{Young2012}
{Young} M.~D., {Dowell} J.~L., {Rhode} K.~L., 2012, \aj, 144, 103

\bibitem[{{Zepf} \& {Ashman}(1993)}]{Zepf1993}
{Zepf} S.~E., {Ashman} K.~M., 1993, \mnras, 264, 611

\end{thebibliography}

\end{document}